\documentclass[useAMS,usenatbib]{mnras}	%Two columns
\usepackage[figuresright]{rotating}
\usepackage{pdflscape} 

\usepackage{graphics}

\usepackage{bigdelim}
\usepackage{bigstrut}

\usepackage{setspace}
\usepackage{soul}

\usepackage{natbib}
\usepackage{mathtools}
\usepackage{amsmath}
\usepackage{amssymb}
\usepackage[subnum]{cases}

\title[CaT strength of GCs]{The WAGGS project - II. The reliability of the calcium triplet as a metallicity indicator in integrated stellar light}

\author[Usher et al.]{Christopher~Usher$^{1}$\thanks{email: c.g.usher@ljmu.ac.uk},
{Thomas Beckwith$^{1,2}$},
{Sabine~Bellstedt$^{3}$},
{Adebusola~Alabi$^{4}$},\newauthor
{Leonie Chevalier$^{3}$},
{Nicola~Pastorello$^{5}$},
{Pierluigi~Cerulo$^{6}$},
{Hannah~S.~Dalgleish$^{1}$},\newauthor
{Amelia~Fraser-McKelvie$^{7}$},
{Sebastian~Kamann$^{1}$},
{Samantha~Penny$^{8}$},
{Caroline~Foster$^{9,10}$},\newauthor
{Richard~McDermid$^{11,12}$},
{Ricardo~P.~Schiavon$^{1}$} 
and {Alexa~Villaume$^{4}$}
\\
$^{1}$Astrophysics Research Institute, Liverpool John Moores University, 146 Brownlow Hill, Liverpool L3 5RF, UK\\
$^{2}$Department of Physics, University of Liverpool, Oliver Lodge, Oxford Street, Liverpool L69 7ZE, UK\\
$^{3}$Centre for Astrophysics \& Supercomputing, Swinburne University of Technology, Hawthorn, VIC 3122, Australia\\
$^{4}$University of California Observatories, 1156 High Street, Santa Cruz, CA 95064, USA\\
$^{5}$Deakin Software and Technology Innovation Laboratory, Deakin University, Burwood, VIC 3125, Australia\\
$^{6}$Departamento de Astronomia, Universidad de Concepcion, Casilla 160-C, Chile\\
$^{7}$University of Nottingham, School of Physics \& Astronomy, Nottingham, NG7 2RD, UK\\
$^{8}$Institute of Cosmology and Gravitation, University of Portsmouth, Dennis Sciama Building, Burnaby Road, Portsmouth PO1 3FX, UK\\
$^{9}$Sydney Institute for Astronomy, School of Physics, A28, The University of Sydney, NSW 2006, Australia\\
$^{10}$ARC Centre of Excellence for All Sky Astrophysics in 3 Dimensions (ASTRO 3D)\\
$^{11}$Department of Physics and Astronomy, Macquarie University, North Ryde NSW 2109, Australia\\
$^{12}$Australian Astronomical Observatory, PO Box 915, North Ryde, NSW 1670, Australia}

\begin{document}

\maketitle

\begin{abstract}
Using data from the WiFeS Atlas of Galactic Globular cluster Spectra we study the behaviour of the calcium triplet (CaT), a popular metallicity indicator in extragalactic stellar population studies.
A major caveat of these studies is that the potential sensitivity to other stellar population parameters such as age, calcium abundance and the initial mass function has not yet been empirically evaluated.
Here we present measurements of the strength of the CaT feature for 113 globular clusters in the Milky Way and its satellite galaxies.
We derive empirical calibrations between the CaT index and both the iron abundance ([Fe/H]) and calcium abundance ([Ca/H]), finding a tighter relationship for [Ca/H] than for [Fe/H].
For stellar populations 3 Gyr and older the CaT can be used to reliably measure [Ca/H] at the 0.1 dex level but becomes less reliable for ages of $\sim 2$ Gyr and younger.
We find that the CaT is relatively insensitive to the horizontal branch morphology.
The stellar mass function however affects the CaT strengths significantly only at low metallicities.
Using our newly derived empirical calibration, we convert our measured CaT indices into [Ca/H] values for the globular clusters in our sample.
\end{abstract}


\section{Introduction}
\label{sec:introduction}

Globular clusters (GCs) have two important roles in the study of stellar populations using integrated galaxy light.
First, GCs in the Local Group provide important tests for stellar population models and analysis techniques as their ages, metallicities and abundances are known independently \citep[e.g.][]{2007ApJS..171..146S, 2010MNRAS.404.1639V, 2018ApJ...854..139C}.
Second, GCs play an important role as observational tracers of galaxy stellar light \citep[e.g.][]{1978ApJ...225..357S, 2010MNRAS.404.1203F, 2014ApJ...796...52B, KPRCB18}.
Present in virtually all galaxies with stellar masses $> 10^{9}$ M$_{\sun}$ \citep[see reviews by ][]{2006ARA&A..44..193B, 2018RSPSA.47470616F}, their high surface brightness allow GCs to be studied at much greater distances than individual stars (GCs have been studied spectroscopically out to a distance of 47 Mpc, \citealt{2011A&A...531A...4M} and with imaging out to a redshift of $z \sim 0.2$ e.g. \citealt{2013ApJ...775...20A}).
Unlike the luminosity weighted means provided by analysing integrated galaxy light, observations of GCs allow the distributions and correlations between stellar population parameters, positions and kinematics to be studied.
Since GCs are essentially single-age, single-metallicity (but not single-abundance pattern, see reviews by \citealt{2012A&ARv..20...50G} and \citealt{2017arXiv171201286B}) stellar populations, it is easier to measure their stellar population parameters than it is for the field star populations of galaxies which contain a range of ages and metallicities.

The calcium triplet (CaT: 8498, 8542 and 8662 \AA) is one of the strongest spectral features in the optical or near-infrared spectra of stars and old stellar populations.
As a triplet of widely spaced lines, it is ideal for kinematic studies of individual stars and galaxies \citep[e.g.][]{2004MNRAS.354.1223K, 2006AJ....132.1645S, 2013MNRAS.428..389P, 2014ApJ...791...80A}.
The CaT wavelength region is less affected by extinction and is closer to the blackbody peak of the cool stars that dominate the luminosity of old or metal rich stellar populations than commonly studied features such as H$\beta$ and Mg$b$ \citep[e.g][]{1994ApJS...94..687W}.
At almost all redshifts, the singly ionised calcium lines of the CaT overlap with a multitude of strong sky emission lines.
Hence, measuring the equivalent width of the CaT requires careful sky subtraction.
Thankfully, the CaT wavelength region is relatively unaffected by telluric absorption lines at low redshift.

In individual stars, the CaT displays complex behaviour with temperature, surface gravity and metallicity.
For both dwarfs and giants of all metallicities the CaT strength peaks at effective temperatures between 4000 and 6000 K \citep[e.g.][]{2002MNRAS.329..863C}.
In giants, the CaT strength increases strongly with decreasing surface gravity.
However, at lower metallicities and in dwarf stars, the relationship between the CaT and surface gravity is weaker \citep[e.g.][]{2002MNRAS.329..863C}.
The CaT is more sensitive to metallicity in giants and at temperatures between 5000 and 6000 K \citep[e.g.][]{2002MNRAS.329..863C}.
As a result, the CaT has long been used to measure metallicities of individual giant stars in GCs \citep[e.g.][]{1991AJ....101.1329A, 1997PASP..109..907R, 2016MNRAS.455..199D}, in open clusters \citep[e.g.][]{2004MNRAS.347..367C, 2015A&A...578A..27C} and in dwarf galaxies in the Local Group \citep[e.g.][]{1991AJ....101..515O, 2004ApJ...617L.119T, 2005AJ....129.1465C, 2008MNRAS.383..183B, 2013ApJ...767..131L}.

For integrated light (i.e. combined spectra of entire stellar populations) the depth of the CaT features increases with metallicity.
The reasons for the CaT's sensitivity to metallicity are two-fold.
First, higher [Ca/H] directly leads to stronger CaT lines.
Second, the effects of higher metallicity on stellar evolution push stars to cooler temperatures and giants to lower surface gravities.
Both lower temperatures and lower surface gravities produce stronger CaT lines.

Early work \citep[e.g.][]{1971ApJS...22..445S, 1978ApJ...221..788C} on integrated light spectra of stellar populations focused on the sensitivity of the CaT to surface gravity to study the initial mass function (IMF).
The dependence of the CaT on metallicity was empirically confirmed by \citet{1987A&A...186...49B} using spectra of 30 star clusters in the Milky Way (MW) and in the Large Magellanic Cloud (LMC).
\citet{1988AJ.....96...92A} were the first to use the CaT as a metallicity indicator.
They measured the strength of the CaT in 27 GCs in the MW and found a strong relationship between the CaT and metallicity.

The CaT has been used in studies of integrated galaxy light \citep[e.g.][]{2008ASPC..390..292C, 2009MNRAS.400.2135F, 2014MNRAS.442.1003P} but its use as a metallicity indicator has been questioned due to the potentially significant effect of the IMF and [Ca/Fe] abundance \citep[e.g.][]{2002ApJ...579L..13S, 2003MNRAS.339L..12C}.
In parallel, interest in the CaT as an IMF indicator has grown in recent years with claims \citep[e.g.][]{2010Natur.468..940V, 2012Natur.484..485C, 2013MNRAS.433.3017L, 2015MNRAS.447.1033M, 2017ApJ...841...68V} that the centres of massive ETGs have relatively more dwarf stars (i.e. a bottom-heavy IMF) than the MW.

Recent studies \citep[e.g.][]{2001MNRAS.322..643G, 2012ApJ...759L..33B, 2013MNRAS.436.1172U} have used the CaT to measure metallicities in unresolved extragalactic GCs.
However, the reliability of the CaT as a GC metallicity indicator has been debated due to its known, but as yet unquantified, dependence on other parameters.
For example, \citet{2010AJ....139.1566F,2011MNRAS.415.3393F} and \citet{2012MNRAS.426.1475U} found different relationships between GC colour and CaT strength in different galaxies.
Different sets of stellar population synthesis models \citep{2012MNRAS.424..157V, 2016ApJ...818..201C} predict different CaT-metallicity relations. 
\citet{2016MNRAS.456..831S} observed a clear correlation between the CaT strength and [Fe/H] in M31 GCs.
However, they also found that [Ca/Fe] affects the CaT strength in GCs.
Despite this, \citet{2012MNRAS.426.1475U} found good agreement between Lick index \citep{1994ApJS...94..687W} based metallicities and their CaT based metallicities.

As discussed by \citet{2016MNRAS.456..831S}, there are advantages and disadvantages to studying GC stellar populations in the MW.
As they are the closest GCs to us, they are the best studied, with measured detailed abundances, ages and mass functions from studies of resolved stars \citep[e.g.][]{2009A&A...508..695C, 2010ApJ...708..698D, 2013ApJ...775..134V, 2010AJ....139..476P, 2015AJ....149..153M, 2017MNRAS.471.3668S}.
However, due to their relative nearness, MW GCs appear more extended on the sky than their extragalactic counterparts with a median half-light radius of one arcmin (the \citet{1996AJ....112.1487H, 2010arXiv1012.3224H} catalogue). It is thus challenging to obtain integrated spectra that sample adequately all phases of stellar evolution.
Additionally, the MW hosts a limited range of GC ages and abundances with no easily observable, massive star cluster younger than $\sim 8$ Gyr (the age of Terzan 7, e.g. \citealt{2010ApJ...708..698D}).
By also studying star clusters in the Milky Way's satellite galaxies, which span a wide range of ages \citep[e.g.][]{1997AJ....114.1920G} and have smaller angular sizes (median half-light radius $= 14$ arcsec for old GCs in the LMC, \citealt{2005ApJS..161..304M}) but still have abundances from individual stars \citep[e.g.][]{2008AJ....136..375M, 2010ApJ...717..277M, 2016ApJ...829...77D} and ages from HST colour-magnitude diagrams \citep[e.g.][]{1998MNRAS.300..665O, 2008AJ....136.1703G, 2015A&A...575A..62N}, we can empirically address concerns about the dependence of the CaT on e.g. horizontal branch morphology, the IMF, age and elemental abundances presented above.

In this paper we use integrated spectra of GCs in the MW and its satellite galaxies (the LMC, Small Magellanic Cloud - SMC and the Fornax dSph) from the WiFeS Atlas of Galactic Globular cluster Spectra (WAGGS, \citealt{2017MNRAS.468.3828U}) to study the behaviour of the CaT spectral feature as a function of a number of stellar population and GC parameters.
As in \citet{2017MNRAS.468.3828U}, we will use the term GC to refer to all massive ($> 10^{4}$ M$_{\sun}$) star clusters irrespective of age.
However, we make a distinction between GCs and objects which likely are or once were the nuclei of galaxies, namely NGC 5139 ($\omega$ Cen) and NGC 6715 (M54) \citep[e.g.][]{2000A&A...362..895H, 1997AJ....113..634I}.
We note that both NGC 5139 and NGC 6715 have extended star formation histories ($> 2$ Gyr, e.g. \citealt{2000AJ....119.1760L, 2006ApJ...647.1075S, 2007ApJ...667L..57S, 2014ApJ...791..107V}) and large metallicity spreads ($> 1$ dex, e.g. \citealt{2010ApJ...722.1373J, 2010A&A...520A..95C}) while GCs have little or no age or metallicity spreads.

This paper is organised as follows.
In Section \ref{sec:sample}, we present our sample and describe our observations and data reduction.
In Section \ref{sec:measurement}, we present our measurement technique and discuss some of the systematics affecting our measurements.
In Section \ref{sec:analysis}, we discuss the effects of metallicity, age, Ca abundance, stochasticity, horizontal branch morphology and the present day mass function on the CaT.
In Section \ref{sec:conclusion}, we summarise our results.

\section{Sample and observations}
The first WAGGS paper \citep{2017MNRAS.468.3828U} provided a detailed discussion of the scientific aims of the project, described the initial WAGGS sample and observations, detailed our data reduction, presented a comparison of repeated observations and described the publicly released spectra.
Here we provide a brief outline of our sample, observations and data reduction, highlighting any additions or differences from what was presented in \citep{2017MNRAS.468.3828U}.

\begin{figure}
\begin{center}
\includegraphics[width=240pt]{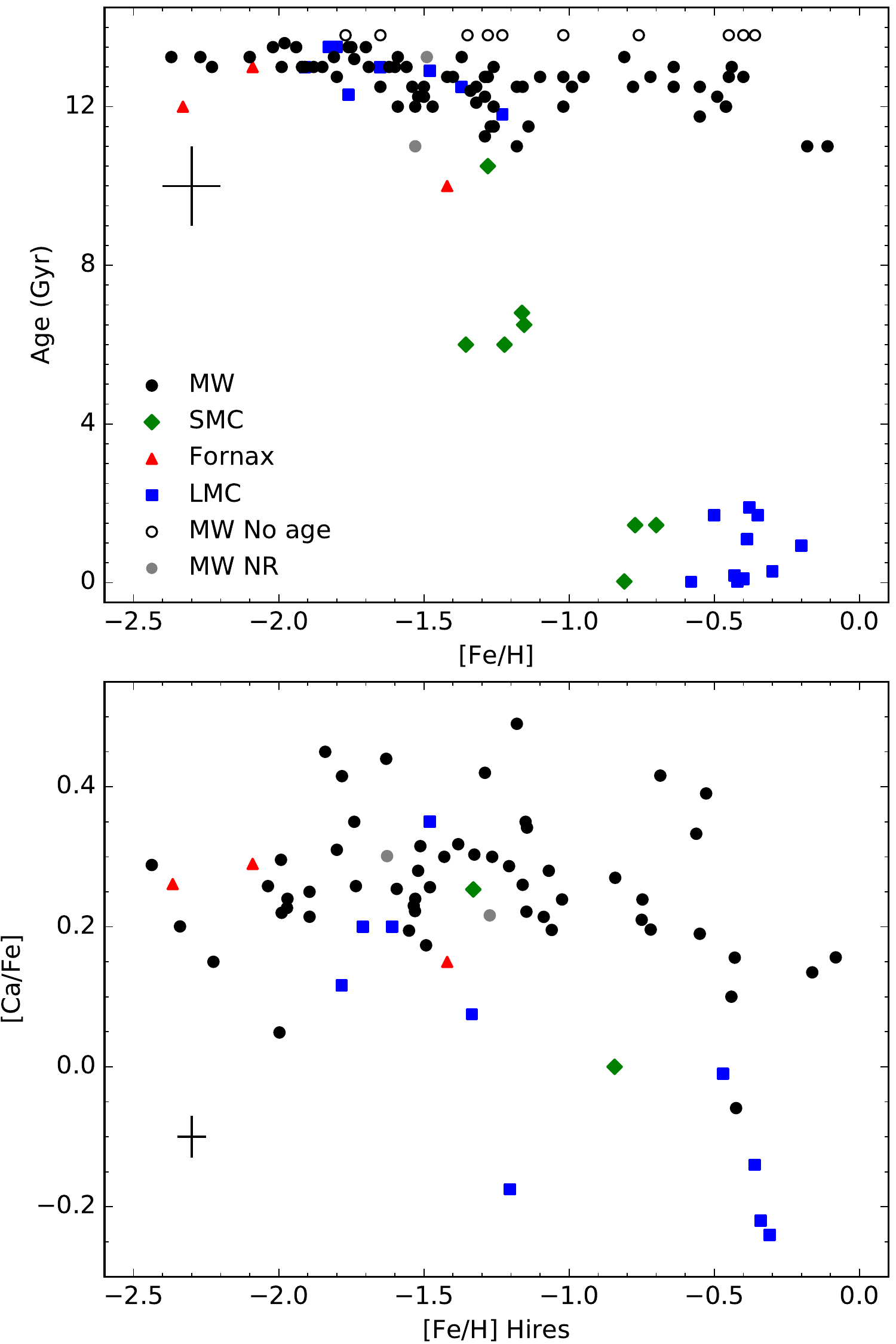}
\caption{Literature stellar population parameters used in this work.
MW GCs are marked as black circles, LMC GCs as blue squares, SMC GCs as green diamonds and Fornax dSph GCs as red triangles.
The nuclear remnants in the MW (NGC 5139 and NGC 6715) are denoted as grey circles.
We note that metallicities, ages and abundances are from a wide range of sources detailed in the text.
Typical uncertainties are provided as a cross in each panel. 
\emph{Top} GC age as a function of metallicity.
MW GCs with no literature ages have been assigned an age of 13.8 Gyr and are marked as hollow circles.
Each galaxy follows its own age-metallicity relation.
\emph{Bottom} [Ca/Fe] versus [Fe/H] from high resolution spectroscopy.
At fixed metallicity our sample covers a wide range of [Ca/Fe] values.
}
\label{fig:sample}
\end{center}
\end{figure}

\label{sec:sample}
We use an expanded version of the WAGGS sample presented in \citet{2017MNRAS.468.3828U}.
As detailed in that paper the sample consists of GCs in the MW, the LMC, the SMC and the Fornax dSph observable from the Siding Spring Observatory with central surface brightness brighter than $\mu_{V} \sim 20$ mag arcsec$^{2}$.
We favoured GCs with high-quality Hubble Space Telescope (HST) photometry  \citep[e.g.][]{2007AJ....133.1658S} and GCs with abundances from high-resolution spectroscopy.
Our sample contains both NGC 5139 and NGC 6715 which are thought to be the nuclear remnants of tidally disrupted (or disrupting) dwarf galaxies accreted by the MW \citep[e.g.][]{2000A&A...362..895H, 1997AJ....113..634I}.
We will refer two these two objects as nuclear remnants through out the rest of the paper.
Additions from the previously presented sample in \citet{2017MNRAS.468.3828U} are 11 GCs in the LMC and SMC and 16 GCs in the MW including a handful of well-studied but lower surface brightness MW GCs (e.g. NGC 288 and NGC 6496) and a number of relatively bright but poorly studied MW GCs (e.g. NGC 6626, NGC 6638 and NGC 6642).
We also observed additional pointings of a handful of nearby GCs (NGC 3201, NGC 6121 and NGC 6397) with low masses in our field-of-view in an attempt to decrease stochastic effects due to the low number of giant stars in the field-of-view.
Our sample spans a wide range of ages and metallicities (see Figure \ref{fig:sample}).
Details of our sample are given in Table \ref{tab:sample}.
In total we analyse 138 spectra of 113 GCs.

\addtolength{\tabcolsep}{-2pt}

\begin{landscape}
\begin{table}
\caption{Sample properties}
\label{tab:sample}
\begin{tabular}{ccccccccccccc}
ID & Galaxy & [Fe/H] & Age & $R_{c}$ & $R_{h}$ & $\mu_{V0}$ & $A_{V}$ & GC & FoV & [Fe/H] & Age & Structural \\
 & & & & & & & & Mass & Mass & Source & Source & Source \\
 &  & [dex] & [Gyr] & [arcsec] & [arcsec] & [mag arcsec$^{-2}$] & [mag] & [$\log$ M$_{\sun}$] & [$\log$ M$_{\sun}$] &  &  &  \\
(1) & (2) & (3) & (4) & (5) & (6) & (7) & (8) & (9) & (10) & (11) & (12) & (13) \\ \hline
    NGC 104 & MW & $-0.72$ & 12.8 & 21.6 & 190 & 14.4 & 0.12 & 6.0 & 4.7 & Harris (2010) & {\citet{2010ApJ...708..698D}} & Harris (2010) \\
    Kron 3 & SMC & $-1.15$ & 6.5 & 21.1 & 39.9 & 20.1 & 0.05 & 5.2 & 4.4 & {\citet{1998AJ....115.1934D}} & {\citet{2008AJ....135.1106G}} & {\citet{2005ApJS..161..304M}} \\
    NGC 121 & SMC & $-1.28$ & 10.5 & 9.6 & 19.0 & 18.3 & 0.45 & 5.6 & 5.2 & {\citet{2016ApJ...829...77D}} & {\citet{2008AJ....135.1106G}} & {\citet{2005ApJS..161..304M}} \\
    NGC 288 & MW & $-1.32$ & 12.5 & 81.0 & 134 & 20.0 & 0.09 & 4.9 & 3.2 & Harris (2010) & {\citet{2010ApJ...708..698D}} & Harris (2010) \\
    NGC 330 & SMC & $-0.81$ & 0.03 & 8.1 & 21.0 & 16.5 & 0.20 & 4.6 & 4.1 & {\citet{1999A&A...345..430H}} & {\citet{2002ApJ...579..275S}} & {\citet{2005ApJS..161..304M}} \\ 
    ... & ... & ... & ... & ... & ... & ... & ... & ... & ... & ... & ... & ... \\ \hline
\end{tabular}

\medskip
\emph{Notes}
Column (1): GC name.
Column (2): Host galaxy.
Column (3): Metallicity in dex.
Column (4): Age in Gyr.
Column (5): Projected core radius in arcsec.
Column (6): Projected half-light radius in arcsec.
Column (7): $V$-band central surface brightness in mag per arcsec$^{-2}$.
Column (8): $V$-band extinction in mag.
Column (9): GC log stellar mass in solar masses calculated from $V$-band luminosity.
Column (10): Log stellar mass in solar masses enclosed by the WiFeS field-of-view calculated from the surface brightness profile.
Columns (11), (12) and (13): Sources for [Fe/H], age and structural parameters respectively.
Harris (2010) refers to the \citet{1996AJ....112.1487H, 2010arXiv1012.3224H} catalogue.
The full version of this table is provided in a machine readable form in the online Supporting Information.

%references for full table
\nocite{2006ApJ...640..801J, 2001AJ....121.2638Z, 2008AJ....135.1106G, 2016A&A...590A..35D, 2002ApJ...579..275S, 2006A&A...456.1085M, 2016ApJ...829...77D, 2014ApJ...785...21M, 2008AJ....136.1703G, 2012ApJ...746L..19M, 2003MNRAS.338...85M, 2013MNRAS.431L.122B, 1991AJ....101..515O, 2016MNRAS.458.3968P, 1997AJ....114.1920G, 2007ApJ...663..296V, 1998MNRAS.300..665O, 2007AJ....133.2053M, 2008AJ....136..375M, 2009AJ....137.4988G, 2006MNRAS.369..697G, 2011ApJ...737....3G, 2015A&A...575A..62N, 1992AJ....104.1086F, 2010ApJ...717..277M, 2007A&A...462..139K, 2011ApJ...735...55C, 2014ApJ...782...50L, 2006AJ....132.1630G, 2005AJ....130..116D, 2010ApJ...708..698D, 2014ApJ...784..157L, 1999A&A...345..430H, 1998AJ....116.2395M, 2011ApJ...738...74D, 2012A&A...546A..53L, 2009AJ....138.1403G, 2005ApJS..161..304M, 1998AJ....115.1934D, 2018ApJ...853...15K, 2014ApJ...797...35G, 2011MNRAS.413..837M}

\end{table}
\end{landscape}

We used the surface brightness profiles calculated from the structural parameters given in Table \ref{tab:sample} using the \textsc{limepy} code \citep{2015MNRAS.454..576G} and a \citet{1966AJ.....71...64K} profile to calculate the $V$-band luminosity within the field-of-view of each datacube.
We converted these enclosed luminosities into masses using the same mass-to-light ratios ($M/L$) used in \citet{2017MNRAS.468.3828U}, namely a constant $M/L = 2$ for all globular clusters older than 10 Gyr and predictions of the \citet{2003MNRAS.344.1000B} stellar population synthesis models for the younger clusters.

\subsection{Observations \& data reduction}

\label{sec:observations}
As described in \citet{2017MNRAS.468.3828U}, we used the WiFeS instrument \citep{2007Ap&SS.310..255D, 2010Ap&SS.327..245D} on the Australian National University 2.3 m telescope at the Siding Spring Observatory.
WiFeS is a dual-arm, 38 by 25 arcsecond image slicer integral field spectrograph.
For this work we only used data observed with the RT615 beam splitter and I7000 grating.
This setup covers $6800 - 9050$ \AA{} with 0.57 \AA{} per pixel and provides a spectral resolution of $\delta \lambda / \lambda \sim 6800$ in the region of the CaT.
To perform reliable sky subtraction, we used WiFeS in nod-and-shuffle mode \citep{2001PASP..113..197G}.
We used the \textsc{PyWiFeS} \citep{2014ascl.soft02034C, 2014Ap&SS.349..617C} pipeline to reduce the data.
We used 2MASS \citep{2006AJ....131.1163S} J-band images to perform astrometry on our datacubes before applying heliocentric velocity corrections.
To create integrated light spectra we simply summed the spatial pixels of each datacube ignoring the first two and last two rows due to their significantly higher noise.
Unlike in \citet{2017MNRAS.468.3828U}, we used the entire datacube for all GCs and did not limit our extracted spectra to within 1 half-light radius for GCs with more extended spatial coverage.
Please see \citet{2017MNRAS.468.3828U} for a detailed description of the observations and data reduction.

\begin{figure}
\begin{center}
\includegraphics[width=240pt]{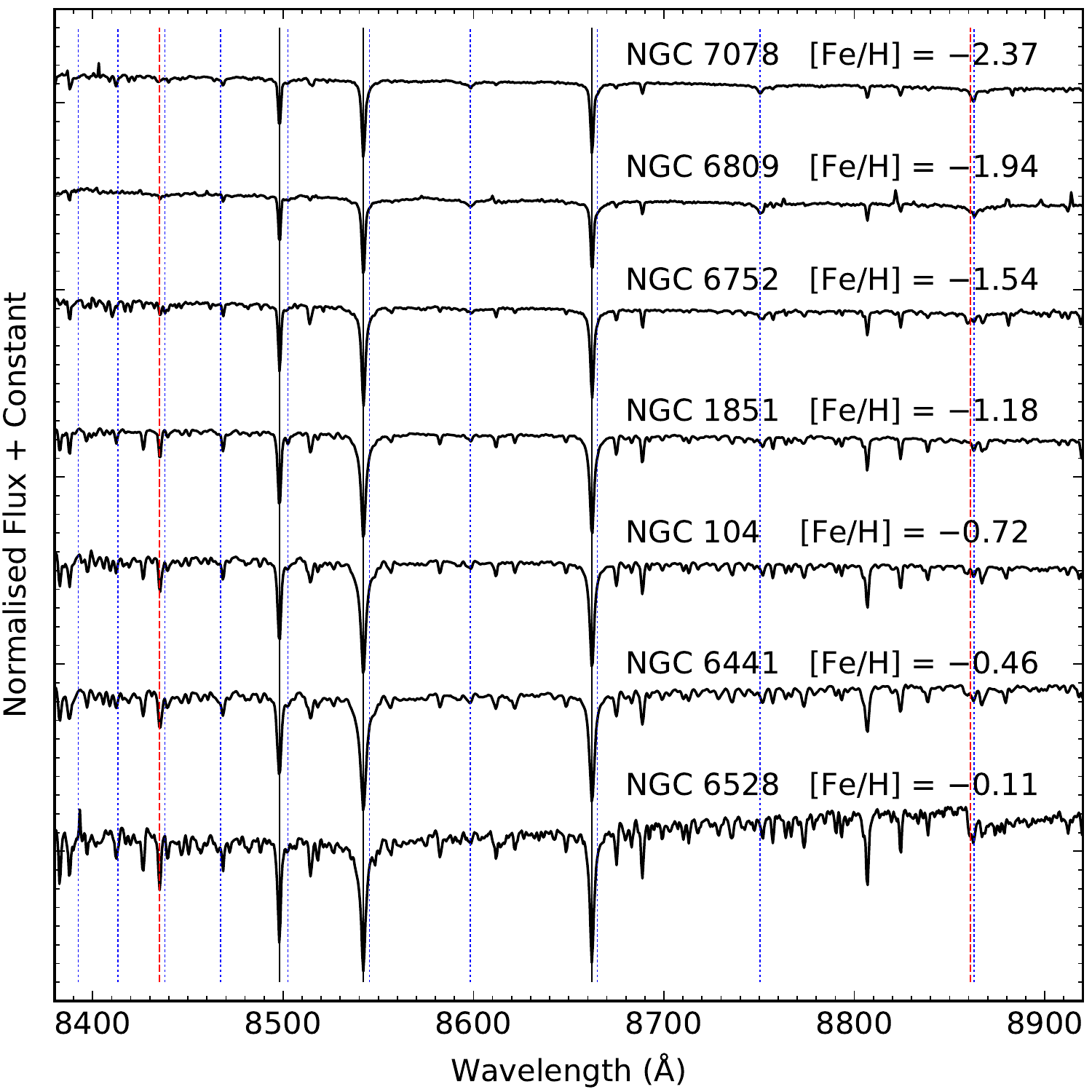}
\caption{Effects of metallicity on integrated GC spectra in the CaT region.
The GCs plotted are all old (age $> 11$ Gyr) and increase in metallicity from top to bottom.
The CaT lines are identified by solid black vertical lines at the top, the hydrogen Paschen lines as dotted blue lines and two TiO bandheads as red dashed lines.
The CaT lines increase with metallicity as do the other metal lines.
Weak Paschen absorption is only present in the most metal poor GCs ([Fe/H] $< -2$) and significant TiO absorption is only seen at near solar metallicity.
We note that fitted velocity dispersions of these GC range from $\sim 5$ km s$^{-1}$ (NGC 6528) to $\sim 17$ km s$^{-1}$ (NGC 6441) which is comparable to the instrument resolution (a velocity dispersion of 19 km s$^{-1}$).}
\label{fig:cat_examples}
\end{center}
\end{figure}

\begin{figure}
\begin{center}
\includegraphics[width=240pt]{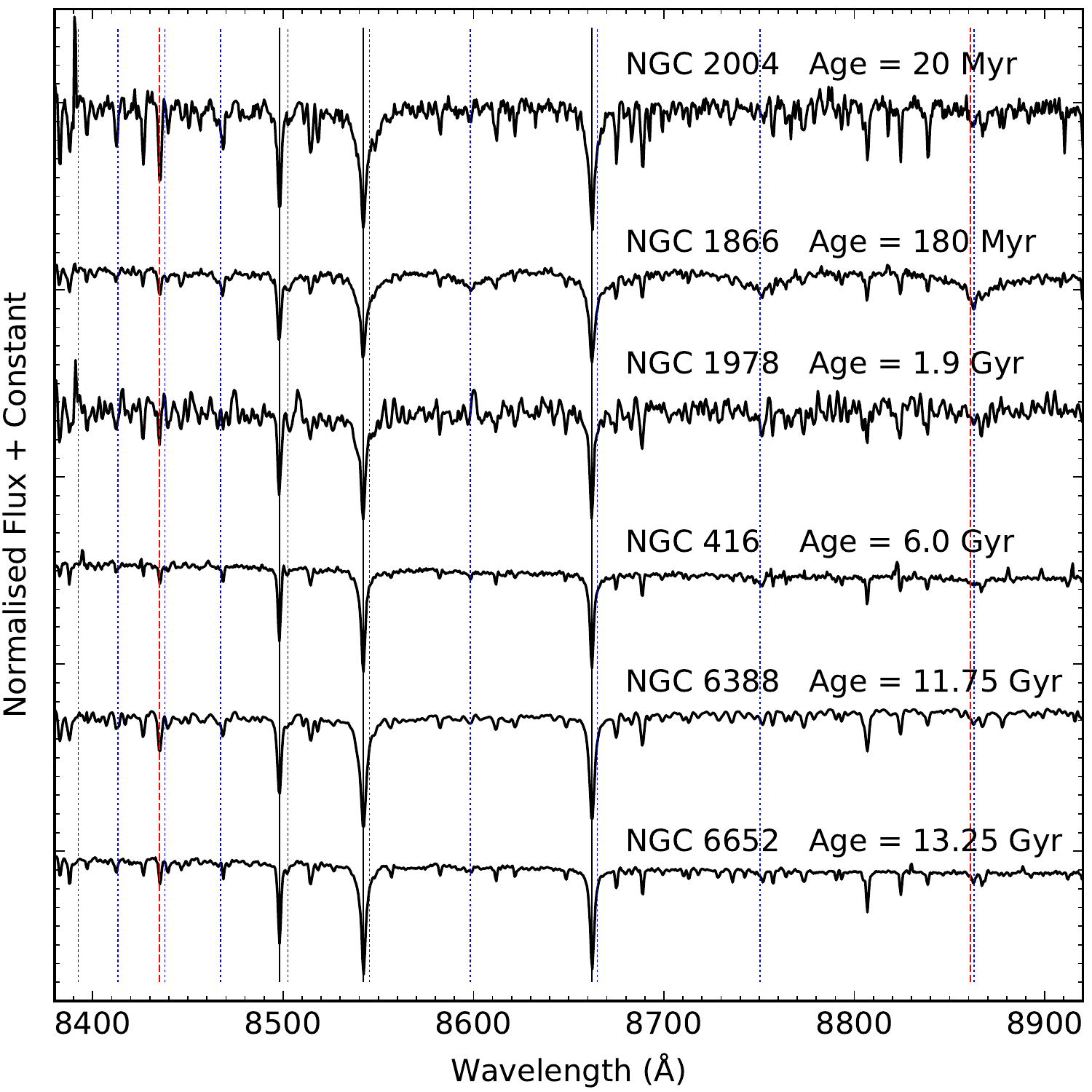}
\caption{Effects of age on integrated GC spectra in the CaT region.
The GCs increase in age from top to bottom.
The GCs are generally LMC and SMC metallicity ([Fe/H] $= -0.8$ to $-0.4$) save for NGC 416 which is significantly more metal poor ([Fe/H] $= -1.2$).
As in Figure \ref{fig:cat_examples}, the CaT lines are identified by solid black vertical lines at the top, the hydrogen Paschen lines as dotted blue lines and two TiO bandheads as red dashed lines.
At old ages ($> 5$ Gyr), age seems to have little effect on the CaT spectral region as the light is dominated by the RGB.
At younger ages ($\sim 2$ Gyr), the thermally pulsing AGB becomes important and the spectra can be dominated by carbon stars.
At even younger ages ($\sim 200$ Myr), hotter stars from the MSTO play an important role and strong Paschen line absorption occurs.
At the youngest ages ($\sim 20$ Myr), the CaT region is dominated by red supergiants which show stronger metal line absorption compared to RGB dominated old populations with the same metallicities.}
\label{fig:cat_age_examples}
\end{center}
\end{figure}

Our observations span nights from 2015 January to 2017 July.
In addition to the observations presented in \citet{2017MNRAS.468.3828U}, we utilized a series of observations of NGC 104 from our 2016 September 29th to October 3rd observing run.
We also utilized observations from our 2017 April 3rd to April 6th run and our 2017 June 29th to July 2nd run.
These observations were conducted using the same instrument set up as those presented in \citet{2017MNRAS.468.3828U} and reduced in an identical manner.
We plot examples of our spectra in the region of the CaT in Figures \ref{fig:cat_examples} and \ref{fig:cat_age_examples}.

\subsection{Literature metallicities and abundances}
We use three sets of metallicities and abundances in this work.
First, we find a literature metallicity for every GC in our sample. 
For the Milky Way, we adopt the metallicities provided by the \citet{1996AJ....112.1487H, 2010arXiv1012.3224H} catalogue.
While the sources of metallicities in the Harris catalogue are heterogeneous, an attempt has been made to place them on the \citet{2009A&A...508..695C} scale.
For the three GCs in the Fornax dSph, we adopt the metallicities derived by \citet{2012A&A...546A..53L}.
For GCs in the LMC and SMC we preferred metallicities from high-resolution spectroscopy to those from the resolved star CaT strengths, and CaT-based metallicities to those based on resolved colour-magnitude diagrams.
We adjusted the metallicity measurements based on the CaT strengths of \text{red giant branch} (RGB) stars to the \citet{2009A&A...508..695C} metallicity scale.
For NGC 1466, we follow the recommendation of \citet{1992AJ....104.1395W} and adopt a metallicity for NGC 1466 based on the more metal-rich star ([Fe/H] $=-1.91$ using the \citealt{2009A&A...508..695C} calibration) from \citet{1991AJ....101..515O}.
\citet{1998AJ....115.1934D} measured the strength of the CaT for four RGB stars in NGC 361 but did not have sufficiently high quality photometry to convert these measurements into metallicities.
Using the HST WFPC2 photometry of \citet{1998AJ....116.2395M} and the CaT-metallicity calibration of \citet{2016MNRAS.455..199D} we derive a metallicity of [Fe/H] $= -1.16$ for NGC 361.
A challenge for all LMC and SMC based metallicities is that their ages and [$\alpha$/Fe] can be quite different to MW GCs of the same metallicity.
Metallicities derived using techniques calibrated on MW GCs (i.e. CaT strengths of RGB stars) may be less reliable.

Second, we assembled a sample of Fe abundances ([Fe/H]) and Ca abundances [Ca/Fe]) based on high resolution ($R > 20$ 000) spectroscopy of individual stars.
We gave preference to studies of the RGB to avoid any biases from comparing measurements from different evolutionary stages.
We also gave preference to studies with large numbers of stars.
We shifted all the abundances to the \citet{2009ARA&A..47..481A} abundance scale.
For GCs with multiple studies, we took the error weighted mean of the available studies.
When no mean abundance for the GC was provided by the authors, we took the simple mean of the abundances of the individual stars as the abundance of the cluster.
If uncertainties for each individual star were provided we used these to calculate the uncertainty in the mean; otherwise the statistical uncertainty in the mean is given by $ \sigma / \sqrt{N-1} $, where $\sigma$ is the standard deviation of the individual star measurements, and $N$ is the number of stars.  
We were careful to add in quadrature the systematic uncertainties provided by the authors to the statistical uncertainty; if no systematic uncertainties were provided we assumed a 0.05 dex systematic uncertainty in [Fe/H] and 0.02 dex in [Ca/Fe] as found by \citet{2009A&A...508..695C} and \citet{2010ApJ...712L..21C}. 
While we aimed to be comprehensive, our high resolution sample should not be seen as complete.

In the case of the \citet{2010ApJ...714L...7C} study of NGC 6715, we use both the stars identified by the authors as members of NGC 6715 and as members of the nucleus of the Sagittarius dwarf galaxy to calculate mean [Fe/H] and [Ca/Fe] abundances as both are captured by our observations.
\citet{2015A&A...573A..92G} studied red HB stars in NGC 6723.
They note that Ca abundances derived from HB stars are systematically higher than those from the RGB.
As such we lower their [Ca/Fe] measurement by their recommended 0.2 dex.
We took APOGEE abundances from \citet{2015AJ....149..153M} and \citet{2017MNRAS.466.1010S}. 
The Gaia-ESO \citep{2012Msngr.147...25G} [Fe/H] measurements are from \citet{2017A&A...601A.112P} while Gaia-ESO [Ca/Fe] measurements were taken from the ESO Science Archive Facility.
We supplement this high-resolution sample with abundances from high resolution integrated studies for Fornax 3, Fornax 4 and Fornax 5 \citep{2012A&A...546A..53L} as well as for NGC 1916 \citep{2011ApJ...735...55C, 2012ApJ...746...29C}.
While we do include the \citet{2010ApJ...722.1373J} study of NGC 5139 and the \citet{2010ApJ...714L...7C} study of NGC 6715 in our plots, we do not include these nuclear remnants in our analysis due to their significant metallicity spreads.
We plot our adopted high-resolution [Fe/H] and [Ca/Fe] values in Figure \ref{fig:sample} and list our adopted values in Table \ref{tab:high_res}.

Third, we use the sample of \citet{2009A&A...508..695C} to provide a homogeneous sample of abundances.
We use the ESO VLT UVES based values for [Fe/H] from \citet{2009A&A...508..695C} and for [Ca/Fe] from \citet{2010ApJ...712L..21C}.
To enlarge the sample, we include subsequent studies by the same authors using the same instrument and analysis.
We also include the \citet{2001AJ....122.1469C} study of NGC 6528 and the \citet{1999ApJ...523..739C} study of NGC 6553 as corrected by \citet{2007A&A...464..967C}.
We note that these two GCs were studied using Keck HIRES observations of red HB stars, rather than the VLT UVES observations of RGB stars used in the rest of the Carretta et al. sample.
We do not include NGC 6715 in our Carretta et al. sample due to the significant metallicity spread in what is likely the nucleus of the Sagittarius dE.

\begin{table}
    \caption{Adopted high resolution abundances}
    \label{tab:high_res}
    \begin{tabular}{cccc}
        Name & [Fe/H] & [Ca/Fe] & Sources \\
             & [dex] & [dex] & \\
        (1) & (2) & (3) & (4) \\ \hline
        NGC 104 & $-0.74 \pm 0.02$ & $0.23 \pm 0.01$ & {\citet{2004A&A...416..925C}} \\
        & & & {\citet{2005A&A...435..657A}} \\
        & & & {\citet{2008AJ....135.1551K}} \\ 
        & & & {\citet{2009A&A...508..695C}} \\
        & & & {\citet{2010ApJ...712L..21C}} \\
        & & & {\citet{2014A&A...572A.108T}} \\
        & & & {\citet{2014ApJ...780...94C}} \\
        & & & {\citet{2017A&A...601A.112P}} \\
        
        NGC 121 & $-1.35 \pm 0.03$ & $0.24 \pm 0.04$ & {\citet{2004oee..sympE..29J}} \\
                & & & {\citet{2016ApJ...829...77D}} \\
NGC 288 & $-1.30 \pm 0.07$ & $0.41 \pm 0.03$ & {\citet{2009A&A...508..695C}} \\
NGC 330 & $-0.84 \pm 0.04$ & $0.00 \pm 0.05$ & {\citet{1999A&A...345..430H}} \\
NGC 362 & $-1.14 \pm 0.03$ & $0.28 \pm 0.04$ & {\citet{2013A&A...557A.138C}} \\
        & & & {\citet{2017A&A...601A.112P}} \\ 
        ... & ... & ... & ... \\ \hline

    \end{tabular}

\medskip
\emph{Notes}
Column (1): GC name.
Column (2): [Fe/H] abundance in dex.
Column (3): [Ca/Fe] abundance ratio in dex.
Column (4): Source(s). 
The full version of this table is provided in a machine readable form in the online Supporting Information.

%references for full table
\nocite{2008A&A...490..625M, 2011ApJ...740...60C, 2016ApJ...829...77D, 2014MNRAS.439.2638Y, 2010ApJ...712L..21C, 2003A&A...411..417M, 2017A&A...601A..31L, 2017ApJ...842...24J, 2015ApJ...810..148C, 2011A&A...532A...8M, 2004AJ....127.2162S, 2002AJ....124.1511L, 2003AJ....125..224R, 2017A&A...605A..12M, 2008ApJ...689.1031Y, 2004AJ....127.3411C, 2011ApJ...735...55C, 2010A&A...520A..95C, 2010ApJ...722L...1C, 2018MNRAS.474.4541M, 1999A&A...345..430H, 2015A&A...578A.116C, 1998AJ....115.1500K, 2001AJ....122.1469C, 2010ApJ...717..277M, 2014A&A...572A.108T, 2008AJ....135.1551K, 2011MNRAS.413..837M, 2013A&A...554A..81K, 2018A&A...614A.109C, 2015MNRAS.450..815M, 1999AJ....118.1273I, 2004A&A...416..925C, 2014MNRAS.445.2994N, 2006A&A...460..269A, 2015A&A...573A..92G, 2015ApJ...800....3C, 2015AJ....150...63J, 2008AJ....136..375M, 1999ApJ...523..739C, 2015AJ....149..153M, 2017A&A...602L..14R, 2004A&A...423..507Z, 2005A&A...435..657A, 2018AJ....155...71J, 2018ApJ...859...81M, 2013A&A...557A.138C, 2010ApJ...722.1373J, 2002AJ....123.3277R, 2017MNRAS.464.2730V, 2017ApJ...846...23O, 2008MNRAS.388.1419O, 2009A&A...508..695C, 2011A&A...533A..69C, 2011A&A...535A..31V, 1997AJ....114.1964S, 2001AJ....122.1438I, 2008ApJ...689.1020Y, 2007RMxAC..28..120L, 2016AJ....152...21J, 2011ApJ...742...37R, 2017A&A...601A.112P, 2002A&A...385..143K, 2011MNRAS.414.2690V, 2014ApJ...780...94C, 2015A&A...583A..69B, 2010AJ....139.2289K, 2013MNRAS.435.3667M, 2012ApJ...746L..19M, 2017MNRAS.466.1010S, 2014AJ....148...67J, 2006ApJ...640..801J, 2015MNRAS.448...42L, 2010ApJ...722L..18V, 2008ApJ...672L..29Y, 2011AJ....141...62L, 2016MNRAS.460.2351V, 2006AJ....132.2346J, 2015A&A...579A.104S, 2014A&A...564A..60C, 2006A&A...453..547L, 2017MNRAS.465...19T, 2017MNRAS.468.1249M, 2005MNRAS.356.1276O, 2006ApJ...646L.119L, 2012ApJ...746...29C, 2005AJ....129..303C, 2009A&A...493..913F, 2009A&A...507..405B, 2013MNRAS.433.2006M, 2007A&A...464..967C, 2004oee..sympE..29J, 2001A&A...369...87G, 2012A&A...546A..53L, 2016MNRAS.458.3968P, 2017A&A...600A.118C}

\end{table}

\section{Measuring the calcium triplet}
\label{sec:measurement}
\subsection{Measurement technique}
\begin{figure}
\begin{center}
\includegraphics[width=240pt]{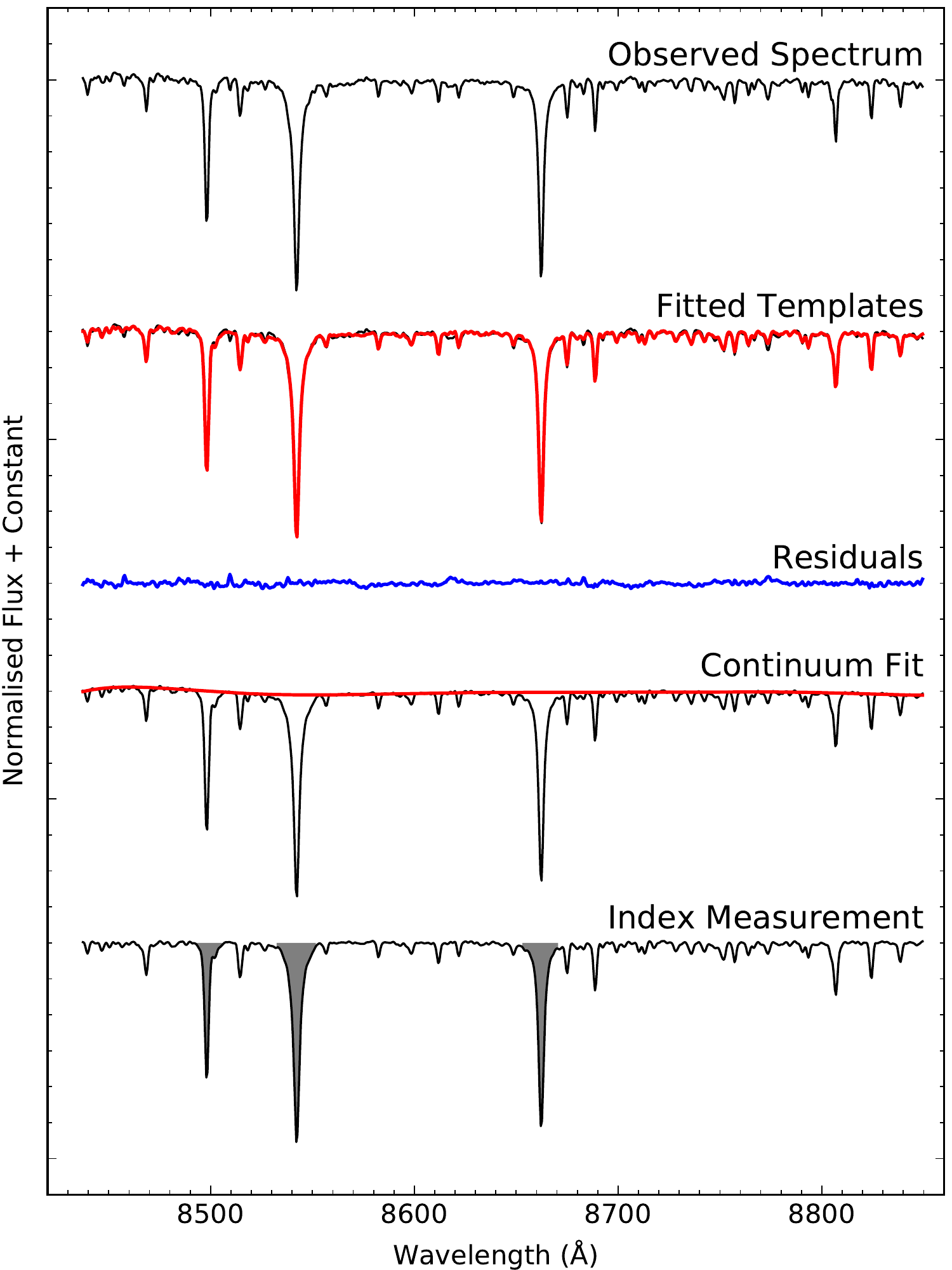}
\caption{Measuring the CaT by template fitting.
First from the top is our observed spectrum of NGC 2808 shifted to the rest frame.
Second from the top, we plot the fitted combination of stellar templates in red over our observed spectrum in black.
Third from the top, in blue we plot the difference of our fitted spectrum and our input spectrum.
Fourth from the top, in red we plot the continuum fitted to the templates in black.
Fifth from the top, we plot the continuum normalised templates in black and the region between the normalised templates and unity in our CaT bandpasses in grey.}
\label{fig:CaT_demo}
\end{center}
\end{figure}

We measured the strength of the CaT using the template fitting method of \citet{2010AJ....139.1566F} and \citet{2012MNRAS.426.1475U}.
In this method, wavelength regions affected by sky emission lines are masked before the observed spectrum is fitted by a linear combination of stellar templates.
The fitted combination of templates is then continuum-normalised and the equivalent width of the CaT is measured on the normalised combination of templates. 
This process allows the CaT to be reliably measured from low signal-to-noise (S/N) spectra with strong residuals from sky subtraction.
We plot a summary of the steps in our CaT measurement method in Figure \ref{fig:CaT_demo}.
As in our previous work we used the same index definition as \citet{1988AJ.....96...92A} (8490.0 to 8506.0, 8532.0 to 8552.0 and 8653.0 to 8671.0 \AA ) to measure the CaT.

As in \citet{2010AJ....139.1566F} and \citet{2012MNRAS.426.1475U} we used the \textsc{pPXF} pixel fitting code \citep[version 5.1.10]{2004PASP..116..138C} to fit the template stars to the observed spectra as well as the radial velocity, velocity dispersion and a continuum normalisation.
As in \citet{2012MNRAS.426.1475U}, we fit a seventh order continuum polynomial and used an initial guess of 10 km s$^{-1}$ for the velocity dispersion.
For the initial guess of the radial velocity we used the literature values presented in Table \ref{tab:sample}.
For a handful of GCs (NGC 416, NGC 1850, NGC 1856 and NGC 2004) in the LMC or SMC, we could not locate literature radial velocities so we used the systemic velocities of these galaxies (278 and 158 km s$^{-1}$ respectively, from NED\footnote{The NASA/IPAC Extragalactic Database is operated by the Jet Propulsion Laboratory, California Institute of Technology, under contract with the National Aeronautics and Space Administration.}).
A major improvement over \citet{2012MNRAS.426.1475U} is using the \textsc{Python} version of \textsc{pPXF} \citep{2004PASP..116..138C} rather than the \textsc{IDL} version.
This change resulted in over a factor of 100 speedup.
The measurement code is available from \url{https://github.com/chrusher/measure_CaT}.

\subsubsection{Effects of skyline masks}
We ran \textsc{pPXF} both using the same sky line masks as in \citet{2012MNRAS.426.1475U} and without masking any pixels affected by sky emission lines.
We note that \textsc{pPXF} performs a weighted fit so the higher variance sky line regions are down-weighted.
Additionally, we allow \textsc{pPXF} to clip pixels more than 3 sigma from the fit.
Another difference from \citet{2012MNRAS.426.1475U} was shifting the lower wavelength limit of the fitted spectra to 8437 \AA{} from 8425 \AA{} to avoid the TiO bandhead at 8420 \AA{} as the presence of this molecular feature complicates the determination of the continuum.
Our upper wavelength limit of our fit, 8850 \AA{}, was chosen to avoid the strong TiO bandhead at 8860 \AA{} and remains unchanged from \citet{2012MNRAS.426.1475U}.

\begin{figure}
\begin{center}
\includegraphics[width=240pt]{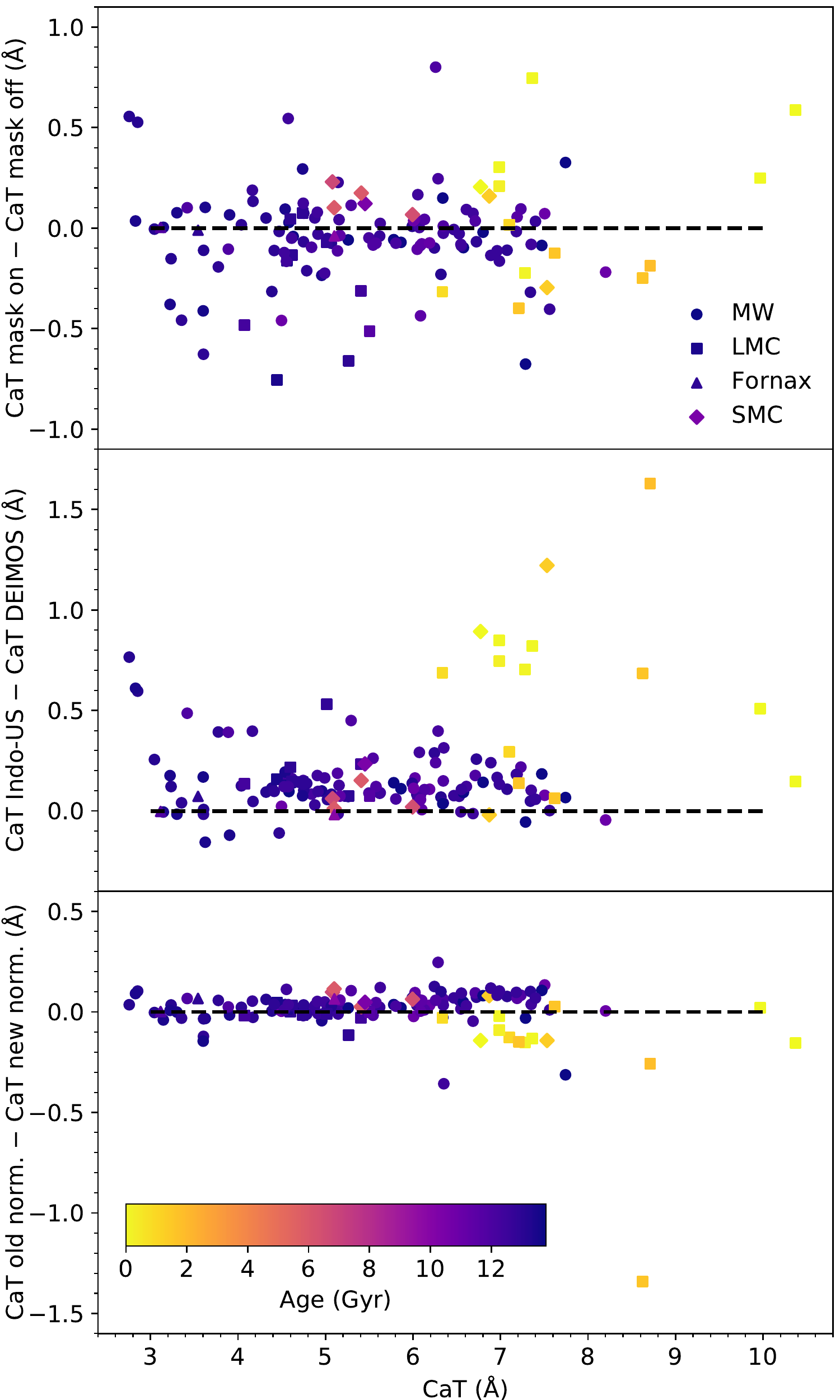}
\caption{Comparison of the CaT measurements using different parameters.
\emph{Top} Difference between CaT measured with and without masking sky emission lines.
\emph{Middle} Difference between CaT measured using the Indo-US templates and the DEIMOS templates.
\emph{Bottom} Difference between the continuum normalisation used in \citet{2012MNRAS.426.1475U} and the one used here.
The points are colour-coded by age with MW GCs denoted by circles, LMC GCs by squares, SMC GCs by diamonds and Fornax dSph GCs by triangles.
We see good agreement between the masked and unmasked CaT measurements but see large scatter.
Modulo a small constant offset, we see agreement between the DEIMOS and Indo-US templates for most old GCs.
However, the most metal poor GCs and the youngest GCs show relatively stronger Indo-US based CaT measurements.
We see little difference between the old and new continuum method save at the highest CaT values where older GCs show slightly stronger CaT values using the new method and young GCs show slightly weaker values.}
\label{fig:compare_templates}
\end{center}
\end{figure}

\begin{figure}
\begin{center}
\includegraphics[width=240pt]{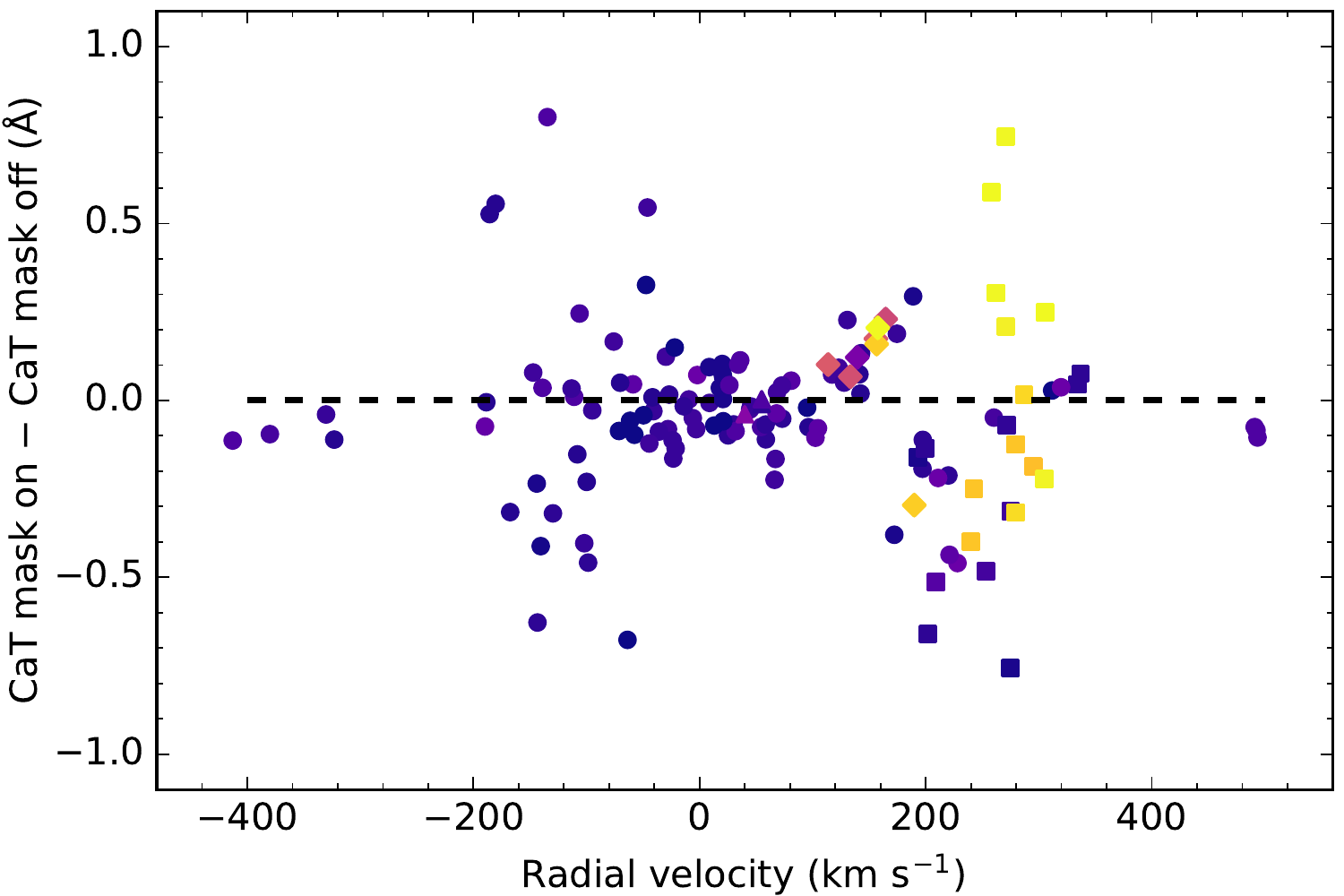}
\caption{Differences between CaT measured with and without masked sky emission lines as a function of radial velocity.
As in Figure \ref{fig:compare_templates}, the points are colour-coded by age with MW GCs denoted by circles, LMC GCs by squares, SMC GCs by diamonds and Fornax dSph GCs by triangles.
The CaT strength difference between the masked and unmasked spectra increases at radial velocities where skylines overlap with the CaT lines.}
\label{fig:mask_on_off}
\end{center}
\end{figure}

In the top panel of Figure \ref{fig:compare_templates} we show comparisons of the CaT strengths measured by fitting masked and unmasked spectra.
There is significant scatter (Root-mean squared - RMS - difference of 0.27 \AA ) in the difference between the CaT strength measured with the skylines masked and unmasked but only a minor median difference (0.04 \AA ).
In Figure \ref{fig:mask_on_off}, we plot this difference as a function of radial velocity.
The difference between the masked and unmasked measurement peaks around radial velocities $-120$ and $+240$ km s$^{-1}$.
This is where the sky line masks overlap with the CaT feature.
GCs with radial velocities between $-200$ and $-40$ km s$^{-1}$ and between 120 and 320 km s$^{-1}$ having a RMS difference of 0.32 \AA{} while GCs outside of these radial velocity ranges have a RMS difference of 0.20 \AA{} between the masked and unmasked measurements.

\subsubsection{Effects of stellar templates}

We fitted our observed spectra using the same 13 stellar templates (observed using the DEIMOS spectrograph by \citealt{2003SPIE.4841.1657F} on the Keck II telescope) as used by \citet{2010AJ....139.1566F} and \citet{2012MNRAS.426.1475U} since the spectral resolution of these templates is similar to that of WiFeS.
Details of the DEIMOS templates are provided in Table \ref{tab:deimos_templates}.
As the DEIMOS templates only cover a limited range of stellar parameters -- for example they lack any metal poor dwarfs -- we also experimented with using stellar templates drawn from the higher resolution (FWHM $\sim 1$ \AA ) Indo-US library \citep{2004ApJS..152..251V}.
We did not broaden the Indo-US templates to match the WiFeS spectral resolution, instead relying on the fitting code to broaden the fitted templates to match the observed spectra.
We selected 40 stars from the Indo-US library that had spectral coverage of the CaT region and covered the range of effective temperatures, surface gravities and metallicities expected for GC stars with a wide range of ages and metallicities.
We also include a carbon star to check whether it would improve the fit to the spectra of intermediate-age GCs.
Details of the Indo-US templates are provided in Table \ref{tab:indo_us_templates}.
For the DEIMOS templates and Indo-US templates we included a constant flux template.
For the DEIMOS templates, this constant flux template is required to fit low metallicity spectra while for the Indo-US templates, the inclusion of a constant flux template has no measurable effect on the fits.
We did not use the \citet{2001MNRAS.326..959C} library of stellar spectra due to their lower spectral resolution (FWHM = 1.5 \AA{} versus 1.26 \AA{} for WiFeS).

\begin{table*}
\caption{DEIMOS templates}
\label{tab:deimos_templates}
\begin{tabular}{lccccc}
\hline
Star & Type & $T_{eff}$ & $\log g$ & [Fe/H] & Reference \\
     & & (K) & ($\log$ cm g s$^{-2}$)  & (dex) & \\
(1) & (2) & (3) & (4) & (5) & (6) \\
\hline
HD 4388 & K3III & & & & \\
HD 35410 & G9III & 4800 & 2.6 & $-0.33$ & {\citet{2014ApJ...785...94L}} \\
HD 36079 & G5II & 5200 & 2.5 & $-0.25$ & {\citet{2014ApJ...785...94L}} \\
HD 44131 & K5III & & & & \\
HD 65934 & G8III & & & & \\
HD 74377 & K3V & 4670 & 4.5 & $-0.37$ & {\citet{2011A&A...531A.165P}} \\
HD 107328 & K1III & 4340 & 1.7 & $-0.52$ & {\citet{2016A&A...587A...2B}} \\
HD 136202 & F8IV & 6050 & 3.8 & $-0.10$ & {\citet{2016A&A...587A...2B}} \\
HD 161096 & K2III & 4590 & 2.6 & $+0.10$ & {\citet{2016A&A...587A...2B}} \\
HD 237903 & K7V & 4080 & 4.7 & $-0.26$ & {\citet{2011A&A...531A.165P}} \\
NGC1904-S71 & & 5140 & 2.0 & $-1.81$ & {\citet{2008ApJ...682.1217K}} \\
NGC1904-S174 & & 4450 & 1.1 & $-1.77$ & {\citet{2008ApJ...682.1217K}} \\
NGC2419-S223 & & 4250 & 0.5 & $-2.09$ & {\citet{2011ApJ...740...60C}} \\
\hline
\end{tabular}

\medskip
\emph{Notes}
Column (1): Name of star.
Column (2): Spectral type according to SIMBAD \citep{2000A&AS..143....9W}.
Column (3): Effective temperature in K.
Column (4): Log surface gravity in cm g s$^{-2}$.
Column (5): [Fe/H] in dex.
Column (6): Source for atmospheric parameters.
\end{table*}

\begin{table}
\caption{Indo-US templates}
\label{tab:indo_us_templates}
\begin{tabular}{lcccc}
\hline
Star & Type & $T_{eff}$ & $\log g$ & [Fe/H] \\
     & & (K) & ($\log$ cm g s$^{-2}$)  & (dex) \\
(1) & (2) & (3) & (4) & (5) \\
\hline
G 37-26 & A4p & 6016 & 4.4 & $-1.95$ \\
G 48-29 & sd:A2 & 6295 & 4.0 & $-2.66$ \\
G 102-27 & G0 & 5423 & 4.0 & $-1.05$ \\
G 163-78 & K0 & 5400 & 4.2 & $+0.45$ \\
G 165-39 & A4 & 6330 & 4.0 & $-1.96$ \\
G 176-11 & M0.5 & 3544 & 4.9 & $+0.00$ \\
G 245-32 & sd:F2 & 6346 & 4.5 & $-1.62$ \\
HD 5916 & G8III & 4755 & 2.0 & $-0.80$ \\
HD 11636 & A5V & 9000 & 4.0 & $+0.16$ \\
HD 19476 & K0III & 4940 & 3.1 & $+0.04$ \\
HD 19510 & A0 & 6109 & 2.6 & $-2.50$ \\
HD 25329 & K1V & 4840 & 4.9 & $-1.68$ \\
HD 39587 & G0V & 5953 & 4.5 & $-0.03$ \\
HD 44007 & G5IV & 4850 & 2.0 & $-1.71$ \\
HD 44478 & M3III & 3450 & 1.0 & $+0.00$ \\
HD 45282 & G0 & 5280 & 3.1 & $-1.52$ \\
HD 78479 & K3III & 4509 & 2.5 & $+0.57$ \\
HD 79028 & F9V & 5881 & 4.2 & $-0.08$ \\
HD 87141 & F5V & 6403 & 4.1 & $+0.04$ \\
HD 92588 & K1IV & 5044 & 3.6 & $-0.10$ \\
HD 92839 & CII & 2847 & & $+0.10$ \\
HD 95849 & K3II & 4430 & 2.3 & $+0.30$ \\
HD 106516 & F5V & 6247 & 4.4 & $-0.70$ \\
HD 109995 & A0p & 8262 & 3.5 & $-1.99$ \\
HD 110281 & K5 & 3950 & 0.2 & $-1.56$ \\
HD 111721 & G6V & 4995 & 2.5 & $-1.26$ \\
HD 116976 & K1III & 4550 & 2.0 & $+0.01$ \\
HD 120933 & K5III & 3820 & 1.5 & $+0.50$ \\
HD 122563 & F8IV & 4500 & 1.3 & $-2.74$ \\
HD 126327 & M7.5 & 3000 & 0.0 & $-0.58$ \\
HD 144585 & G5V & 5831 & 4.0 & $+0.23$ \\
HD 148783 & M6III & 3250 & 0.2 & $+0.00$ \\
HD 158148 & B5V & & & \\
HD 160933 & F5V & 5765 & 4.0 & $-0.39$ \\
HD 165195 & K3p & 4450 & 1.1 & $-2.24$ \\
HD 175545 & K2III & 4429 & 2.9 & $+0.29$ \\
HD 180928 & K4III & 4000 & 1.2 & $-0.55$ \\
HD 187111 & G8 & 4429 & 1.2 & $-1.54$ \\
HD 218935 & G8III & 4819 & 2.5 & $-0.31$ \\
HD 338529 & B5 & 6100 & 3.6 & $-2.41$ \\
\hline
\end{tabular}

\medskip
\emph{Notes}
Column (1): Name of star.
Column (2): Spectra type.
Column (3): Effective temperature in K.
Column (4): Log surface gravity in cm g s$^{-2}$.
Column (5): [Fe/H] in dex.
All parameters are from \citet{2004ApJS..152..251V}.
\end{table}

In the middle panel of Figure \ref{fig:compare_templates} we show a comparison of the DEIMOS and Indo-US templates.
The CaT measurements using the Indo-US templates are larger by a median difference 0.12 \AA{} (RMS 0.31 \AA ) although the difference and scatter is smaller for GCs 3 Gyr and older (median difference of 0.11, RMS difference of 0.19 \AA ).
However, at the lowest CaT strengths, the Indo-US based CaT strengths are stronger than the DEIMOS ones by $\sim 0.6$ \AA .
Most younger GCs ($< 2$ Gyr) also show stronger Indo-US based values than DEIMOS based ones.
For both the lowest metallicity GCs and for most of the younger GCs, this difference is due to the Indo-US templates better fitting the Paschen lines of hotter stars since the DEIMOS templates lack any stars hotter than the 6100 K HD 136202 \citep{2015A&A...579A..20M}.

\begin{figure}
\begin{center}
\includegraphics[width=240pt]{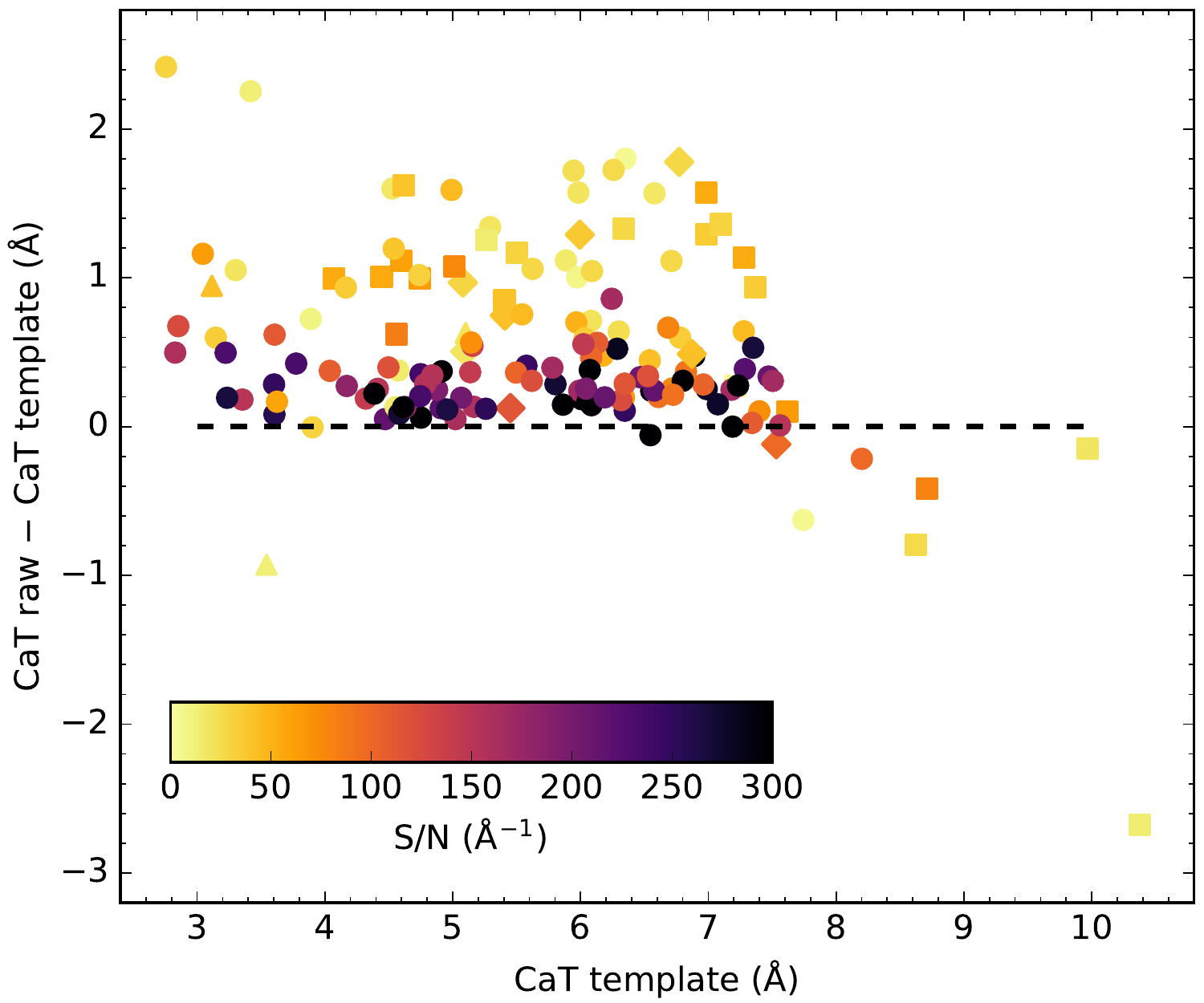}
\caption{Differences between CaT measurement using templates and direct measurement using same normalisation and index definition.
The points are colour-coded by S/N with MW GCs denoted by circles, LMC GCs by squares, SMC GCs by diamonds and Fornax dSph GCs by triangles.
At high S/N the directly measured CaT strengths are offset by a constant $0.21 \pm 0.03$ \AA{} while at low S/N the directly measured CaT values are stronger as the continuum normalisation procedure clips off statistical fluctuations as if they were absorption lines.}
\label{fig:raw_CaT}
\end{center}
\end{figure}

To assess the effects of the template fitting process on our measurements, we also directly ran our normalisation code on the raw spectra.
We plot a comparison of the CaT measured from the directly normalised spectra with the CaT values from the fitted templates in Figure \ref{fig:raw_CaT}.
At high S/N ($> 250$ \AA $^{-1}$), the CaT values directly measured from the spectra are offset to larger CaT values by $0.21 \pm 0.03$ \AA{} independent of CaT strength.
At lower S/N the directly measured CaT values are offset to even larger values since at lower S/N our continuum fitting process overestimates the height of the continuum when applied to noisier spectra.
This is not an issue for our template fitting process as it is the high S/N fitted templates that are continuum-normalised.

\subsubsection{Effects of continuum determination}

In this work we refined the continuum determination method used in \citet{2012MNRAS.426.1475U}.
We note that at the spectral resolution considered in this work, there are no true regions of continuum in the spectral range we study.
As such we fit a pseudo-continuum to regions of low line opacity by first masking wavelength regions with strong absorption before fitting a linear combination of polynomials with least squares to the masked templates.
Pixels more than 0.4 per cent below or 2 per cent above the fitted polynomials were rejected and the polynomials were refitted.
This continuum fitting process was repeated until no more pixels were rejected.
We note that the wavelength mask is only used in the first iteration of the fitting.

In \citet{2012MNRAS.426.1475U}, a wavelength mask based on the \citet{2003MNRAS.340.1317V} stellar population models was used.
We defined a new wavelength mask in the following manner.
We used synthetic spectra calculated with \textsc{ATLAS12} \citep{1970SAOSR.309.....K, 2013ascl.soft03024K} and \textsc{SYNTHE} \citep{1993sssp.book.....K} of a $T_{\rm eff} = 4300$, $\log g = 1.5$ [Fe/H] $= -0.5$ giant and a $T_{\rm eff} = 4240$, $\log g = 4.7$ [Fe/H] $= -0.5$ dwarf to define a new wavelength mask, stellar parameters typical of an old (12.6 Gyr) stellar population with this metallicity \citep[e.g.][]{2016ApJ...823..102C}. 
As in \citet{2008ApJ...682.1217K}, we identified continuum regions by smoothing the model spectra to the resolution of WiFeS ($R = 6800$) and selecting wavelength regions where the ratio of the smoothed flux to the continuum was greater than 0.95.
To ensure that the continuum regions, and the sections between them, were at least 1.2 \AA{} wide (2 pixels for WiFeS), we adjusted their limits. 
We then compared the new continuum mask with our integrated spectra of NGC 2808, NGC 6388, NGC 6528, NGC 7078 and NGC 7089 to identify absorption features either present in the observed spectra but not the model spectra or vice versa.
Most notably, we masked the cores of the hydrogen Paschen lines which are visible in metal poor GCs.
We change the order of the polynomial used to fit the continuum to 7 to match the order of the polynomial used by \textsc{pPXF} for fitting the continuum.
We also changed the limits for pixel rejection during normalisation to include pixels up to 2 per cent above the model rather than 1 per cent above as in \citet{2012MNRAS.426.1475U}.
This resulted in faster and more stable determinations of the continuum level.

In the bottom panel of Figure \ref{fig:compare_templates} we show comparison \citet{2012MNRAS.426.1475U} normalisation with the updated normalisation adopted here.
Most GCs show almost no difference between the CaT strengths measured using old and new normalisations (median difference of 0.03 \AA , RMS difference of 0.14 \AA ).
However, at high CaT strengths ($> 6.5$ \AA ) the old normalisation shows slightly higher CaT values for old GCs (median difference 0.07 \AA , RMS difference 0.10 \AA ) and slightly lower CaT strengths for young GCs (median difference $-0.13$ \AA , RMS difference 0.37 \AA ).

\subsection{CaT measurement systematics}
\label{sec:systematics}
\begin{figure}
\begin{center}
\includegraphics[width=240pt]{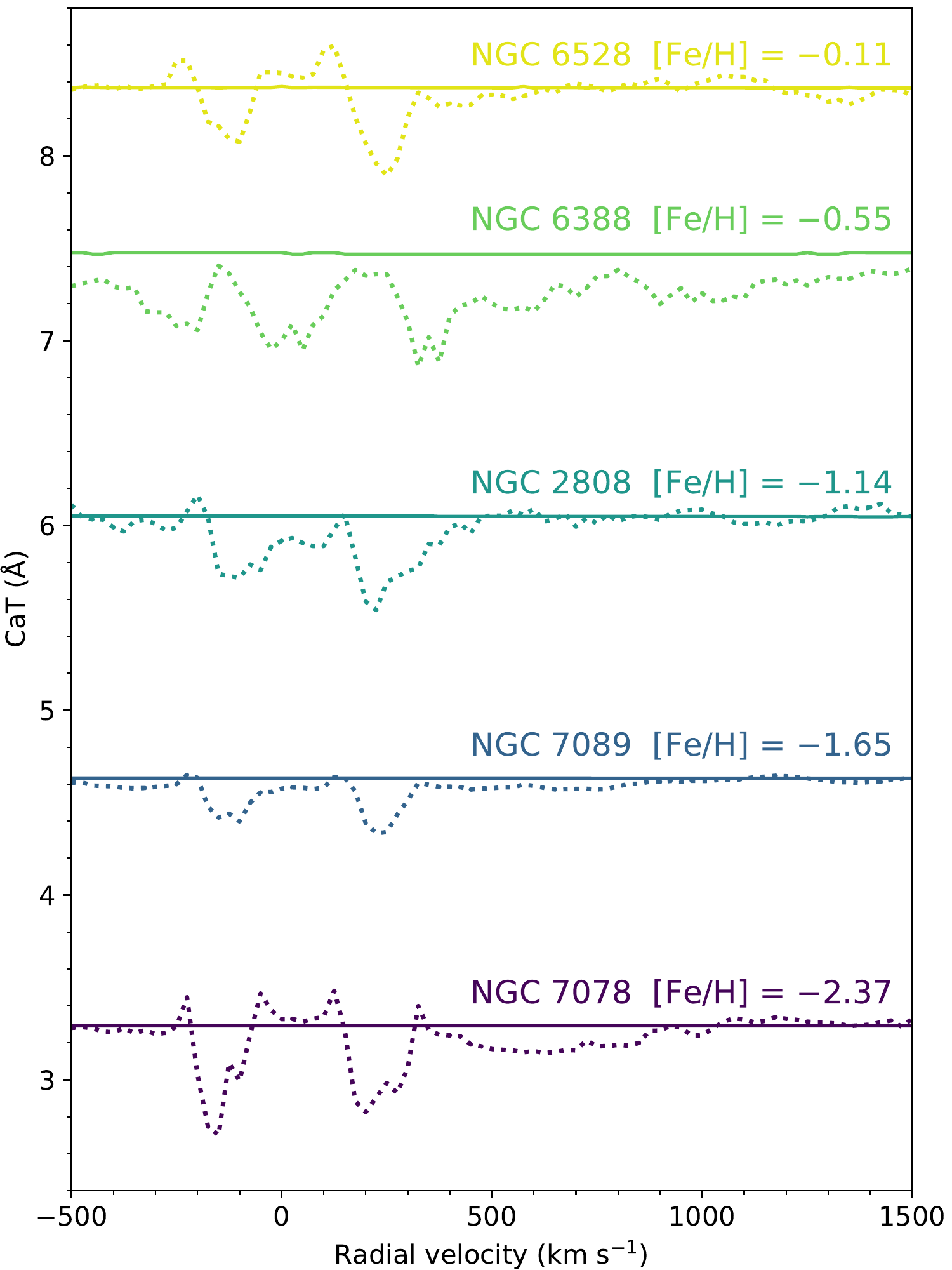}
\caption{Effects of radial velocity on CaT measurement.
The solid lines show the CaT strength of 5 GCs measured without masking sky emission lines when shifted to a range of radial velocities while the dotted lines show the CaT strengths measured when masking the sky lines. 
The lines are colour-coded by the metallicity of the GCs.
Masking sky lines introduces significant biases in the CaT measurements at radial velocities such as $\pm 200$ km s$^{-1}$ where the sky lines overlap with the CaT lines.}
\label{fig:rv_CaT_modelling}
\end{center}
\end{figure}

We investigated the effects of radial velocity, velocity dispersion and S/N on our measurement process by using our spectra of NGC 2808, NGC 6388, NGC 6528, NGC 7078 and NGC 7089.
These GCs were selected as the highest S/N spectra spanning the range of observed metallicities.
To simulate the effect of observing GCs at different radial velocities, we shifted these GC spectra to radial velocities between $-500$ and 1500 km s$^{-1}$ in steps of 25 km s$^{-1}$.
At each of these velocities we repeated our CaT measurement process.
The effects of these radial velocity shifts on our CaT measurements are shown in Figure \ref{fig:rv_CaT_modelling}.
When the radial velocity shifts the sky line wavelength mask onto the CaT lines, such as at $-125$ and $+225$ km s$^{-1}$, the CaT measurement becomes biased. 
Our choice of templates has no effect on this radial velocity bias.
As can be seen in Figure \ref{fig:mask_on_off}, we observe this bias in the difference between our masked and unmasked CaT measurements.
These variations are smaller at radial velocities above 500 km s$^{-1}$ where extragalactic GCs are typically studied.
Even when we do not mask the spectra, we still expect the CaT measurement to be noisier at radial velocities where the CaT spectral features overlap with skylines.

\begin{figure}
\begin{center}
\includegraphics[width=240pt]{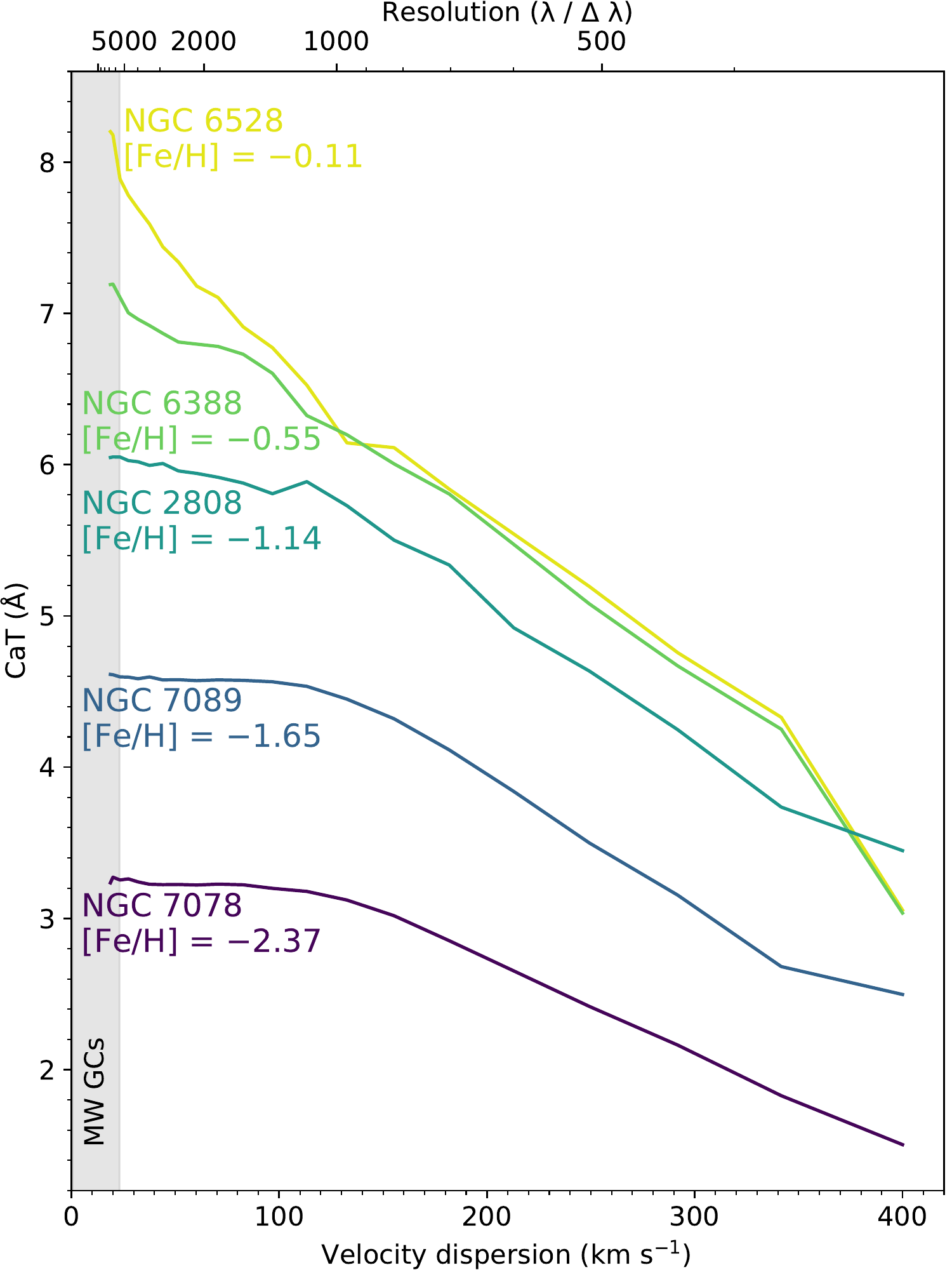}
\caption{Effects of velocity dispersion (or equivalently spectral resolution) on CaT measurement.
Like Figure \ref{fig:rv_CaT_modelling}, the lines are colour-coded by the metallicity of the GCs.
Higher velocity dispersions lead to weaker CaT values with more metal rich GCs being more strongly affected as well as being affected at lower velocity dispersions.
Our WiFeS spectra have a spectral resolution of $R = 6800$ (19 km s$^{-1}$) in the region of the CaT while literature central velocity dispersion measurements for MW GCs range from 0.6 to 23 km s$^{-1}$ with a median of 6 km s$^{-1}$ \citep{2018MNRAS.478.1520B}.
The lowest velocity dispersion plotted is the WiFeS resolution and the range of MW velocity dispersion is given by the shaded region.}
\label{fig:sigma_CaT_modelling}
\end{center}
\end{figure}

We velocity broaden these GC spectra to a range of 20 velocity dispersions evenly spaced in logarithmic space between 20 and 400 km s$^{-1}$ accounting for the resolution of our spectra (18.8 km s$^{-1}$ in velocity dispersion).
This is equivalent to lowering the spectral resolution of our data from $R = 6800$ to a range of resolutions between $R = 6400$ and $R = 320$.
This velocity dispersion range spans the observed range of not just GCs and nuclear remnants but also galaxies \citep[e.g.][]{2014MNRAS.443.1151N}.
The effects of velocity dispersion (or equivalently spectral resolution) are shown in Figure \ref{fig:sigma_CaT_modelling}.
For low and intermediate metallicities, the CaT strength is relatively insensitive to velocity dispersion up to $\sim 100$ km s$^{-1}$ ($R \sim 1200$) before declining with velocity dispersion.
For higher metallicity GCs, the CaT declines at all velocity dispersions.
For velocity dispersions above $\sim 100$ km s$^{-1}$ we can not effectively distinguish between the CaT strengths of [Fe/H] $= -0.55$ (NGC 6388) and [Fe/H] $= -0.11$ (NGC 6528) using our technique.
Whether we mask the sky lines or not and our choice of templates has virtually no influence on the effect of velocity dispersion on CaT measurements.
We note that the velocity dispersion of GCs -- 0.6 to 23 km s$^{-1}$ with a median of 6 km s$^{-1}$ in the MW \citep{2018MNRAS.478.1520B} -- is smaller or comparable to the instrument resolution.
Thus the velocity dispersion of our GCs has no effect on our measured CaT strengths except for our most metal-rich and most massive GCs where the effect is still relatively small ($\sim 0.1$ \AA ).
The effects of velocity dispersion would be important in studies of the galaxy light and the effects of spectral resolution would be important for studies carried out with a significantly lower resolution spectrograph.
 
\begin{figure}
\begin{center}
\includegraphics[width=240pt]{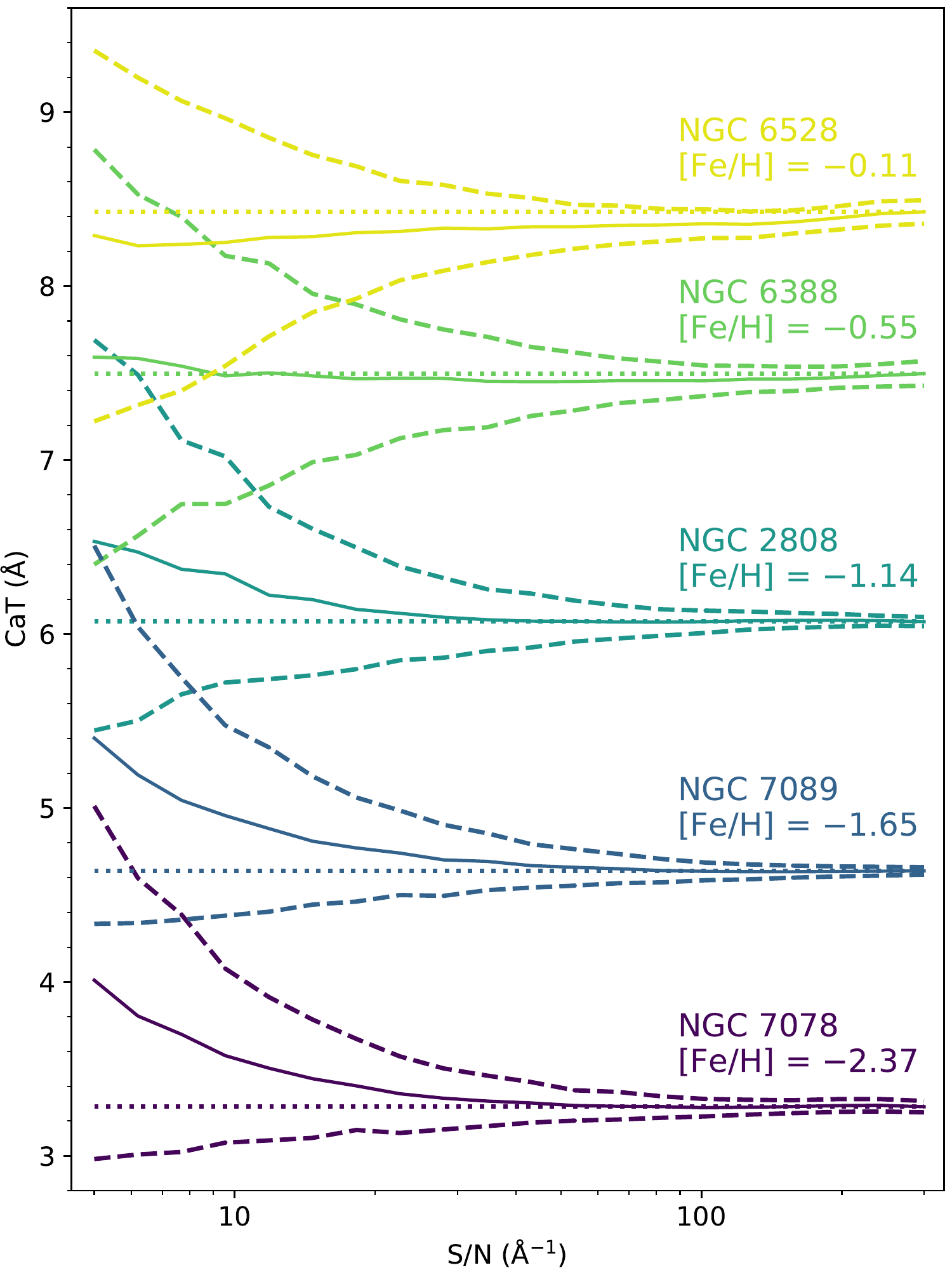}
\caption{Effects of S/N on CaT measurement.
For each GC, the solid line is the mean CaT value measured as a function of S/N, the dashed lines give the 68 percentiles of the measured CaT and the dotted line gives the CaT strength measured on the original spectra.
Like Figure \ref{fig:rv_CaT_modelling}, the lines are colour-coded by the metallicity of the GCs.
At low S/N, measurements of low metallicity GCs are biased towards higher values while the most metal-rich GCs are biased towards lower metallicities.
At fixed S/N, measurements of higher metallicity GCs have higher uncertainties.
As the median S/N of our sample is 116 \AA $^{-1}$, this S/N bias does not have a major effect on our measurements.}
\label{fig:s2n_CaT_modelling}
\end{center}
\end{figure}

We added Gaussian noise to each of these GC spectra to obtain a range of 20 spectra with S/N evenly spaced in logarithmic space with between 5 and 300 \AA $^{-1}$.
The effects of S/N on these five spectra are plotted in Figure \ref{fig:s2n_CaT_modelling}.
As previously reported by \citet{2012MNRAS.426.1475U}, at low S/Ns, template-based CaT measurements of low metallicity GCs are biased to higher values while the most metal-rich GC has its CaT strength biased to slightly smaller values.
This effect is less than 0.2 \AA{} for GCs with S/N $> 20$ \AA $^{-1}$ making it unimportant for this study as the median S/N of our spectra is 116 \AA $^{-1}$ and 127 of our 138 spectra have S/N $> 20$ \AA $^{-1}$.
However, it is important for extragalactic studies such as those of \citet{2012MNRAS.426.1475U} where most spectra have low S/N.

We repeated this S/N analysis for both masked and unmasked sky lines and for the Indo-US templates.
Measurements with masked sky lines showed larger biases with S/N (by up to a factor of 2) and larger uncertainties at fixed S/N (by up to 70 \%) compared to the unmasked CaT measurements.
Using the Indo-US templates also results in larger biases and uncertainties although to a lesser extent than the effects of masking sky lines.
Given the biases introduced by masking sky lines due to their overlap with the CaT lines at some radial velocities and the larger S/N based biases and uncertainties with both masking sky lines and the Indo-US templates, we use the unmasked, DEIMOS template-based CaT measurements in our analysis.
We give our unmasked, DEIMOS template-based CaT measurements in Table \ref{tab:measurements}.

\begin{table}
    
    \caption{CaT measurements}
    \begin{tabular}{ccccc}
        Name & rv & CaT & [Ca/H] & Notes \\
             & [km s$^{-1}$] & [\AA ] & [dex] \\
        (1) & (2) & (3) & (4) & (5) \\ \hline

NGC 104 & $-21.4_{-0.6}^{+0.8}$ & $6.90_{-0.10}^{+0.11}$ & $-0.55_{-0.04}^{+0.04}$ &  \\
Kron 3 & $133.3_{-1.8}^{+0.0}$ & $5.99_{-0.22}^{+0.17}$ & $-0.91_{-0.08}^{+0.07}$ &  \\
NGC 121 & $139.4_{-0.6}^{+0.5}$ & $5.45_{-0.29}^{+0.34}$ & $-1.12_{-0.11}^{+0.14}$ &  \\
NGC 288 & $-44.9_{-0.9}^{+0.8}$ & $4.53_{-0.14}^{+0.19}$ & $-1.49_{-0.06}^{+0.08}$ & LM \\
NGC 330 & $157.6_{-5.3}^{+4.5}$ & $6.77_{-0.21}^{+0.43}$ & $-0.60_{-0.08}^{+0.17}$ & Y  \\
        ... & ... & ... & ... & ... \\ \hline
    \label{tab:measurements}
    \end{tabular}

\medskip
\emph{Notes}
Column (1): GC name.
Column (2): Radial velocity in km s$^{-1}$ measured using DEIMOS templates on the unmasked spectra. 
Column (3): CaT strength in \AA{} measured using DEIMOS templates on the unmasked spectra. For both the radial velocity and the CaT strength the uncertainty has been estimated using a block bootstrap of the datacubes.
Column (4): [Ca/H] abundance in dex measured from the CaT strength using Equation \ref{eq:CaT_Ca_H_hires}.
Column (5): Notes on [Ca/H] measurement. `LM' indicates that the [Ca/H] measurement should be considered unreliable due to the low ($< 5 \times 10^{3}$ M$_{\sun}$) mass within the observed field-of-view while `Y' indicates that the [Ca/H] measurement should be considered less reliable since the GC is younger than 3 Gyr.
The full version of this table is provided in a machine readable form in the online Supporting Information.
\end{table}

\subsection{Classical indices}
\label{sec:classic}
\begin{figure*}
\begin{center}
\includegraphics[width=504pt]{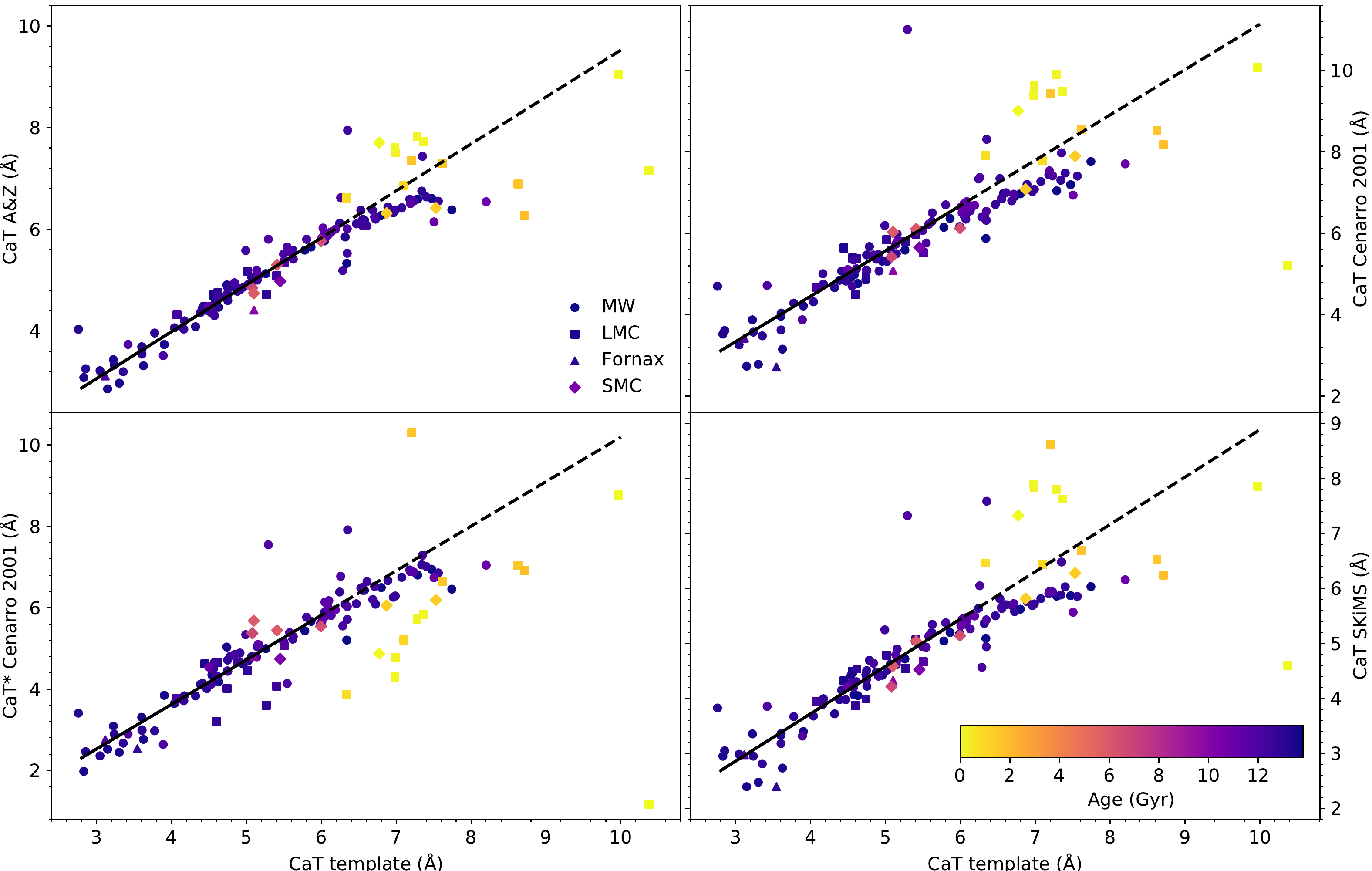}
\caption{Comparison of classical CaT index measurements and the template-based CaT measurements.
We compare our template-based CaT measurement with the CaT strength measured using the index definition of \citet{1988AJ.....96...92A} (top left), the CaT index of \citet{2001MNRAS.326..959C} (top right), the CaT* index of \citet{2001MNRAS.326..959C} (bottom left) and the CaT SKiMS index used by \citet{2009MNRAS.400.2135F} and \citet{2014MNRAS.442.1003P} (bottom right).
As in Figure \ref{fig:compare_templates}, in each panel the points are colour-coded by age with MW GCs denoted by circles, LMC GCs by squares, SMC GCs by diamonds and Fornax dSph GCs by triangles.
In each panel the black line is a fit between the classical index and the template-based index for old GCs with template-based CaT strengths of less than 6 \AA{} with a solid line for the range of CaT values fitted and a dashed line for the extrapolation to higher CaT values.
While each of the classical CaT indices follows the behaviour of the template  based CaT at low CaT strengths, at high CaT strengths old GCs show weaker classical CaT strengths relative to the trend at lower CaT strengths.
Younger GCs have relatively stronger classical CaT values compared to  template based strengths for all indices except the CaT* one.}
\label{fig:index_comparison}
\end{center}
\end{figure*}

We measured the CaT strengths on the observed spectra using a number of classical index definitions from the literature.
We used the \citet{1988AJ.....96...92A} definition (CaT A\&Z), the \citet{2001MNRAS.326..959C} definition (CaT Cenarro), the \citet{2001MNRAS.326..959C} definition with the correction for Paschen line absorption (CaT* Cenarro) and the \citet{2014MNRAS.442.1003P} SKiMS definition.
Besides the CaT indices, we also measured the \citet{2001MNRAS.326..959C} PaT Paschen line index.
For each index we adopted each author's index and continuum definitions and calculated the index strength and associated error using the formula provided in the appendix of \citet{2001MNRAS.326..959C}.
We plot our classical CaT index measurements in Figure \ref{fig:index_comparison}.

At low to intermediate CaT strengths ($< 6$ \AA ), all four of the classical indices show linear relationships with the template-based CaT measurements.
However, at high CaT strengths all of the classical CaT measurements are lower than extrapolations of the linear relationship between the classical indices and the template.
This is due to the stronger effects of line blanketing on the classical indices continuum passbands.
We note that the feature passbands are the same for the \citet{1988AJ.....96...92A} as for our template-based measurements so much of the difference in behaviour is due to the differences in the way the continuum is determined.
Most younger GCs have relatively stronger classical indices compared to template-based indices except for the CaT* index of \citet{2001MNRAS.326..959C} which shows weaker indices.
This is mostly due to the effects of the stronger Paschen line absorption at young ages on the feature passbands.
The CaT* index was designed to correct for Paschen line absorption but it appears to over correct the absorption from these H lines.
Unlike the template fitting method that can handle bad pixels such as those caused by poor sky line subtraction, all of the classical indices can be easily affected.

We performed the same tests on the effects of radial velocity, velocity dispersion and S/N on the classical CaT indices as we did for our template-based CaT measurements.
Unsurprisingly, measuring classical indices on spectra that have not been shifted to the rest frame can lead to catastrophically wrong index measurements.
As shown by previous authors \citep[e.g.][]{2001MNRAS.326..959C}, higher velocity dispersions lead to weaker classical CaT indices and less sensitivity to metallicity although the broader \citet{2001MNRAS.326..959C} indices show less dependence on the velocity dispersion (or spectral resolution).
Unlike the template-based measurements, the classical indices show no bias with S/N but show larger uncertainties at fixed S/N by at least 40 \% compared to the DEIMOS template-based CaT measurements.

\begin{figure}
\begin{center}
\includegraphics[width=240pt]{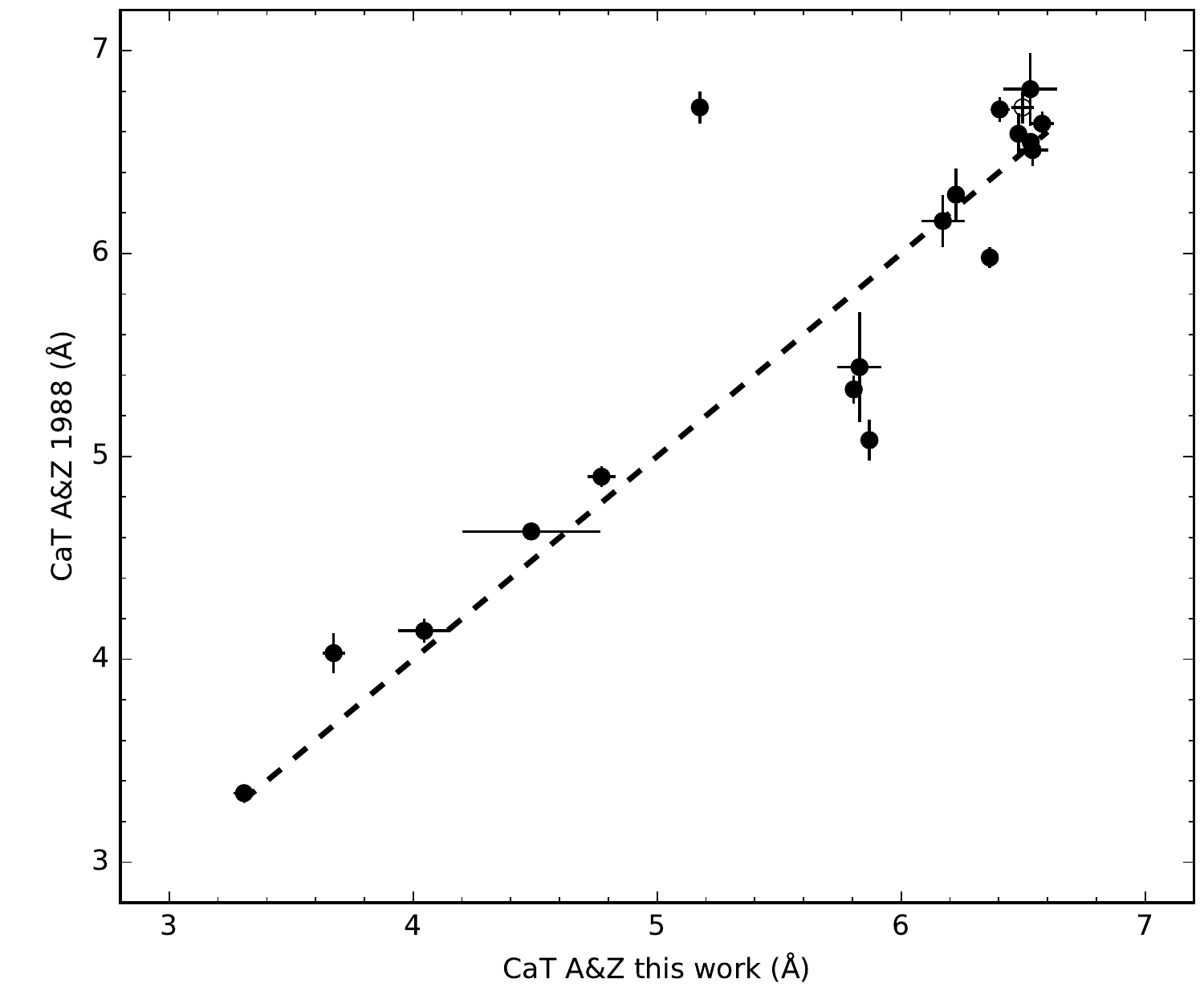}
\caption{Comparison between our CaT measurements and those of \citet{1988AJ.....96...92A} using the \citet{1988AJ.....96...92A} index definition.
For this measurement we have smoothed our spectra to match that of \citet{1988AJ.....96...92A} ($R = 2000$).
The empty point is the measurement of NGC 5927 with the long period variable NGC 5927 V3 removed from the extracted spectra by masking it in the datacube.
The dashed line is the 1-to-1 line.
While there is general agreement between the two studies, the difference between the two is larger than the mean uncertainties.}
\label{fig:AZ_CaT_comparison}
\end{center}
\end{figure}

The only previous study of the CaT strength of the integrated light of MW GCs is \citet{1988AJ.....96...92A}.
We smoothed our spectra to match their spectral resolution of $R = 2000$ and measured the index according to their definition on the smoothed spectra.
In Figure \ref{fig:AZ_CaT_comparison}, we show the comparison of the CaT measurements by \citet{1988AJ.....96...92A} with our measurements using the same index definition on the smoothed spectra.
We note that while the agreement is generally good, the root-mean-squared difference between the two studies (0.3 \AA ) is significantly larger than the mean uncertainty in the differences between the studies (0.1 \AA ) indicating that the uncertainties of one or both of the studies are likely underestimated.
These underestimated uncertainties are likely at least partially due to stochastic effects as the drift scanning by \citet{1988AJ.....96...92A} and our WiFeS observations likely covered different radial ranges (\citealt{1988AJ.....96...92A} do not specify how far they drift scanned nor over what range of their long slit observations they extracted their spectra other than saying that their scans covered at least the core radius).

\subsection{Uncertainties and repeated measurements}
\label{sec:repeats}
As in \citet{2012MNRAS.426.1475U} we used a Monte Carlo technique to estimate the uncertainties our CaT measurements.
Using the uncertainties provided by the \textsc{PyWiFeS} pipeline and the best fit combination of templates, we created 1024 realisations of the spectra and performed the template fitting, continuum normalisation and index measurement procedures on each. 
This is the same process as in \citet{2012MNRAS.426.1475U} although we use a factor of 10 more realisations resulting in a factor of 3 improvement in the precision of the uncertainty estimate.

We also used a block bootstrap technique to estimate the uncertainties in our CaT measurements.
We spatially divided each datacube into 48 blocks and extracted a spectrum for each of the blocks.
We then drew 1024 samples with replacement from these 48 block spectra.
For each sample we summed the sampled block spectra together and performed the template fitting, continuum normalisation and index measurement procedures on each. 
We then used the distribution of sample measurements to calculate the uncertainties. 

\begin{figure}
\begin{center}
\includegraphics[width=240pt]{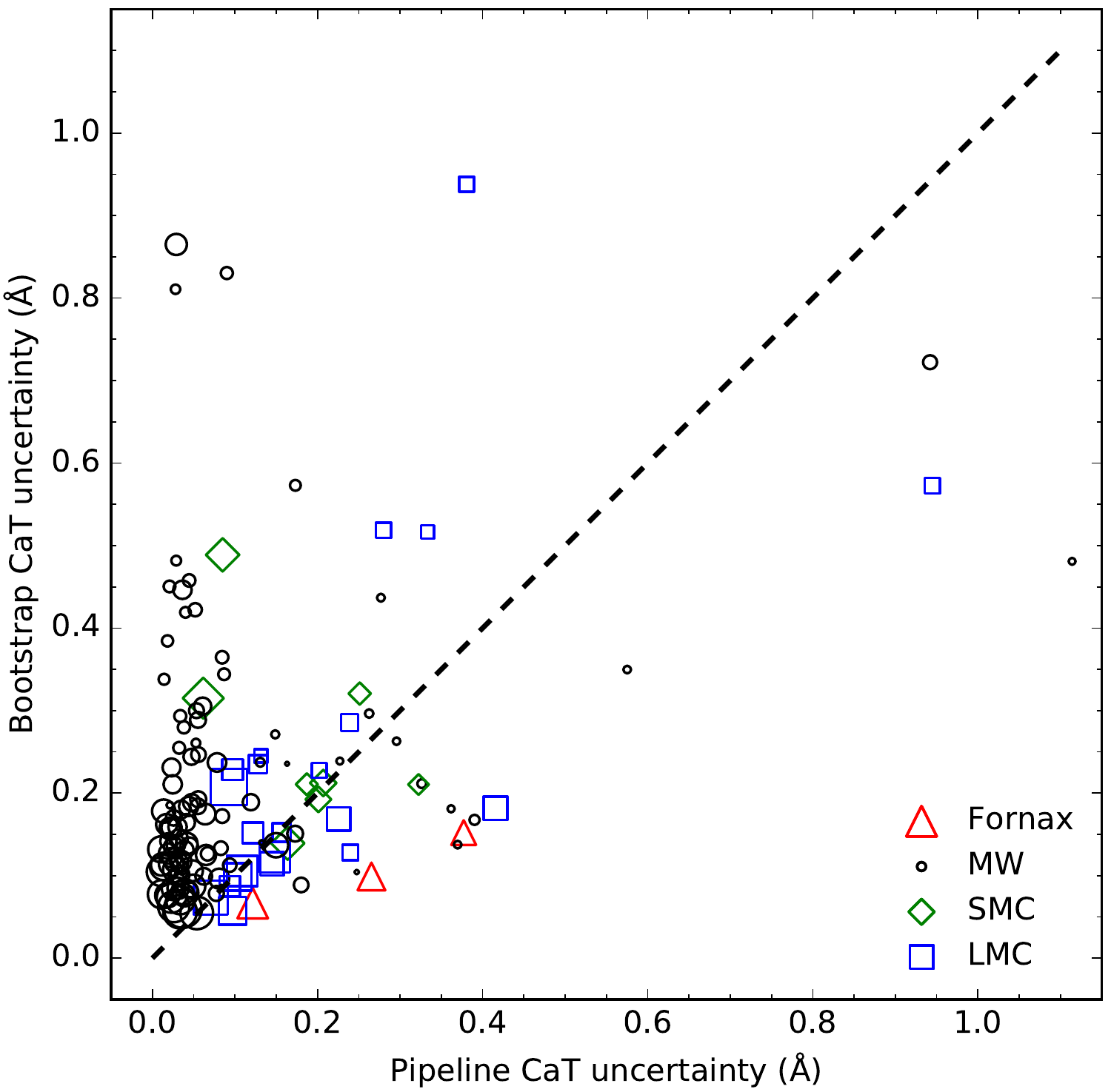}
\caption{Comparison of the CaT strength uncertainties derived using the block bootstrap technique with the uncertainties calculated from the uncertainties provided by the data reduction pipeline.
MW GCs are shown as black circles, LMC GCs as blue squares, SMC GCs as green diamonds and Fornax GCs as red triangles.
The size of the points is proportional to the mass enclosed within the field-of-view.
Most of our GCs show significantly larger bootstrap uncertainties compared to the pipeline based uncertainties.}
\label{fig:boot_stat_CaT}
\end{center}
\end{figure}

A comparison of the uncertainties provided by these two methods is given in Figure \ref{fig:boot_stat_CaT}.
In general, the bootstrap uncertainties are larger than those calculated using the Monte Carlo technique from the pipeline provided flux uncertainties at high S/N but smaller at low S/N (the pipeline uncertainties strongly correlate with the inverse of the S/N).
This effect is stronger for GCs with lower stellar masses in the field-of-view.
A challenge to studying the CaT in the integrated light of GCs in the Milky Way is that the CaT strength of individual stars depends strongly on surface gravity and hence luminosity.
Since the brightest giants have CaT strengths significantly stronger than dwarfs, stochastic variations in the number of bright giants within the observed field-of-view can noticeably change the measured CaT strength.
This significant stochastic variation within the datacubes was already noted in \citet{2017MNRAS.468.3828U} (see their figure 15).
The bootstrap uncertainties include this statistical effect, with observations of GCs such as NGC 5927 whose integrated light are dominated by a small number of bright giants showing the largest bootstrap uncertainties.
However, for observations such as some fields in NGC 6121 and NGC 6397 where there are no bright giants in the observed field-of-view, the bootstrap uncertainties underestimate the true uncertainties on the measurement.

Another way of estimating the uncertainties of our measurements is by comparing CaT indices in repeat spectra that are available for some of the GCs in our sample.
Our repeated observations fall into two classes.
First, we have cases where there is substantial spatial overlap between the repeated observations of the same GCs.
Eight GCs (NGC 1466, NGC 1846, NGC 2210, NGC 2808, NGC 6284, NGC 6304, NGC 6342 and NGC 6717) were observed twice and two (NGC 104 and NGC 7099) were observed three times in this manner.
Second, we have cases where the luminosity surface density is low and we deliberately observed multiple pointings of the same GC to obtain a more representative spectrum.
We note that the two different classes represent different regimes of enclosed mass since the first class has a median observed mass of $2.7 \times 10^{4}$ M$_{\sun}$ while the second class has a median observed mass of 800 M$_{\sun}$.

\begin{figure}
\begin{center}
\includegraphics[width=240pt]{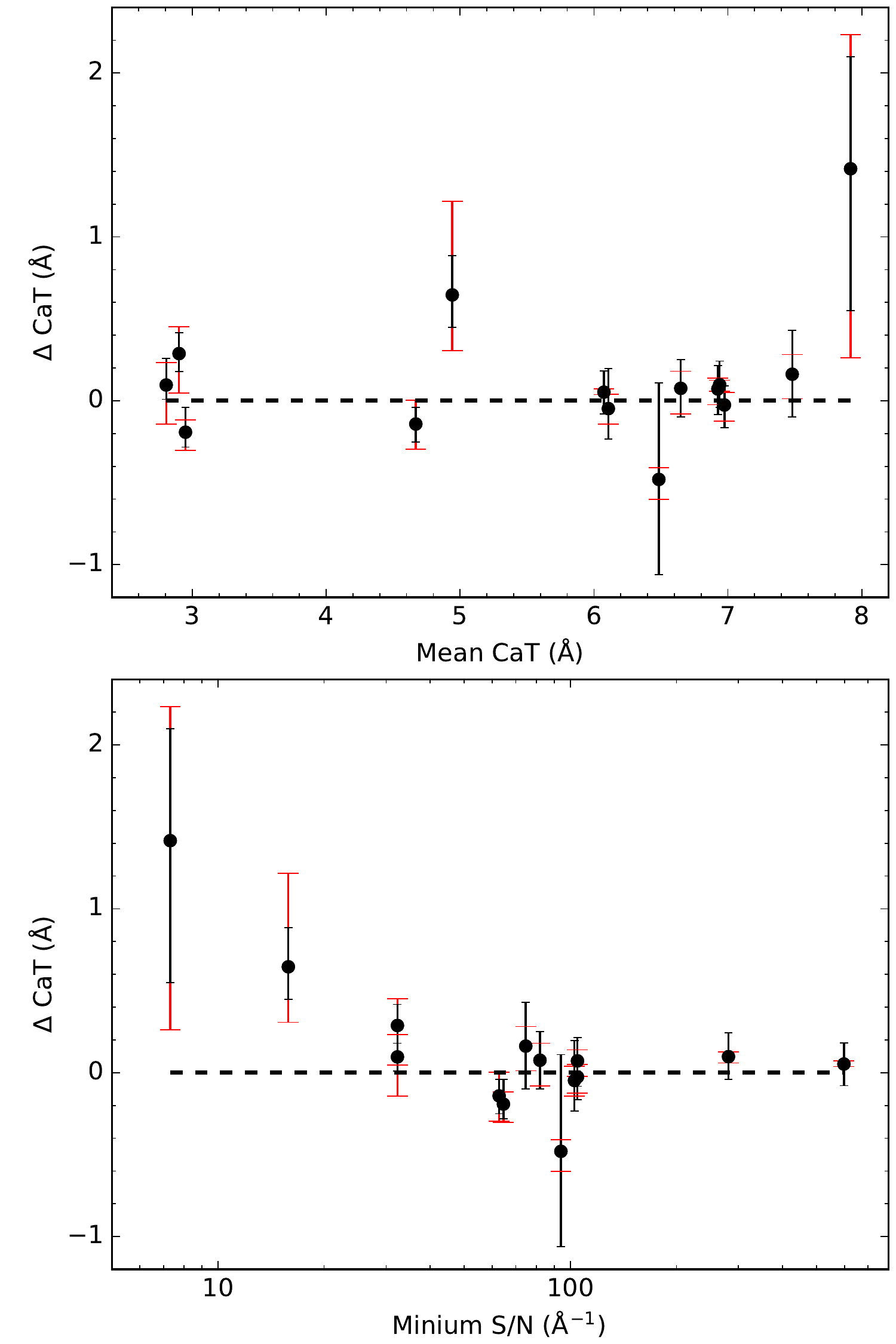}
\caption{Differences between repeated CaT measurements for observations with substantial spatial overlap.
In the top panel we show the difference as a function of the average CaT measurement of the two observations while in the bottom panel we show the difference as a function of the lower S/N of the two observations.    
The red error bars were calculated from the uncertainties provided by the data reduction pipeline while the black error bars were calculated using the block bootstrap technique.
The pipeline uncertainties generally underestimate the differences between repeated measurements while the bootstrap uncertainties are consistent.}
\label{fig:CaT_both_repeats}
\end{center}
\end{figure}

We show the difference of the repeated measurements in the first class in Figure \ref{fig:CaT_both_repeats}.
We find a median absolute difference of 0.12 \AA{} between repeated measurements and median absolute difference of 0.10 \AA{} for those measurements with at least a S/N of 25 \AA $^{-1}$.
We find a $\chi^{2}$ value of 57.2 for the 14 repeated measurements using the pipeline-based uncertainties and a $\chi^{2}$ value of 25.2 using the bootstrap uncertainties.
These $\chi^{2}$ values reduce to 53.2 and 13.1 for the 12 repeated measurements with S/N $> 25$ \AA $^{-1}$.
The difference in repeated CaT measurements is consistent with the bootstrap-based uncertainties but is underestimated by a factor of 2 by the pipeline-based uncertainties.
For subsequent analysis of the GCs in the first class we combined the repeated observations using a S/N weighted sum.

\begin{figure}
\begin{center}
\includegraphics[width=240pt]{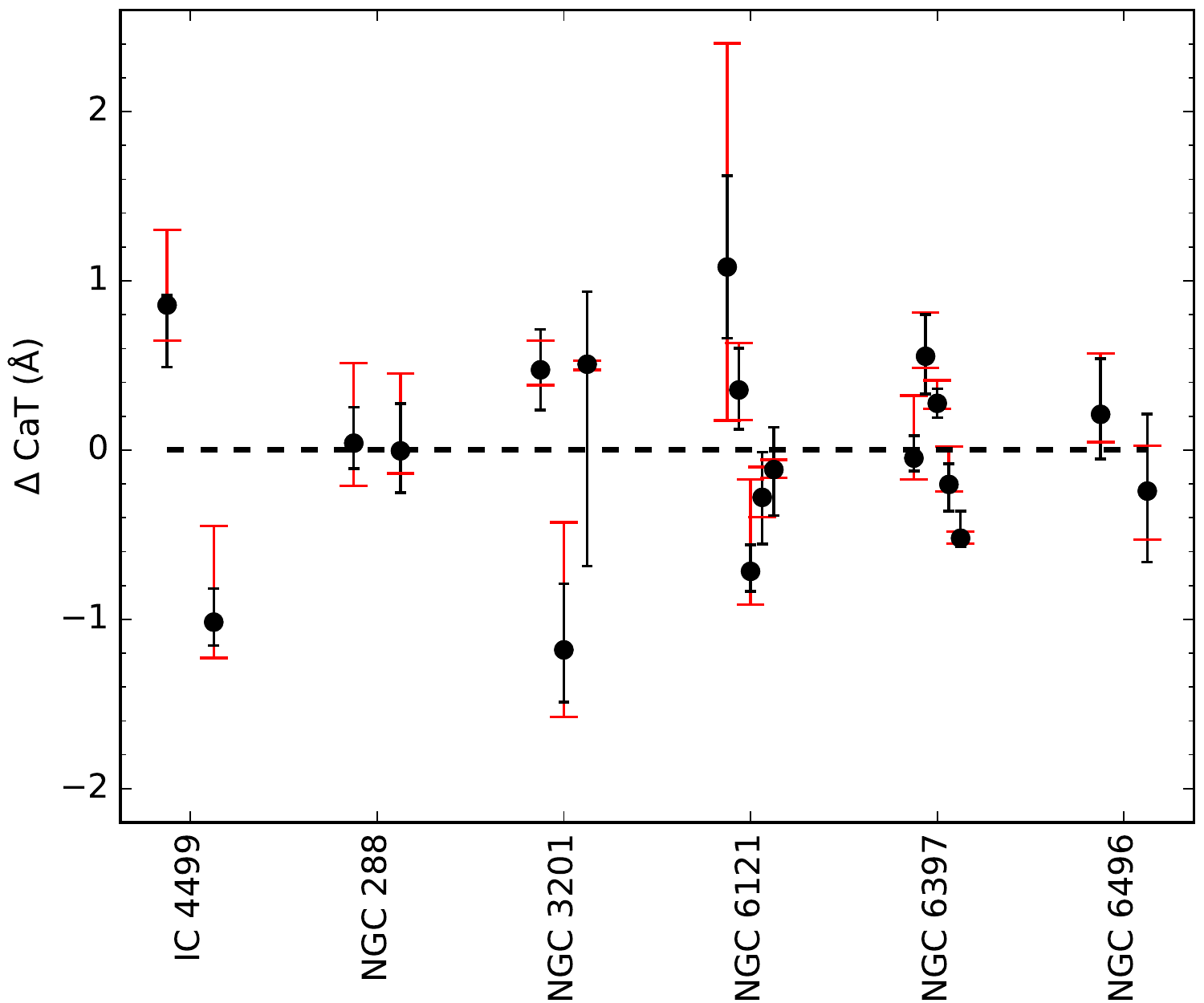}
\caption{Differences of CaT measurements between individual pointings and the combination of all pointings for GCs with multiple spatially distinct pointings.
The red error bars were calculated from the uncertainties provided by the data reduction pipeline while the black error bars were calculated using the block bootstrap technique.
Comparisons of different pointings of the same GC have been offset in the x-axis for legibility.
In most cases, the differences between pointings of the same GC are larger than the uncertainties.
These significant CaT strength spreads are due to stochastic variations in the number of bright giants in the WiFeS field.}
\label{fig:CaT_multi_pointing}
\end{center}
\end{figure}

For the second class we summed spectra of the individual pointings together with equal weights before performing our full measurement process on the summed spectra.
In Figure \ref{fig:CaT_multi_pointing}, we compare the CaT values of individual pointings with the measurements of the summed spectra.
In most GCs there are significant differences between the pointings due to the small number of giant stars in each pointing.
The case of NGC 6121 is instructive as an example of this.
By using catalogues from the HST ACS Globular Cluster Treasury Survey \citep{2008AJ....135.2055A} we identified what stars contribute to our datacubes.
The \textsc{NGC6121\_field1} pointing is dominated by a single bright ($M_{I} \sim -1.3$) RGB star and has a CaT measurement of 5.63$_{-0.23}^{+0.25}$ \AA{} while the \textsc{NGC6121\_field2} pointing, which has no stars brighter than $M_{I} \sim 2.1$ (corresponding to the base of the RGB), has CaT strength of 4.55$_{-0.12}^{+0.16}$ \AA .
We note that these two pointings each contain stellar masses of about 750 M$_{\sun}$ whereas our median pointing covers $2.5 \times 10^{4}$ M$_{\sun}$.

Our observations of NGC 5927 represent an extreme case.
We measure a CaT strength of $6.29_{-0.34}^{+1.39}$ \AA{} from our single pointing which is lower but not significantly so than other GCs with similar metallicities ([Fe/H] $= -0.49$ in the Harris catalogue).  
Inspection of our NGC 5927 datacube reveals that our integrated spectrum is dominated by the long period variable NGC 5927 V3.
We observed NGC 5927 just after the maximum of NGC 5927 V3 \citep{2010ApJ...719.1274S}.
The flux from within 2.5 pixels of NGC 5927 V3 contributes $\sim 25$ \% of the integrated flux of the datacube in the spectral region of the CaT while if the datacube was spatially uniform, only 2.3 \% of the flux would be within 2.5 pixels.
Our spectrum of NGC 5927 V3 reveals the strong TiO absorption bands and insignificant CaT absorption characteristic of a very cool giant.
The large bootstrap uncertainties of NGC 5927 show the large effect this single star has on our observations.
Re-extracting the spectrum of NGC 5927 while masking spaxels within 2.5 arcsec of NGC 5927 V3, we find a CaT strength consistent with GCs with similar metallicities ($7.52_{-0.19}^{+0.17}$ \AA ).
This large effect of a single star still occurs in the unmasked cube despite there being $4.6 \times 10^{4}$ M$_{\sun}$ of stellar mass within the WiFeS field-of-view.
Stochastic sampling of the IMF appears to be the dominant source of uncertainty for most of our observations.
We return to the impact of stochastic effects in Section \ref{sec:stochastic}.

\subsection{Kinematics}

As part of our template fitting process the radial velocity and velocity dispersion of each spectra are fit.
We give our radial velocity measurements in Table \ref{tab:measurements}.
For our repeated measurements we find an RMS difference of 5.2 km s$^{-1}$ which is significantly larger than either the pipeline-based or bootstrap-based uncertainties (both $\sim 1$ km s$^{-1}$).
We find similar sized differences in radial velocities of the same GC measured from different WiFeS gratings (Dalgleish et al. in prep.).
The origin of these systematic differences is currently under investigation.
As the radial velocity is a nuisance parameter in our analysis, these systematic differences have no effect on our index measurements. 

\begin{figure}
\begin{center}
\includegraphics[width=240pt]{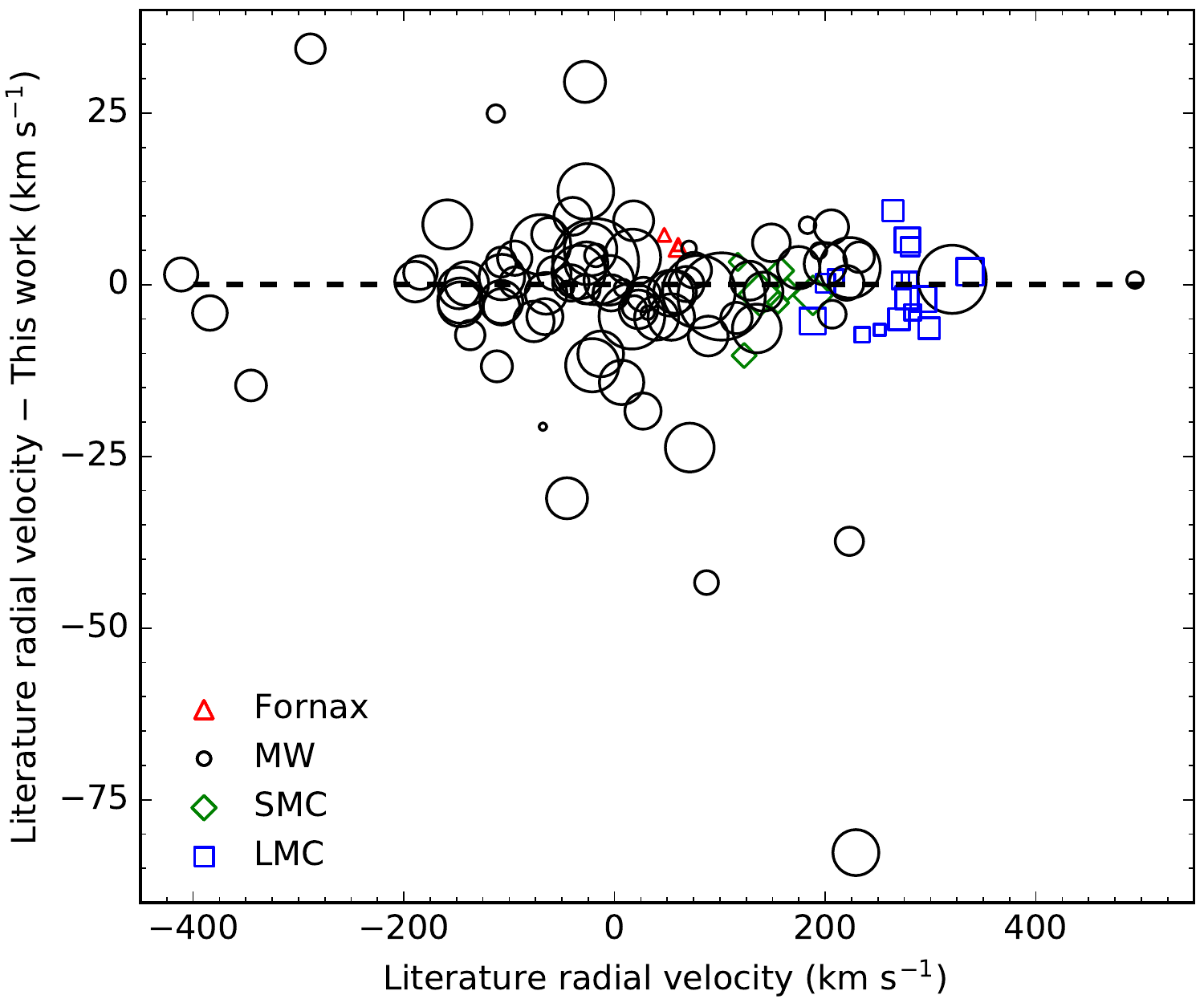}
\caption{Comparison between our mean radial velocity measurements and literature measurements.
The size of the points is inversely proportional to the S/N, and as in Figure \ref{fig:boot_stat_CaT}, MW GCs are shown as black circles, LMC GCs as blue squares, SMC GCs as green diamonds and Fornax GCs as red triangles.
While differences between our measurements and literature measurements are generally small (a median absolute difference of 4 km s$^{-1}$), there are a number of MW GCs with significant discrepancies.}
\label{fig:rv_comparison}
\end{center}
\end{figure}

We compare our radial velocity measurements with literature in Figure \ref{fig:rv_comparison}.
We are unaware of literature radial velocity measurements of NGC 411, NGC 1850, NGC 1856 and NGC 2004.
In general our measurements are in good agreement with literature values (the median of the absolute velocity differences is 4 km s$^{-1}$) but for a small number of MW GCs  we find significant differences ($> 25$ km s$^{-1}$) between our measurements and those in the \citet{1996AJ....112.1487H, 2010arXiv1012.3224H} catalogue.
As also noted by \citet{2018MNRAS.478.1520B}, several of the radial velocity measurements in the Harris catalogue are based on low resolution spectroscopy of small numbers of stars.
Of the GCs with larger than a 15 km s$^{-1}$ difference between our work and the Harris catalogue, two of our measurements agree with the values from high resolution spectroscopy (NGC 5634, \citealt{2015A&A...579A.104S} and NGC 6287, \citealt{2002AJ....124.1511L}), one is consistent within uncertainties (Pal 11) and the remainder (NGC 6235, NGC 6316, NGC 6333, NGC 6356, NGC 6496, NGC 6569 and NGC 6584) have values in the Harris catalogue based on older, low resolution work \citep[e.g.][]{1981ApJS...45..259W, 1984ApJS...55...45Z, 1986PASP...98..403H}.
We note that for these GCs our measurements are in general agreement with the recent low resolution measurements of \citet{2012A&A...540A..27S}, \citet{2016A&A...590A...9D} and \citet{2018arXiv180803834V}.
We find a median absolute difference of 3 km s$^{-1}$ between our measurements and those of \citet{2018MNRAS.478.1520B}.
The two GCs with more than a 15 km s$^{-1}$ difference between our work and  \citet{2018MNRAS.478.1520B} either are consistent within uncertainties (Pal 11) or have a velocity based on only 13 stars in \citet[][NGC 6642]{2018MNRAS.478.1520B}.

Our velocity dispersion measurements are in general agreement with literature values (mostly from the \citealt{1996AJ....112.1487H, 2010arXiv1012.3224H} catalogue) for velocity dispersions greater than $\sim 5$ km s$^{-1}$.
A detailed study of the systematic uncertainties affecting our kinematics measurements is beyond the scope of this work and will be considered in future work.

\section{Analysis}
\label{sec:analysis}
\begin{figure*}
\begin{center}
\includegraphics[width=504pt]{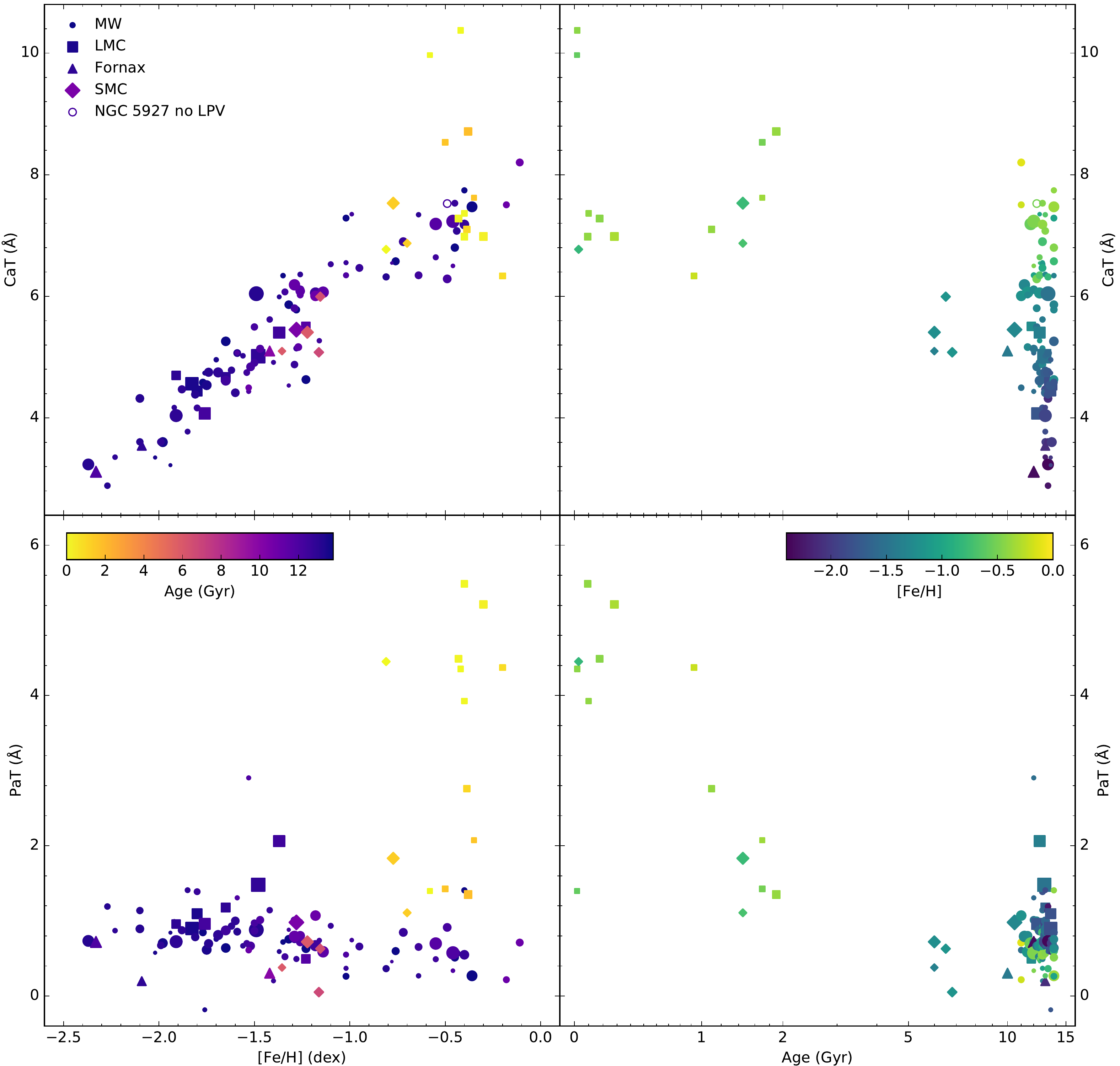}
\caption{\emph{Top left} CaT strength as a function of metallicity.
The points show our measurements colour coded by age with sizes proportional to the mass enclosed within the WiFeS field-of-view.
\emph{Top right} CaT strength as a function of age.
The points and models are colour coded by metallicity.
\emph{Bottom left} Strength of the Paschen lines using the index definition of \citet{2001MNRAS.326..959C} as a function of metallicity.
\emph{Bottom right} Strength of the Paschen lines as a function of age.
In both panels, the empty circle is the CaT measurement of NGC 5927 with the long period variable NGC 5927 V3 removed from the extracted spectra.
CaT strength increases with metallicity and is largely independent of age apart from at ages younger than 2 Gyr.
PaT strength decreases with increasing age.
The PaT indices of old GCs slightly decrease with increasing metallicity.
The PaT indices of $\sim 6$ Gyr GCs in the SMC are similar to old GCs with the same metallicities.}
\label{fig:CaT_PaT}
\end{center}
\end{figure*}

In Figure \ref{fig:CaT_PaT}, we plot our CaT measurements of all 113 GCs as a function of both age and metallicity.
Example spectra ordered by metallicity may be seen in Figure \ref{fig:cat_examples} and ordered by age in Figure \ref{fig:cat_age_examples}.
GCs 6 Gyr and older follow a roughly linear relationship between our template-based CaT strength and metallicity in line with previous studies \citep{1988AJ.....96...92A, 2012MNRAS.426.1475U, 2016MNRAS.456..831S}.
GCs younger than 2 Gyr, however, show more complicated behaviour. 
GCs with ages of 1.4 to 2 Gyr in the LMC and SMC show a range of CaT strengths with some showing elevated CaT strengths compared to old MW GCs with the same metallicity while other GCs show similar CaT strengths.
GCs in the age range 100 Myr to 1.1 Gyr also show a range of CaT strengths with some (but not all) GCs showing significantly weaker CaT strengths than GCs with the same metallicity.
However, the youngest GCs ($\sim 20$ Myr) in our sample show the strongest CaT strengths of any GC.
Before discussing the effects of age on the CaT in more detail, we will discuss the effects of metallicity and Ca abundance on the CaT.

\subsection{An empirical CaT-metallicity relationship}
\label{sec:CaT_Z}
Due to the lack of strong spectral features sensitive to their abundances in the optical, it is challenging (e.g. O) or impossible (e.g. Ne) to measure from stellar spectra the abundances of all elements that significantly contribute to the total metallicity Z.
Thus the abundance of a single element is often used as a proxy for the total metallicity.
Typically, the Fe abundance ([Fe/H]) is used as a proxy for stellar metallicity but [Fe/H] is only a direct proxy for metallicity for populations with the same abundance pattern.

\begin{figure}
\begin{center}
\includegraphics[width=240pt]{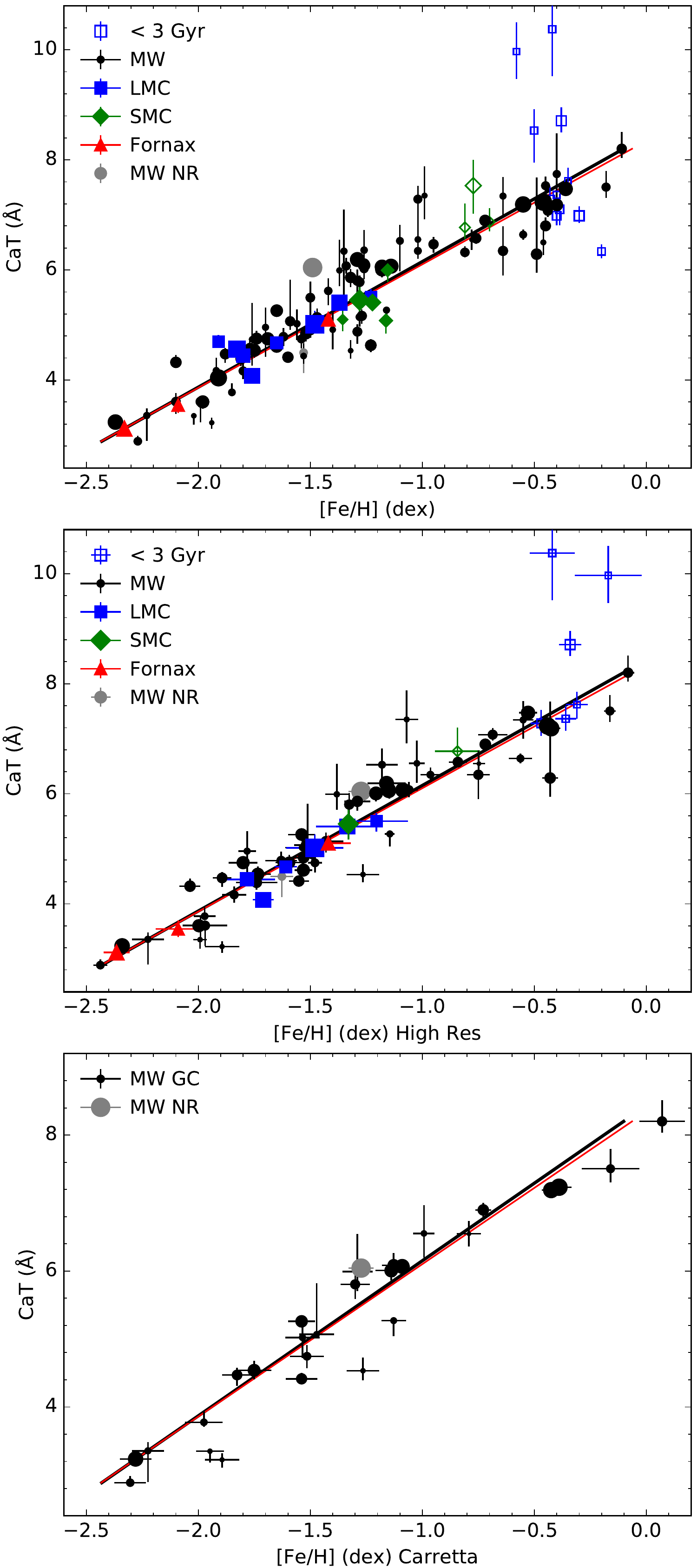}
\caption{CaT-[Fe/H] relationship.
\emph{Top} CaT strength versus all [Fe/H] measurements as in the upper left panel of Figure \ref{fig:CaT_PaT}.
\emph{Middle} CaT strength as a function of [Fe/H] from high resolution spectroscopy.
\emph{Bottom} CaT strength versus [Fe/H] from the Carretta et al. group.
In all three panels, black circles are MW GCs, grey circles are MW nuclear remnants, blue squares are LMC GCs, green diamonds are SMC GCs and red triangles are Fornax dSph GCs while empty points are GCs younger than 3 Gyr.
The size of each point is proportional to the mass observed within the WiFeS field-of-view.
The thick black solid line is the CaT-[Fe/H] relationship fit to the high resolution [Fe/H] measurements of old GCs (Equation \ref{eq:CaT_Fe_H_hires}) while the thin red solid line is the CaT-[Fe/H] fit to the Carretta et al. [Fe/H] measurements (Equation \ref{eq:CaT_Fe_H}).}
\label{fig:CaT_Fe_H}
\end{center}
\end{figure}

Given the common use of [Fe/H] to measure metallicity, we first compare our CaT measurements with our three samples of literature [Fe/H] measurements in Figure \ref{fig:CaT_Fe_H}.
As mentioned before, many of the GCs younger than 3 Gyr show a larger range in CaT strengths than GCs with the same [Fe/H].
We defer a detailed discussion of the effects of age on the CaT to Section \ref{sec:age} and focus only on GCs older than 10 Gyr in this subsection.
Unsurprisingly, there is less scatter in the CaT-[Fe/H] relation when using high resolution abundances.

We used the Markov Chain Monte Carlo (MCMC) code \textsc{pymc3} \citep{10.7717/peerj-cs.55} to fit [Fe/H] as a linear function of CaT strength accounting for the uncertainty in both variables.
We used weakly informative Gaussian priors on both the slope and the intercept ($\mu = 0$ and $\sigma = 10$ for both). 
We performed fits for both our sample of high resolution [Fe/H] measurements
\begin{equation}
\text{[Fe/H]}_{\text{High Res}} = 0.438_{-0.007}^{+0.007} \text{CaT} - 3.696_{-0.040}^{+0.037}
\label{eq:CaT_Fe_H_hires}
\end{equation}
and for the sample of [Fe/H] measurements by Carretta et al.  
\begin{equation}
\text{[Fe/H]}_{\text{Carretta}} = 0.446_{-0.013}^{+0.013} \text{CaT} - 3.720_{-0.072}^{+0.080}.
\label{eq:CaT_Fe_H}
\end{equation}
To reduce stochastic effects (see Section \ref{sec:stochastic}) we restricted our fits to GCs with stellar masses greater than $5 \times 10^{3}$ M$_{\sun}$ within the field-of-view.
This gives us 54 GCs in our high resolution sample and 18 GCs in the Carretta sample.
Using the high observed mass, high resolution sample we find an RMS difference of 0.14 dex between the literature [Fe/H] and those calculated using Equation \ref{eq:CaT_Fe_H_hires} and a $\chi^{2}$ value of 131.5 for 52 degrees-of-freedom; for the high observed mass Carretta sample we find an RMS difference of 0.12 dex and a $\chi^{2}$ value of 36.7 for 16 degrees-of-freedom.
The $\chi^{2}$ value is calculated as
\begin{equation}
\chi^{2} = \sum \frac{(\text{[Fe/H]}_{\text{Calc.}} - \text{[Fe/H]}_{\text{Lit.}})^{2}}{\sigma^{2}_{\text{Calc.}} + \sigma^{2}_{\text{Lit.}}}
\end{equation}
and the degrees-of-freedom as the difference between the number of GCs and the number of model parameters (2 for a linear fit, 1 for a constant value).
This compares to median uncertainties in our calculated [Fe/H] of 0.05 dex and 0.07 dex for the high resolution and Carretta samples respectively.
The high $\chi^{2}$ value for the high resolution sample is partially driven by the unrealistically low uncertainties on some of the literature [Fe/H] measurements.
Given the similarities between the two relations, we adopt the high resolution sample for the rest of this work due to its lower uncertainty.

\subsubsection{Effects of calcium abundance}
\label{sec:Ca_Fe}
Most but not all GCs in the MW and other galaxies show enhanced [Ca/Fe] abundance ratios (e.g. Figure \ref{fig:sample}, \citealt{2005A&A...439..997P}, \citealt{2013ApJ...773L..36C}, \citealt{2016ApJ...829..116S}).
Since the CaT spectral feature is dominated by atomic Ca lines, we expect that CaT strength should depend on the Ca abundance as well as on the metallicity.
Unlike the light $\alpha$-elements (O, Ne and Mg), which are primarily formed via hydrostatic burning in high mass stars, but like the other heavy $\alpha$-elements (Si, S, Ar and Ti), Ca is mainly formed by explosive nucleosynthesis, mostly in core collapse supernovae but with a significant contribution from type Ia supernovae \citep[e.g.][]{1995ApJS..101..181W, 2009MNRAS.399..574W, 2018MNRAS.480..800H}.
As such the Ca abundance should more closely trace the total metallicity than the Fe abundance but not as well as the O or Mg abundances.
Stellar population synthesis modelling by \citet{2012ApJ...759L..33B} showed that the relationship between Z and CaT has little dependence on [$\alpha$/Fe] (equivalent to the relationship between [Fe/H] and CaT strongly depending on [$\alpha$/Fe]) while the models of \citet{2012ApJ...747...69C} also predict a significant dependence on [Ca/Fe] at fixed [Fe/H].
Observations by \citet{2016MNRAS.456..831S} showed that GCs in M31 with low [Ca/Fe] have lower CaT strengths compared to GCs with the same [Fe/H] but higher [Ca/Fe].
We also note that Ca does not show abundance spreads within GCs unlike the spreads in the lighter $\alpha$-process elements O and Mg observed in almost all GCs \citep[e.g.][]{2010ApJ...712L..21C}.

\begin{figure}
\begin{center}
\includegraphics[width=240pt]{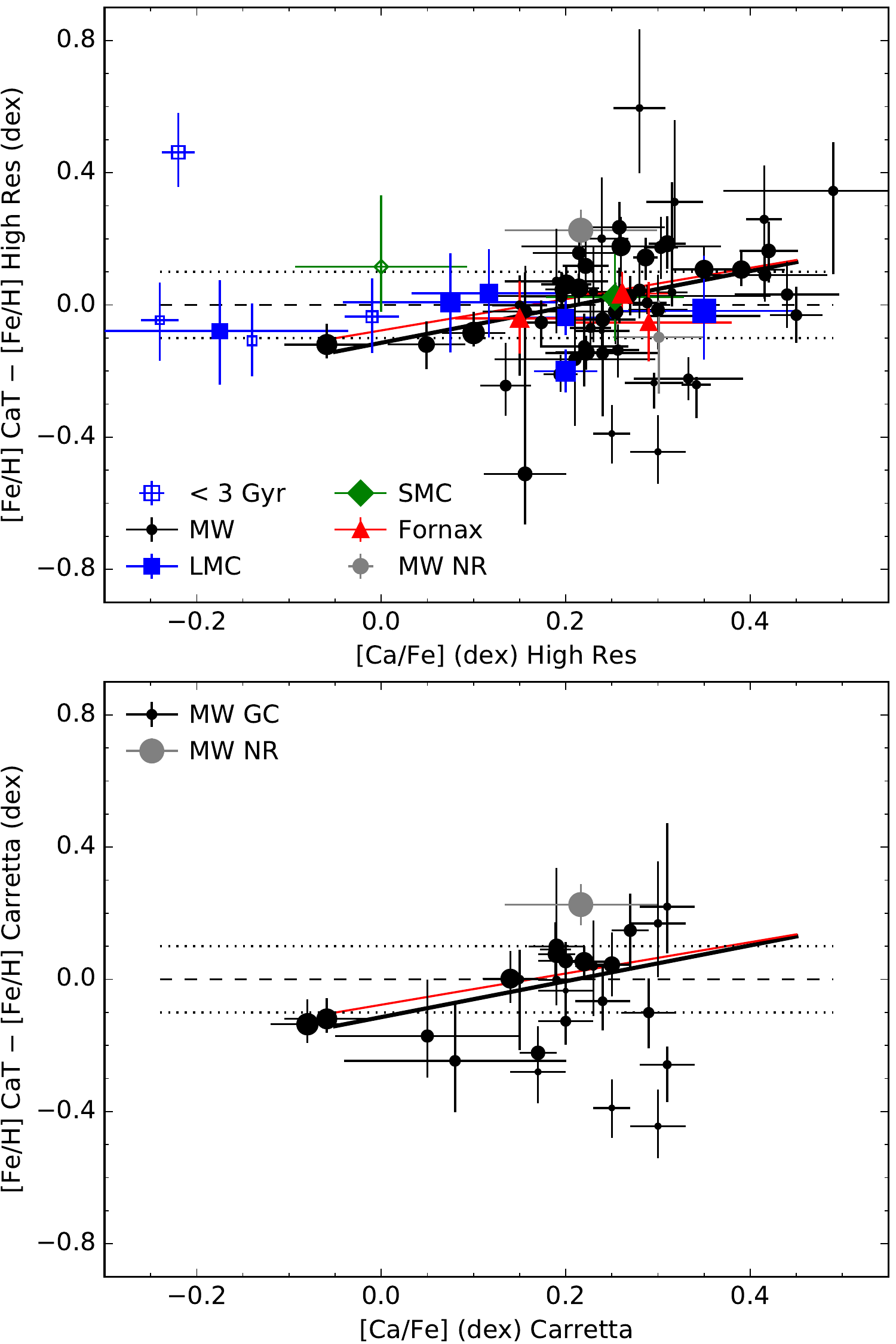}
\caption{Comparison of the [Fe/H]$_{\text{CaT}} - $[Fe/H]$_{\text{lit.}}$ difference and [Ca/Fe].
\emph{Top} [Fe/H] as a function of [Ca/Fe] based on high resolution spectroscopy.
\emph{Bottom} [Fe/H] strength versus [Ca/Fe] based on abundances from the Carretta et al. group.
The colours, shapes and size of the points are the same as in Figure \ref{fig:CaT_Fe_H}.
The thin red line is the $\Delta$[Fe/H]-[Ca/Fe] relationship fit with the Carretta et al. abundances while the thick black line is the $\Delta$[Fe/H]-[Ca/Fe] fit to the high resolution abundances of old GCs.
The dashed line shows $\Delta$[Fe/H] $= 0$ while the dotted lines show $\Delta$[Fe/H] $\pm 0.1$.
GCs with lower [Ca/Fe] ratios have weaker CaT strengths than GCs with higher [Ca/Fe].
The majority of GCs with high [Ca/Fe] ratios but weaker than expected CaT strengths have low stellar masses within the observed field-of-view and hence less reliable CaT measurements due to stochastic effects.}
\label{fig:delta_Fe_H_Ca_Fe}
\end{center}
\end{figure}

In Figure \ref{fig:delta_Fe_H_Ca_Fe}, we show the difference between [Fe/H] measurements from high resolution spectroscopy and the [Fe/H] calculated from our CaT triplet measurements as a function of the [Ca/Fe] ratio.
Although there is a lot of scatter, we do see a trend in that GCs with lower [Ca/Fe] values have lower CaT based [Fe/H] values and GCs with higher [Ca/Fe] have higher CaT based [Fe/H].
Using the high observed mass, high resolution sample, Kendell's correlation test gives evidence for a significant correlation between $\Delta$[Fe/H] and [Ca/Fe] (a correlation coefficient of $\tau = 0.30$ corresponding to a significance of $p = 0.001$), while the smaller, high mass Carretta sample does not possess a significant correlation ($\tau = 0.25$, $p = 0.16$). 

To further assess whether a relation exists between [Ca/Fe] and the CaT strength, we used \textsc{pymc3} to fit the difference between the CaT based [Fe/H] and the literature [Fe/H] as a function of [Ca/Fe] for the high observed mass, high resolution sample and for the high mass Carretta sample.
As before we used weakly informative Gaussian priors.
We find significant slopes for both our high resolution sample ($0.54_{-0.13}^{+0.10}$ from 52 GCs) and the Carretta sample ($0.47_{-0.19}^{+0.19}$ from 17 GCs).
We plot these relations in Figure \ref{fig:delta_Fe_H_Ca_Fe}.
The Bayesian information criterion (BIC) strongly favours a linear relationship between the [Fe/H] difference and [Ca/Fe] with a difference in BIC of 32.1 for the high mass, high resolution sample and a difference of 5.8 for the high mass Carretta sample.
Both favour the linear relationship over a constant difference.
We note that the relation between the residuals about the CaT-[Fe/H] fit and [Ca/Fe] is not driven by the decrease in [Ca/Fe] with increasing metallicity.
Our sample shows a range of [Ca/Fe] at most metallicities (Figure \ref{fig:sample}) and if the sample is split by metallicity, weak evidence for a relationship with [Ca/Fe] is seen at low, intermediate and high metallicities.
Our relationship between the difference in CaT based [Fe/H] and high resolution based [Fe/H] and [Ca/Fe] is broadly similar to the relationship that \mbox{\citet{2016MNRAS.455..199D}} found for the CaT strengths of individual giant stars.

Given that the CaT strength depends on the Ca abundance, we performed fits using \textsc{pymc3} of [Ca/H] as a linear function of CaT strength for both our high resolution sample
\begin{equation}
\text{[Ca/H]}_{\text{High Res}} = 0.434_{-0.009}^{+0.008} \text{CaT} - 3.384_{-0.044}^{+0.047}
\label{eq:CaT_Ca_H_hires}
\end{equation}
using the 52 high observed mass GCs with [Ca/H] measurements and for the Carretta sample
\begin{equation}
\text{[Ca/H]}_{\text{Carretta}} = 0.394_{-0.013}^{+0.014} \text{CaT} - 3.274_{-0.083}^{+0.080}
\label{eq:CaT_Ca_H}
\end{equation}
using the 17 high observed mass Carretta GCs with [Ca/H] measurements.
As for the [Fe/H]-CaT relationships, we use weakly informative Gaussian priors and only fit GCs with at least a mass of $5 \times 10^{3}$ M$_{\sun}$ in the observed field-of-view.
Using the high observed mass, high resolution sample we find an RMS difference of 0.14 dex between the literature [Ca/H] and those calculated using Equation \ref{eq:CaT_Fe_H_hires} with a $\chi^{2}$ value of 94.9 for 50 degrees-of-freedom; for the high observed mass Carretta sample we find a RMS difference of 0.10 dex and a $\chi^{2}$ value of 17.0 for 15 degrees-of-freedom.
These compare with median uncertainties in [Ca/H] of 0.07 dex for both literature samples respectively.
The lower $\chi^{2}$ values for [Ca/H] compared to [Fe/H] show that there is a  stronger relationship between [Ca/H] and CaT strength than between [Fe/H] and CaT.
This conclusion is further supported by the fact that [Ca/H] abundances are more uncertain than the [Fe/H] ones.
Unsurprisingly, the BICs for the [Ca/H] fits are significantly smaller than for the [Fe/H] fits (differences of 15.3 and 12.6 for the high resolution and Carretta samples respectively).
This is also supported by the slightly lower RMS value for [Ca/H] for the Carretta sample although the high resolution sample shows similar RMS differences for [Fe/H] and [Ca/H].
Using bootstrapping we can estimate the uncertainty on these differences.
Drawing 1024 samples from the high observed mass GCs, the $\chi^{2}$, BIC and RMS values are smaller for [Ca/H] relation than for the [Fe/H] relation 96.6, 85.6 and 54.4 \% of the time for the high resolution sample and 97.3, 96.4 and 82.3 \% for the Carretta sample.

If we restrict our high mass, high resolution sample to the MW, we see larger differences between the [Ca/H] and [Fe/H] in RMS (0.15 dex for [Fe/H] versus 0.14 dex for [Ca/H]), reduced $\chi^{2}$ (a difference in $\chi^{2}$ of 29.9 for 40 degrees-of-freedom) and BIC (a difference of 20.0) compared to the entire high mass, high resolution sample.
The sample of 10 GCs in satellite galaxies is consistent with both the [Ca/H] and [Fe/H] relations but does not show a preferences for either relation.
We note that the median uncertainties in literature [Ca/H] measurements in the satellite galaxies are twice as large in the satellite galaxy GCs as in the MW GCs (0.06 versus 0.13 dex).
If we split the MW sample in half by [Fe/H], by the observed stellar mass or by the fraction of the GC observed, we see a preference for a relationship with [Ca/H] for each of the subsets at lower significance than for the entire MW sample.
Given the similarities between the two relations, we adopt the high resolution sample relation for the rest of this work due to its lower uncertainty.
We give [Ca/H] values based on Equation \ref{eq:CaT_Ca_H_hires} for each of our GCs in Table \ref{tab:measurements}.

\begin{figure}
\begin{center}
\includegraphics[width=240pt]{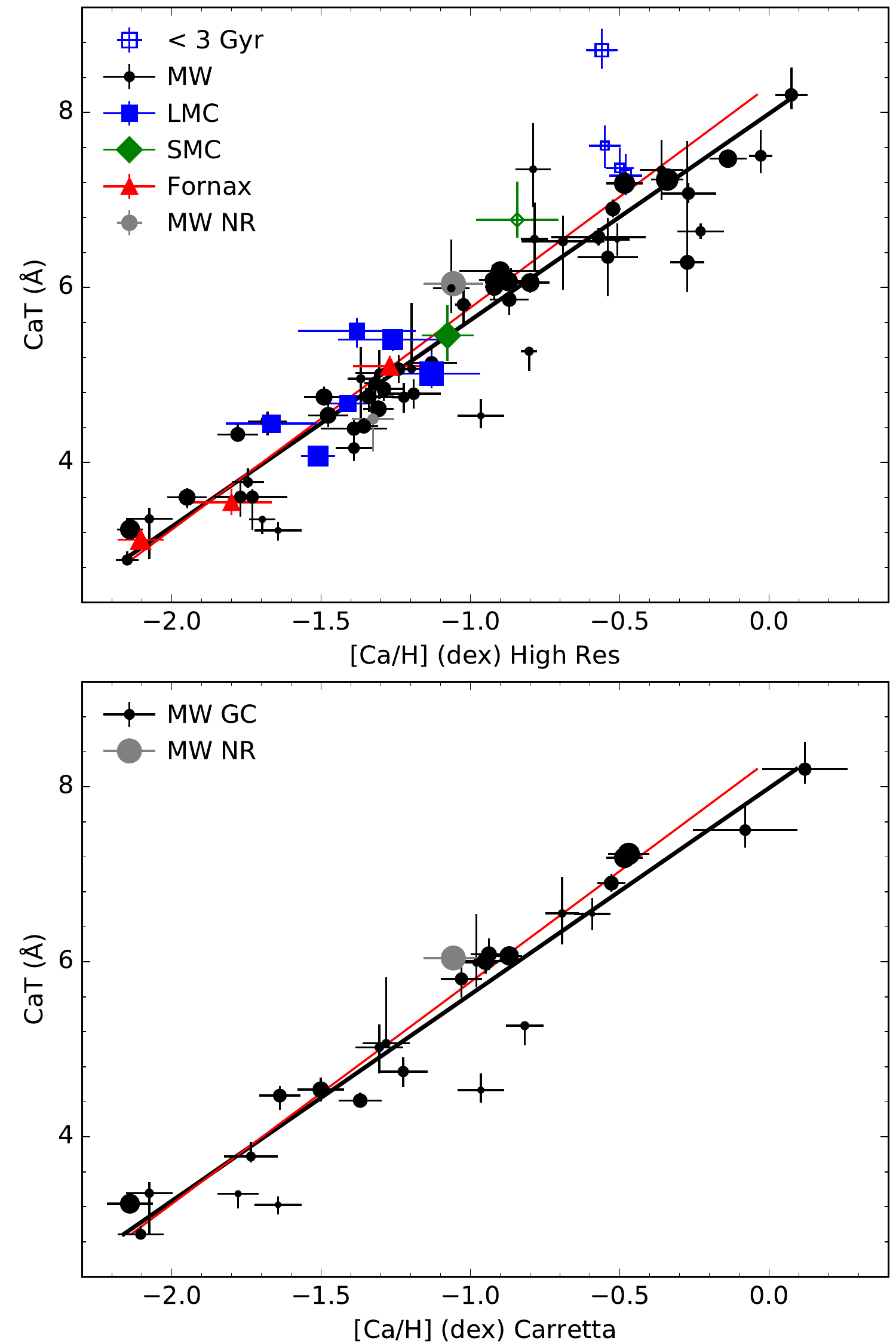}
\caption{CaT-[Ca/H] relationship.
\emph{Top} CaT strength as a function of [Ca/H] from high resolution spectroscopy.
\emph{Bottom} CaT strength versus [Ca/H] from the Carretta et al. group.
The colours, shapes and size of the points are the same as in Figure \ref{fig:CaT_Fe_H}.
The thick black solid line is the CaT-[Ca/H] relationship fit to the high resolution [Fe/H] measurements of old GCs (Equation \ref{eq:CaT_Ca_H_hires}) while the thin red solid line is the CaT-[Ca/H] fit to the Carretta et al. [Fe/H] measurements (Equation \ref{eq:CaT_Ca_H}).}
\label{fig:CaT_Ca_H}
\end{center}
\end{figure}

The two nuclear remnant in our sample, NGC 5139 and NGC 6715, show significant metallicity spreads ($> 1$ dex, e.g. \citealt{2010ApJ...722.1373J, 2010ApJ...714L...7C}). 
Both of these objects have CaT strengths in general agreement with their mean [Fe/H] and [Ca/H] abundances suggesting that the CaT can be used to measure the mean metallicities of stellar populations with significant metallicity spreads.

The CaT also depends on more than just the total metallicity and the Ca abundance.
The spectral range we fit with stellar templates and use to determine the continuum contains atomic lines of a number of elements including Mg, Si, Al, Ti and Fe as well as CN and TiO molecular bands.
As such the abundances of all these elements can have effects our CaT strength measurements.
We note that the effects of these other elements on the CaT spectral region are stronger at high metallicity than at low metallicity \citep[see figures 22 and 23 of][]{2018ApJ...854..139C}.
Abundance variations may also have minor effects the CaT strength via the effects of changing opacity on stellar structure and evolution \citep[e.g.][]{2007ApJ...666..403D, 2012ApJ...755...15V} or via the dependence of the singly ionised CaT on the electron pressure \citep[e.g.][]{2012ApJ...747...69C}.
However, the effects of varying abundances on spectral indices is usually dominated by effects of varying line opacity within the index definition rather than via the effects of varying abundances on the isochrones \citep{2007ApJS..171..146S, 2009ApJ...694..902L}.
We note that the dependence of the CaT on the abundance of other elements is a common problem for spectral indices \citep[e.g.][]{2005A&A...438..685K, 1995AJ....110.3035T, 2009ApJ...694..902L, 2012ApJ...747...69C}.

As noted earlier, the abundances of $\alpha$-elements do not always vary in lockstep due to the different nucleosynthetic origins.
For example, stars in dwarf galaxies and some stars in the MW halo follow different [Mg/Fe]-[Ca/Fe] relationships than the MW disc or bulge \citep[e.g.][]{2013A&A...560A..44V, 2013ApJ...778..149M, 2014A&A...572A..88L, 2018ApJ...852...49H}.
However, any abundance effect on the CaT by the [Mg/Ca] ratio is likely small since the Fornax GCs in our sample, which show low [Mg/Ca] ratios \citep{2006A&A...453..547L, 2012A&A...546A..53L} compared to MW or LMC GCs with similar [Ca/Fe] ratios \citep[e.g.][]{2010ApJ...717..277M, 2010ApJ...712L..21C}, follow the same CaT-[Fe/H] and CaT-[Ca/H] relations as the MW or LMC GCs within uncertainties.
Given the lack of a significant difference between CaT strengths of the Fornax GCs and MW GCs with the same [Fe/H] or [Ca/H] abundances, we expect the CaT strengths of extragalactic GCs to follow the same CaT-[Fe/H] and CaT-[Ca/H] relations as for our sample.

Finally, we note that all our GCs equal in age or older than NGC 1978 ($\sim 1.9$ Gyr, \citealt{2007AJ....133.2053M}) that have been studied show evidence for multiple populations \citep{2017arXiv171201286B, 2018MNRAS.473.2688M} such that a significant fraction of stars in the GCs in our sample show enhanced abundances of He, N, Na or Al and depleted C, O or Mg.
Changes in the abundances of these elements could affect the CaT via their effects on the horizontal branch morphology (He), on the pseudo-continuum (C, N and O) or on electron pressure (Na, Mg and Al).
These abundance changes also implies the abundances of GCs in our sample is not the same as field stars of the same age, metallicity or $\alpha$-element enhancement.
We return to the effects of multiple populations in Section \ref{sec:hb} where we study the effects of horizontal branch morphology on the CaT.

\subsubsection{The CaT at high metallicity}

The behaviour of the CaT at high metallicity has been debated.
The relationship between CaT strength and [Fe/H] observed by \citet{1988AJ.....96...92A} flattens at high metallicity.
The model predictions of \citet{2003MNRAS.340.1317V} and \citet{2016ApJ...818..201C} reproduce this behaviour.
Both authors explained this flattening as the effect of a cooler RGB at higher metallicity as CaT strength decreases rapidly with declining temperature for stars cooler than 3600 K \citep[e.g.][]{2002MNRAS.329..863C}.
\citet{2003MNRAS.340.1317V} and \citet{2016ApJ...818..201C} both explain this weakening of the CaT in cooler stars as due to the lower fraction of ionised Ca at lower temperatures.
However, the lower temperatures and higher metallicities also produce stronger line opacity in the CaT spectral region.
This lowers the flux in the pseudo-continuum passbands used to calculate the CaT index, weakening the measured CaT strength.
As seen in the top left panel of Figure \ref{fig:index_comparison}, this effect is stronger for the classical indices observed by \citet{1988AJ.....96...92A} and modelled by \citet{2003MNRAS.340.1317V} and \citet{2016ApJ...818..201C} than for our template based CaT measurements where the pseudo-continuum is calculated by iteratively fitting a polynomial to wavelength regions with low line opacity.

On the basis of a non-linear colour--CaT relation for the GCs in NGC 1407, \citet{2010AJ....139.1566F} questioned whether the CaT saturates at high metallicity.
However, the GC system of NGC 4494 shows a linear relationship between colour and the CaT \citep{2011MNRAS.415.3393F}.
\citet{2012MNRAS.426.1475U} found a wide range of colour-CaT relationships for their galaxies but found good agreement between metallicities calculated from CaT strengths measured using the template fitting technique and a linear fit to the \citet{2003MNRAS.340.1317V} stellar population models and metallicities calculated using Lick indices \citep{1994ApJS...94..687W, 1997ApJS..111..377W} and models such as those of \citet{2003MNRAS.339..897T} up to [Z/H] $= 0.2$ dex once the effects of [$\alpha$/Fe] in the \citet{2001MNRAS.326..959C} spectral library (used by the \citet{2003MNRAS.340.1317V} models) were accounted for.
\citet{2015MNRAS.446..369U} using stacked spectra of extragalactic GCs found a linear relationship between CaT strength and the sum of weak Fe lines in the CaT region in the range $-1.7 <$ [Z/H] $< -0.2$ suggesting that the CaT strength is a reliable metallicity indicator up to near solar metallicities. 
In their study of M31 GCs \citet{2016MNRAS.456..831S} found a clear linear relationship between CaT and [Fe/H] for GCs less metal-rich than [Fe/H] $= -0.4$ but their single higher metallicity GC (B193-G244, [Fe/H] $= -0.16$) showed weaker than expected CaT absorption, though they note that they had difficulty determining the continuum for this GC.

Our study shows a clear, linear relationship between CaT strength and [Ca/H] up to solar [Ca/H].
The GCs more metal rich than [Fe/H] $= -0.5$ show weaker CaT strengths than expected for their [Fe/H] due to the effects of lower [Ca/Fe] ratios.
The mean difference of the [Fe/H] abundances calculated from the CaT and literature [Fe/H] is $-0.12 \pm 0.05$ dex for the four GCs with [Fe/H] $= -0.5$ (excluding NGC 5927 due to its large observational errors) while the difference in [Ca/H] is $0.00 \pm 0.07$ dex for the same GCs.
The Carretta sample of abundances shows this more dramatically due to the higher [Fe/H] value for NGC 6528 ([Fe/H] $= 0.07$ for \citealt{2001AJ....122.1469C} versus [Fe/H] $= -0.08$ for average of literature studies) and the low [Ca/Fe] ratio for NGC 6388 and NGC 6441 \citep{2007A&A...464..967C}.
The two most metal-rich GCs in our study, NGC 6528 ([Fe/H] $= -0.11$, \citealt{1996AJ....112.1487H, 2010arXiv1012.3224H} catalogue) and NGC 6553 ([Fe/H] $= -0.18$, \citealt{1996AJ....112.1487H, 2010arXiv1012.3224H} catalogue), have CaT strengths consistent with our [Fe/H]-CaT and [Ca/H]-CaT relationships within uncertainties.
The saturation of the CaT at high [Fe/H] in observations of \citet{1988AJ.....96...92A} and in the models of \citet{2003MNRAS.340.1317V} and \citet{2016ApJ...818..201C} is likely due to the effects of line blanketing at high metallicity on the classical indices used as well as the effects of lower [Ca/Fe] ratios at high metallicity in the MW.

We note that GC stellar population studies in the Milky Way at high metallicity are a challenge as only NGC 6528 and NGC 6553 have near solar metallicities and manageable amounts of foreground reddening.
There is still considerable uncertainty in the abundances for these two GCs, with [Ca/Fe] measurements for NGC 6528 ranging from [Ca/Fe] $= -0.4$ \citep{2004A&A...423..507Z} to [Ca/Fe] $= 0.34$ \citep{2005MNRAS.356.1276O}.
Observing these bulge GCs is also a challenge due to the high field star surface density and reddening in the bulge.
It is more difficult to reliably determine the continuum at solar metallicity than at lower metallicities and the CaT measurement is significantly more sensitive to the spectral resolution (or velocity dispersion) at high metallicity (Figure \ref{fig:sigma_CaT_modelling}).
As is highlighted by the effect of a single cool giant in NGC 5927 (Section \ref{sec:repeats}), stochastic effects are likely stronger at higher metallicities due to the higher temperature sensitive of the CaT at the lower temperatures reached by the most luminous metal rich giants.
Taken together with observations in M31 and in more distant galaxies, our observations show that the CaT can be used as indicator of [Ca/H] up to at least solar metallicity, although the reliability of the CaT declines at near solar metallicity and higher.
The effects of Ca abundance limits the usefulness of the CaT in measuring [Fe/H] in populations with low [Ca/Fe] ratios such as those found at high metallicities.

\subsection{Behaviour of the CaT with age}
\label{sec:age}

\begin{figure}
\begin{center}
\includegraphics[width=240pt]{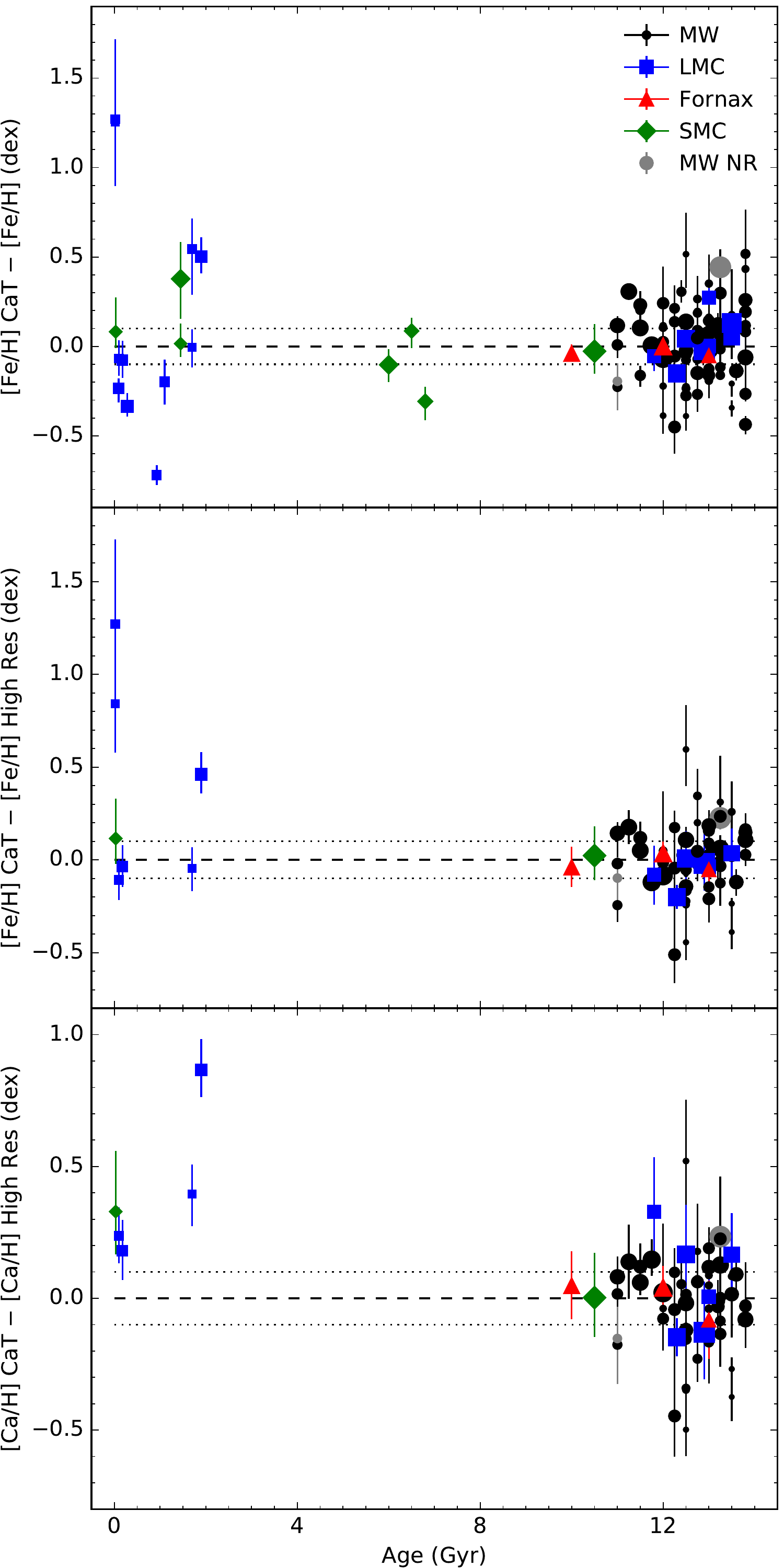}
\caption{\emph{Top} Differences in CaT based [Fe/H] with all literature [Fe/H] as a function of age.
\emph{Middle} Differences in [Fe/H] between CaT based and high resolution spectroscopy values.
\emph{Bottom} Differences in [Ca/H] between CaT based and literature values.
The dashed line shows $\Delta$[Fe/H] or $\Delta$[Ca/H] $= 0$ while the dotted lines show $\Delta$[Fe/H] or $\Delta$[Ca/H] $= \pm 0.1$.
For ages older than 3 Gyr, the strength of the CaT is independent of age.
GCs between 1.4 and 2 Gyr with significant contributions from C stars show significantly stronger CaT strengths than predicted by our CaT-metallicity relations but GCs without significant numbers of C stars in this age range show the same CaT strengths as old GCs with the same [Fe/H].
GCs younger than 1.4 Gyr show CaT strengths slightly weaker than old GCs with the same [Fe/H] except for the youngest GCs ($< 25$ Myr) which are dominated by RSGs.
The CaT values of young ($< 3$ Gyr) GCs are larger than their [Ca/H] would predict.}
\label{fig:delta_Fe_H_age}
\end{center}
\end{figure}

On the basis of our observations and stellar evolution calculations \citep[e.g.][]{2012MNRAS.427..127B, 2016ApJ...823..102C}, we can break the behaviour of the CaT into four age ranges.
To estimate the contribution of different stellar evolutionary phases to the CaT wavelength region, we used the predictions of the MIST isochrones \citep{2016ApJ...823..102C} with a metallicity of [Fe/H] $= -0.4$ (similar to the young and intermediate age GCs in the LMC) for the HST Wide Field Camera 3 F845M filter and a \citet{2001MNRAS.322..231K} IMF.
In the oldest age range, with GCs older than $\sim 3$ Gyr, the CaT is essentially insensitive to age (top right panel of Figure \ref{fig:CaT_PaT}, Figure \ref{fig:delta_Fe_H_age}).
As seen in Figure \ref{fig:CaT_PaT}, no significant change in Paschen absorption as measured by the PaT index with age is seen in this age range.
Here, the CaT spectral region is dominated by red giant branch (RGB) stars with smaller contributions from the main sequence (MS), the core He burning phase (CHeB - the red clump or horizontal branch) and the asymptotic giant branch (AGB).
Since the effective temperatures and surface gravities of the post-MS stellar evolution phases vary little within this age range and the relative contributions of different stellar evolution phases is nearly constant with age in this range, the CaT has little or no age dependence for populations older than $\sim 3$ Gyr.

The age range of $\sim 1.3$ to $\sim 3$ Gyr is one of transitions.
The importance of the RGB declines as the CHeB comes to dominate the CaT spectral region while the contribution of the AGB, especially the thermally pulsing AGB, reaches its maximum in this age range \citep{1998MNRAS.300..533G, 2013ApJ...777..142G}.
Additionally, the temperature of the MS turnoff increases towards younger ages.
Higher temperatures lead to weaker CaT absorption and stronger Paschen line absorption, especially above $\sim 6500$ K \citep{2002MNRAS.329..863C}, so as the contribution of MS turnoff stars hotter than this increases, the Paschen line strength of the integrated population increases and the CaT strength decreases.
This increase in Paschen line absorption at younger ages can be seen via the increase of the PaT index in Figure \ref{fig:CaT_PaT}.

\begin{figure}
\begin{center}
\includegraphics[width=240pt]{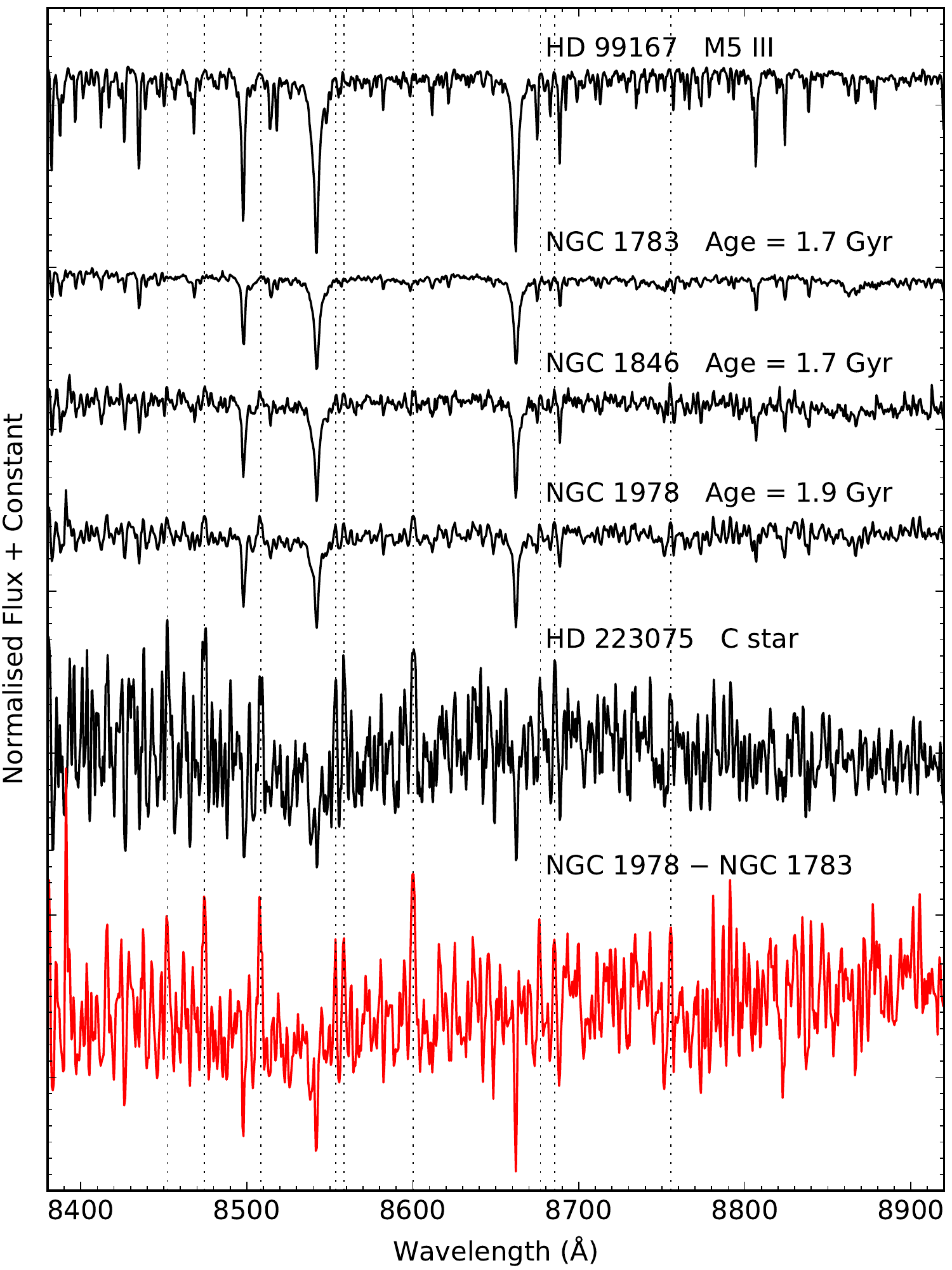}
\caption{Effects of C stars on the spectra of intermediate age GC spectra in the CaT region.
From top to  bottom we show the spectrum of an M giant from the Indo-US library \citep{2004ApJS..152..251V}, the spectra of three intermediate age GCs with similar metallicities, the Indo-US spectrum of a C star and the scaled difference of NGC 1978 and NGC 1783 in red.
We note that the S/N of the Indo-US C star spectrum is $\sim 450$ \AA $^{-1}$; the spectrum is dominated by real molecular features and not noise.
The dotted vertical lines show some of the stronger C star spectral features.
The spectrum of NGC 1978 clearly shows the molecular absorption characteristic of a carbon star while the NGC 1846 spectrum shows weaker carbon star features.
The spectrum of NGC 1783, however, shows no evidence for carbon stars.
Despite similar ages and metallicities for all three GCs, we measure significantly higher CaT strengths for NGC 1846 and NGC 1978 ($8.53_{-0.58}^{+0.39}$ and $8.71_{-0.21}^{+0.25}$ \AA{} respectively) compared to NGC 1783 ($7.62_{-0.26}^{+0.23}$ \AA ) due our inability to correctly place the continuum when C star molecular features are present.}
\label{fig:cat_inter_age_examples}
\end{center}
\end{figure}

The SMC GCs NGC 411 and NGC 419 have similar ages and metallicities (1.45 Gyr, [Fe/H] $= -0.7$, \citealt{2014ApJ...797...35G}) but different CaT values ($6.87_{-0.17}^{+0.25}$ \AA{} versus $7.53_{-0.51}^{+0.46}$ \AA ).
Likewise the three LMC GCs NGC 1783, NGC 1846 and NGC 1978 have similar ages (1.7 to 2.0 Gyr \citealt{2014ApJ...797...35G, 2007AJ....133.2053M}) and metallicities ([Fe/H] $\sim -0.4$, \citealt{2006AJ....132.1630G, 2008AJ....136..375M}) but have dramatically different CaT strengths ($7.62_{-0.26}^{+0.23}$, $8.53_{-0.58}^{+0.39}$ and $8.71_{-0.21}^{+0.25}$ \AA{} respectively).
Careful inspection of the spectra of these GCs (Figure \ref{fig:cat_inter_age_examples}) reveals that the GCs with stronger than expected CaT strengths show the molecular spectral features distinctive to carbon stars (C stars).
If we subtract a scaled spectrum of NGC 1783 from NGC 1978, we get a spectrum that resembles that of a C star.
The presence of more C stars in NGC 1978 and to a lesser extent NGC 1846 (compared to NGC 1783) is supported by \citet{1990ApJ...352...96F} who found that 7 of their 16 AGB stars in NGC 1978 were C stars, 9 out of the 21 stars in NGC 1846 and 5 out of the 17 stars in NGC 1783.
As can be seen in Figure \ref{fig:delta_Fe_H_age}, the GCs in this age range without major contributions of light from C stars have CaT based [Fe/H] values consistent with literature values but NGC 1783, the only such GC with abundances from high resolution spectroscopy, shows a stronger CaT based [Ca/H] than expected for its literature [Ca/H] abundance by 0.4 dex.
The GCs with C stars show CaT based [Fe/H] $\sim 0.5$ dex higher and CaT based [Ca/H] $\sim 1$ dex than their literature values.

The strong C dominated molecular absorption present in C stars is unlike the atomic and TiO dominated absorption present in M giants (Figure \ref{fig:cat_inter_age_examples}).
The strong molecular absorption present in C stars lowers the pseudo-continuum but for our template-based CaT measurements, the extra absorption in the CaT feature passbands leads to stronger CaT measurements than expected from old GCs with the same metallicity.
Since our set of Indo-US templates includes a C star but our DEIMOS templates do not, our Indo-US template-based measurements are more strongly affected by C stars.
For the four classical indices we consider, the presence of C stars leads to lower CaT indices strengths compared to the GCs with similar ages and metallicities.
We note that there is an observed lack of C stars at ages younger than 300 Myr or older than 4 Gyr in the LMC and SMC \citep[e.g.][]{1990ApJ...352...96F} in line with the stellar evolution models of \citet{2017ApJ...835...77M} who predict that the number of C stars per unit stellar mass peaks at an age of 1.6 Gyr, declining to virtually none for ages younger than 200 Myr or older than 2.5 Gyr.
Both observations \citep{2013ApJ...774...83B} and models \citep{2017ApJ...835...77M} find relatively few C stars at supersolar metallicities.

In the age range of $\sim 100$ Myr to $\sim 1.2$ Gyr, the CaT spectral region is dominated by CHeB and the MS turnoff is hot (effective temperatures of $\gtrsim 10000$ K).
Our GCs in this age range have slightly weaker CaT strengths than old GCs with the same metallicity and strong Paschen absorption.
The two GCs at 100 to 200 Myr with high resolution abundances however have [Fe/H] values consistent with those calculated from their CaT values (Figure \ref{fig:delta_Fe_H_age}); their calculated [Ca/H] are too high by $\sim 0.2$ dex.
Using the classical indices to measure the CaT strength in this age range results in stronger measured CaT strengths due to the contribution from the Paschen 13, 15 and 16 lines which overlap in wavelength with the CaT lines.
The CaT* index of \citet{2001MNRAS.326..959C} attempts to correct for this Paschen absorption but appears to over-correct at ages less than 1 Gyr as the CaT* indices in this age range are weaker than those of older GCs with similar metallicities. 
The lack of hot stars in the DEIMOS templates means that only the CaT lines themselves are fit and thus our DEIMOS based CaT measurements are less affected by Paschen absorption.
The Indo-US templates, however, do include hot stars and hence the associated CaT measurements are affected by the Paschen lines in a similar manner to the classical index measurements.
The smaller effect of Paschen absorption on the DEIMOS templates motivates our use of them rather than the Indo-US templates.

In the youngest age bin, with ages of only a couple tens of million years, the light in the CaT spectral region is dominated by supergiant stars.
The two youngest GCs in our sample - NGC 2004 and NGC 2100 - have ages of 20 Myr \citep{2015A&A...575A..62N}.
At this age $> 90$ \% of the light in the CaT region \citep{2014ApJ...787..142G} comes from red supergiant stars.
The light of these cool, very low surface gravity stars gives these GCs the strongest CaT values in our sample.
The light of the slightly older NGC 330 \citep[30 Myr,][]{2002ApJ...579..275S} is dominated by a mix of blue and red supergiants.
The presence of the blue supergiants gives NGC 330 similar Paschen absorption to older (100 Myr to 1 Gyr) GCs.
For NGC 330 we measure a broadly similar CaT strength to an old population with a similar [Fe/H] using the DEIMOS templates, but a stronger CaT strength than expected from its [Ca/H].

The only previous study to consider the effects of age on the CaT is that of \citet{2016MNRAS.456..831S} who found no strong effect down to ages of $\sim 2$ Gyr.
While this is in line with our observations, we note that \citet{2016MNRAS.456..831S} only studied three GCs younger than 8 Gyr and the ages used by \citet{2016MNRAS.456..831S} are mostly based on integrated light spectroscopy that provides much less reliable ages than those derived by fitting isochrones to photometry of resolved stars.

We find no age effect on our CaT strengths for GCs 6 Gyr old and older.
For GCs 2 Gyr old and younger, the CaT is a less reliable metallicity indicator.
While intermediate-age GCs with significant numbers of C stars and young GCs dominated by red supergiants have significantly stronger ($> 0.5$ dex) CaT based [Fe/H] values compared to their literature abundances, the difference between CaT based and literature [Fe/H] values is generally less than 0.3 dex for the other young and intermediate age GCs in our sample.
The difference in [Ca/H] is higher due to the low [Ca/Fe] abundance in these GCs. 
We note that we are limited in our ability to quantify how little the CaT changes between 10 Gyr and 6 Gyr by the lack of abundances from high resolution spectroscopy for the $\sim 6$ Gyr SMC GCs. 
We also note that we are limited to a relatively small range of ages and metallicities by the availability of high apparent surface brightness GCs in the MW and its satellite galaxies.
The CaT is much less sensitive to age than spectral indices commonly used to study metallicity such as the Mg$_{b}$ and Fe5270 Lick indices \citep[e.g.][]{1994ApJS...95..107W, 2007ApJS..171..146S, 2010MNRAS.404.1639V, 2012ApJ...747...69C}.

\subsection{Effects of the horizontal branch morphology}
\label{sec:hb}
The horizontal branch (HB) morphology can have a significant effect on the integrated light of a stellar population.
In particular, an old population with a particularly blue HB can mimic the appearance of a younger population in the optical \citep[e.g.][]{2000AJ....120..998L, 2004ApJ...608L..33S, 2005MNRAS.362....2T}.
GCs in the MW and other galaxies show a wide range of horizontal branch morphologies even at constant metallicity \citep[the second parameter effect, e.g.][]{1960ApJ...131..598S, 1965JRASC..59..151V}.
Recent work \citep{2010ApJ...708..698D, 2014ApJ...785...21M} suggest that much of the variation can be explained by age differences between GCs but a third parameter is also required to explain the range of observed HB morphologies.

In principle, a hot horizontal branch can affect the CaT through the effect of stronger Paschen lines contaminating the CaT.
In their study of GCs in NGC 1407, \citet{2010AJ....139.1566F} measured significant Paschen absorption in their fitted spectra and suggested that extremely blue HBs would lead to stronger CaT measurements.
However, the reanalysis of the same spectra by \citet{2012MNRAS.426.1475U} found no evidence for significant Paschen absorption while \citet{2015MNRAS.446..369U} found no evidence for strong Paschen absorption in their stacked GC spectra.

\begin{figure}
\begin{center}
\includegraphics[width=240pt]{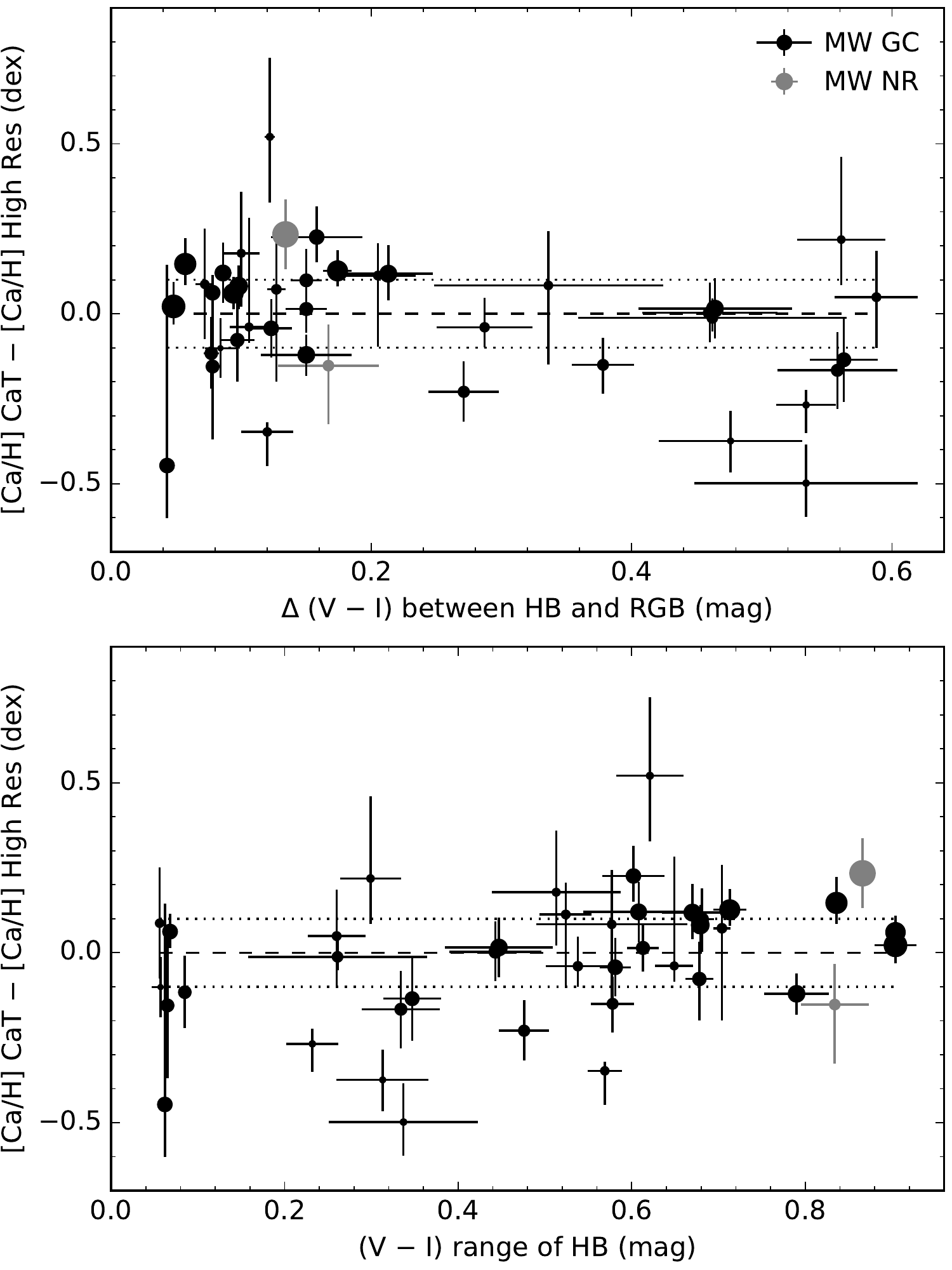}
\caption{Effects of HB morphology on the CaT.
\emph{Top} Difference between literature and our CaT based [Ca/H] as a function of the $(V - I)$ colour difference between the red edge of HB and the RGB \citep[The L1 parameter from][]{2014ApJ...785...21M}.
\emph{Bottom} Difference between literature and our CaT based [Ca/H] as a function of the $(V - I)$ colour range of the HB \citep[The L2 parameter from][]{2014ApJ...785...21M}.
The colours, shapes and size of the points are the same as in Figure \ref{fig:CaT_Fe_H}.
As in Figure \ref{fig:delta_Fe_H_Ca_Fe}, the dashed line shows $\Delta$[Ca/H] $= 0$ while the dotted lines show $\Delta$[Ca/H] $= \pm 0.1$.
The strength of the CaT does not significantly depend on the HB morphology.}
\label{fig:milone_hb}
\end{center}
\end{figure}

In our study we do not see clear evidence for old GCs with enhanced Paschen line absorption relative to other GCs at the same metallicity.
The handful of old GCs with PaT index measurements in Figure \ref{fig:CaT_PaT} above others with similar metallicity are due to low S/N or poorly subtracted sky lines.
To study whether there is a more subtle effect of the HB morphology on the CaT, in Figure \ref{fig:milone_hb} we compare the difference between our CaT based [Ca/H] with the [Ca/H] from high resolution spectroscopy with two measures of HB morphology from \citet{2014ApJ...785...21M}.
The first, the L1 parameter of \citet{2014ApJ...785...21M} measures the colour difference between the HB and the RGB.
This parameter depends on the age and metallicity of a GC.
The lack of a relationship between this parameter (Kendell's $\tau = -0.07$, $p = 0.63$) and the [Ca/H] difference indicates that we are capturing the behaviour of the CaT with age and metallicity with our fitted [Ca/H]-CaT relation at old ages.
The second, the L2 parameter of \citet{2014ApJ...785...21M} measures the colour range of the HB and correlates with the GC mass.
A weak correlation between the L2 parameter and the [Ca/Fe] difference is observed (Kendell's $\tau = 0.35$, $p = 0.01$).
However, this relation is driven by the highly uncertain measurements of NGC 5927 and NGC 6637.
Without GCs with uncertainties in calculated [Ca/H] larger than 0.2 dex, the correlation is non-significant (Kendell's $\tau = 0.25$, $p = 0.09$).

We can use the colour range of the HB as a proxy for the size of light element abundance variations within a GC since the colour range of the HB strongly correlates with the size of the He spread \citep{2014ApJ...785...21M} and with the range of the Na-O anti-correlation \citep{2007ApJ...671L.125C, 2010A&A...517A..81G}.
The lack of any relationship between the strength of the CaT and the colour range of the HB indicates that the CaT does not strongly vary with the strength of the multiple population phenomena.

\subsection{Stochastic and aperture effects}
\label{sec:stochastic}
Due to the limited field-of-view of WiFeS (25 by 38 arcsec) we only observe a fraction of each GC.
While for our most distant GCs -- those in the Fornax dwarf -- we observe $\sim 75$ \% of the light, for the median MW GC we only observe 10 \% of the light (see figure 2 in \citealt{2017MNRAS.468.3828U}).
This is a concern if the observed (central) part of the GC is not representative of the whole GC.

The observed part of the GC can be unrepresentative for two reasons.
First is stochastic sampling of all stellar evolutionary stages.
If the stellar mass sampled by the observations is relatively low, then the number of stars from short lived but bright stellar evolutionary phases can vary dramatically.
This is especially a concern for any spectral features in the far red, because of the important contribution to the integrated light by bright short-lived stars.
This effect is illustrated in Section \ref{sec:repeats} and in Figure \ref{fig:CaT_multi_pointing} through the difference in CaT strength in different pointings as well as by the effects of the long period variable NGC 5927 V3 on the spectrum of NGC 5927 and of variable numbers of C stars on the spectra of intermediate age GCs (Figure \ref{fig:cat_inter_age_examples}).
We note that since stochasticity is a function of the total stellar mass sampled by the observations, regardless of what fraction of GC mass that represents,  observations of low mass GCs observed in their entirety are affected in the same way as those sampling small fractions of massive GCs.

Secondly, if there is a radial stellar population gradient in the GC, the centre is no longer representative of the whole.
While normal GCs do not possess significant age \citep{2012A&A...541A..15M, 2015MNRAS.451..312N} or metallicity spreads \citep[e.g.][]{2009A&A...508..695C}, GCs do show star-to-star variations in the abundances of various light elements including He, N, O and Na \citep[See reviews by][]{2012A&ARv..20...50G, 2017arXiv171201286B}.
Radial gradients in abundances of these elements are observed in several GCs \citep[e.g.][]{2009A&A...505..117C, 2011A&A...525A.114L, 2015ApJ...804...71L}.
More importantly for the CaT, GCs are mass segregated due to dynamical effects \citep[e.g.][]{1999ApJ...523..752R, 2003AJ....126..815L, 2004A&A...425..509A, 2015ApJ...814..144B}.
Since the strength of the CaT strongly depends on surface gravity, a radial variation in the ratio of giant-to-dwarf stars should introduce a radial gradient in the CaT strength.

\begin{figure*}
\begin{center}
\includegraphics[width=504pt]{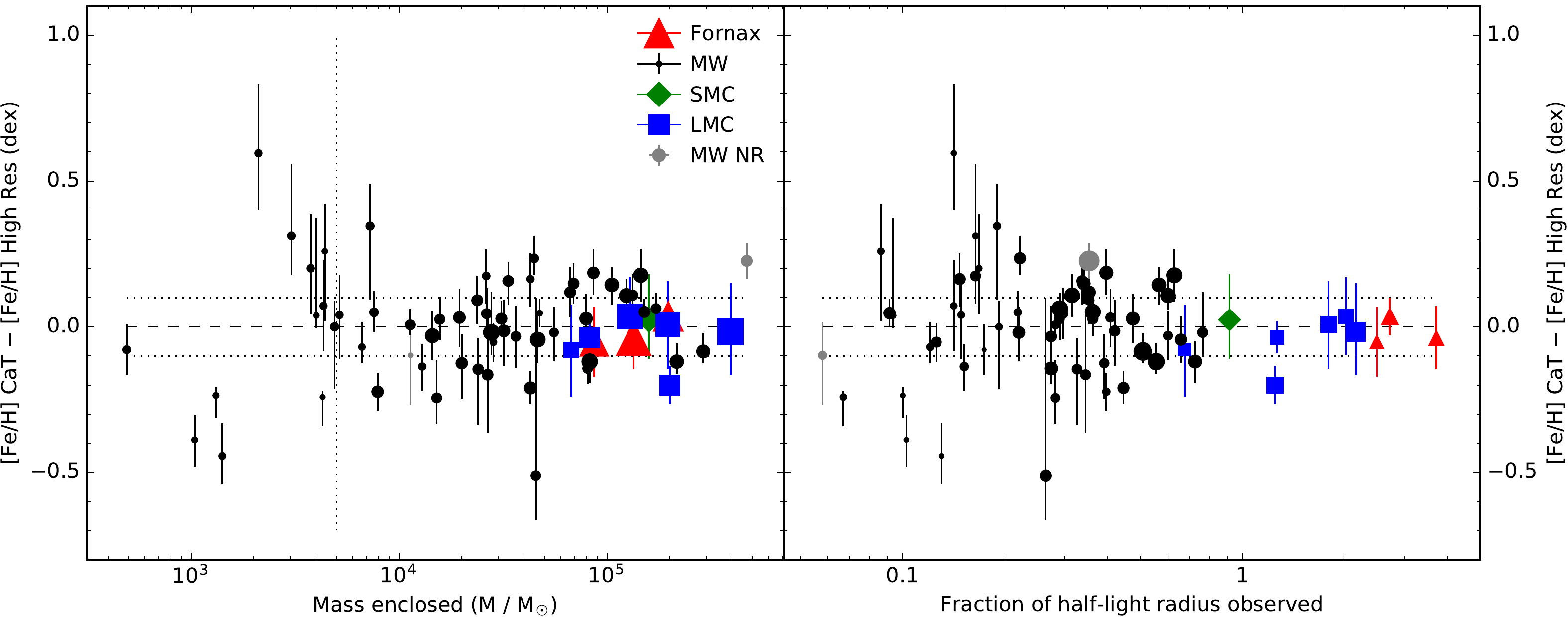}
\caption{Observational effects on the CaT measurement for old GCs.
\emph{Left} Difference between literature and our CaT based [Fe/H] as a function of the stellar mass within the field-of-view of WiFeS.
\emph{Right} Difference between literature and our CaT based [Fe/H] as a function of the fraction of the half-light radius observed with WiFeS.
The colours, shapes and size of the points are the same as in Figure \ref{fig:CaT_Fe_H} save for the left-hand panel where the point size is proportional to the fraction of the half-light radius within the field-of-view.
As in Figure \ref{fig:delta_Fe_H_Ca_Fe}, the dashed line shows $\Delta$[Fe/H] $= 0$ while the dotted lines show $\Delta$[Fe/H] $= \pm 0.1$.
For old GCs, the scatter in $\Delta$[Fe/H] increases below an enclosed mass of $\sim 10^{4}$ M$_{\sun}$.
For this reason we restrict our fits to GCs with more than $5 \times 10^{3}$ M$_{\sun}$ (the dotted vertical line).
$\Delta$[Fe/H] is largely independent of the fraction of the GC observed save for an increased scatter at small fractions due to the low enclosed mass.}
\label{fig:coverage_delta_fe_h}
\end{center}
\end{figure*}

In Figure \ref{fig:coverage_delta_fe_h}, we plot the difference between our CaT based [Fe/H] and that from literature as both a function of the mass within the observed field-of-view and as a function of the fraction of the half-light radius within the field-of-view.
The scatter in $\Delta$[Fe/H] is higher at low enclosed mass than at high enclosed mass as well as for GCs with poor radial coverage.
As noted earlier, to try to reduce stochastic effects of low stellar mass, we only include GCs more massive than $5 \times 10^{3}$ M$_{\sun}$ in our fits and comparisons.
This limit is broadly similar to the predicted mass for stochastic effects to dominate as calculated by \citet{2004A&A...413..145C} for the $I$-band.
We do not see any clear evidence for a relationship between the fraction of the half-light radius observed and the difference between the CaT based [Fe/H] and the literature values.
Kendall's correlation test finds no evidence for a correlation between either the observed mass ($\tau = 0.06$, $p = 0.49$) or the fraction of the half-light radius observed ($\tau = -0.05$, $p = 0.59$) and [Fe/H].
This indicates that the CaT value we measure from the centres are representative of the GCs overall.

Extragalactic observations of GCs should be less affected by both of these effects.
Due to their much larger distances, it is much easier to observe the entire extent of a GC (in M31 the median GC half light radius is 0.7 arcsec e.g. \citealt{2010MNRAS.402..803P}, at the distance of the Virgo cluster the median radius is 0.03 arcsec e.g. \citealt{2010ApJ...715.1419M}).
Additionally, extragalactic studies typically target GCs in the brighter half of the GC luminosity function and as such observe GCs more massive than $\sim 2 \times 10^{5}$ M$_{\sun}$\citep[e.g.][]{2007ApJS..171..101J, 2010ApJ...717..603V}.
While most spectroscopic observations of galaxies enclose enough mass for stochastic effects not to be an issue, care should be taken with observations of nearby low surface brightness galaxy light.
We encourage the effects of properly sampling all stellar evolutionary phase to be considered in both stellar population modelling and observations, especially when testing these methods on GCs in the Local Group.
Indeed, valuable insights into stellar populations can be gained by studying how the spectral energy distribution varies from spatial resolution element to element \citep[e.g.][]{2001MNRAS.320..193B, 2003ApJ...583..712J, 2014ApJ...797...56V, 2016ApJ...827....9C}.

\subsection{Effects of stellar mass function and dynamical evolution}
\label{sec:smf}

As mentioned earlier, GCs are mass-segregated, with higher mass stars being more centrally concentrated \citep[e.g.][]{1958ApJ...127..544S}.
GCs lose mass via tidal interactions and two body relaxation \citep[e.g.][]{2003MNRAS.340..227B} with stars at larger radii being preferentially lost.
Thus GCs more quickly lose their lower-mass stars giving the present-day mass function of GCs, both globally and in their centres, fewer low mass stars than their IMFs \citep[e.g.][]{2007ApJ...656L..65D, 2010AJ....139..476P, 2015MNRAS.453.3278W}.

Since stellar population models predict that the CaT is sensitive to the IMF \citep[e.g.][]{2003MNRAS.340.1317V, 2012ApJ...747...69C}, we must consider to what extent the dynamical evolution of GCs affects our measurements.
Using the stellar mass function slopes from \citet{2017MNRAS.471.3668S}, Kendall's correlation test gives little evidence for a correlation between the mass function slope $\alpha$ (defined such that a \citealt{1955ApJ...121..161S} IMF has a slope of $\alpha = -2.35$) and the difference between our CaT based [Ca/H] and literature [Ca/H] for the high mass, high resolution sample ($\tau = -0.29$, $p = 0.10$)
As for the relationship between [Ca/Fe] and CaT (Section \ref{sec:Ca_Fe}), we use \textsc{pymc3} to fit  linear relationship between the [Ca/H] difference as a function of $\alpha$ for the 17 GCs in our high mass, high resolution sample with mass function slope measurements.
We find a slope of $-0.225_{-0.060}^{+0.059}$ and a difference in the BIC of 12, both supporting significant relationship between the slope of the mass function and the strength of the CaT.
However, the sign of this relation is in the opposite direction to what is commonly expected for the behaviour of the CaT with the slope of the mass function (the CaT is usually expected to be stronger with a higher fraction of giants).

\begin{figure}
\begin{center}
\includegraphics[width=240pt]{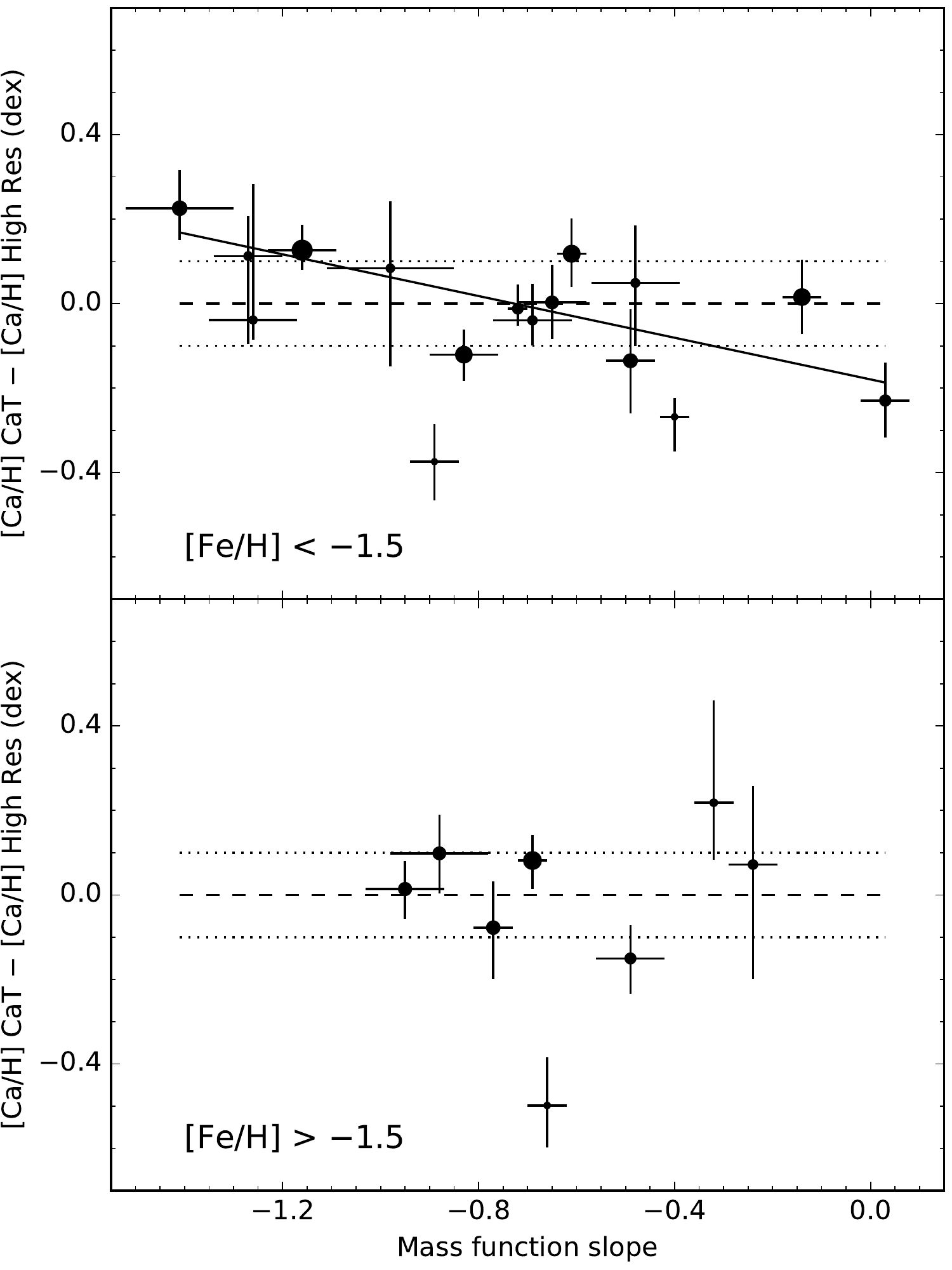}
\caption{Differences between literature and our CaT based [Ca/H] as a function of the stellar mass function slope from \citet{2017MNRAS.471.3668S}.
The top panel shows GCs with [Fe/H] $< -1.5$ while the bottom panel shows GCs with [Fe/H] $> -1.5$.
The black line in the top panel shows our fitted relationship between $\Delta$[Ca/H] and the mass function slope for the lowest metallicity GCs.
The colours, shapes and size of the points are the same as in Figure \ref{fig:CaT_Fe_H}.
As in Figure \ref{fig:delta_Fe_H_Ca_Fe}, the dashed line shows $\Delta$[Ca/H] $= 0$ while the dotted lines show $\Delta$[Ca/H] $= \pm 0.1$.
A \citet{1955ApJ...121..161S} IMF would have a slope of $\alpha = -2.35$.
The most metal poor GCs with steeper mass functions (more low mass stars) show higher CaT values than GCs with similar metallicities but flatter mass functions.
No relationship between CaT strength and mass function slope is seen for the more metal rich GCs though we note the most metal-rich GC in common between our sample and that of \citet{2017MNRAS.471.3668S} has [Fe/H] $= -1.19$ (NGC 1851).}
\label{fig:mass_func_ca_h}
\end{center}
\end{figure}

If we split the sample with mass function measurements up by metallicity, a more complicated picture emerges. 
In Figure \ref{fig:mass_func_ca_h}, we plot the difference between our CaT [Ca/H] and literature [Ca/Fe] measurements versus the stellar mass function slope from \citet{2017MNRAS.471.3668S} for GCs with [Fe/H] $< -1.5$ and $> - 1.5$.
We note that the most metal-rich GC in common with our high mass, high resolution sample and that of \citet{2017MNRAS.471.3668S} only has a metallicity of [Fe/H] $= -1.2$ (NGC 1851).
Using Kendall's correlation test, we see no evidence for an anticorrelation ($\tau = -0.31$, $p = 0.19$) in the 11 high enclosed-mass GCs in the low metallicity subsample nor in the higher metallicity subsample ($\tau = -0.20$, $p = 0.57$) which admittedly only contains 6 high-mass GCs.
Using \textsc{pymc3} to fit relations between $\alpha$ and the [Ca/H] difference we find a significant slope for the low-metallicity subsample ($-0.254_{-0.045}^{+0.045}$) but not for the higher-metallicity subsample ($0.06_{-0.40}^{+0.36}$).
Likewise, the difference in BIC favours a linear relationship over a constant value for the low-metallicity subsample ($\Delta$BIC $= 11.2$) but not for the higher-metallicity GCs ($\Delta$BIC $= -0.5$).
The correlation between $\alpha$ and the CaT strength seems to be driven by the lower-metallicity GCs.

Unfortunately, direct measurements of the stellar mass function are only available for a small subset of our GCs.
Since there is a strong anti-correlation between the half-mass radius relaxation time and the stellar mass function \citep{2017MNRAS.471.3668S}, we also consider how the CaT depends of the half-mass radius relaxation time.
Following \citet{1996AJ....112.1487H}, we calculate the half-mass radius relaxation time using equation 11 of \citet{1993ASPC...50..373D} from the mass and half light radii in Table \ref{tab:sample}.
For our entire high enclosed mass, high resolution abundances sample Kendall's correlation test finds a significant correlation between the log relaxation time and the [Ca/H] difference ($\tau = 0.26$, $p = 0.005$).
Using \textsc{pymc3}, we find a significant slope ($0.146_{-0.037}^{+0.036}$ for a linear log relaxation time-$\Delta$ [Ca/H] relation and strong evidence that a linear relation is favoured over a constant value ($\Delta$BIC $= 11.8$).
This behaviour is consistent with what is seen with the direct measurements of the mass function slope.

\begin{figure}
\begin{center}
\includegraphics[width=240pt]{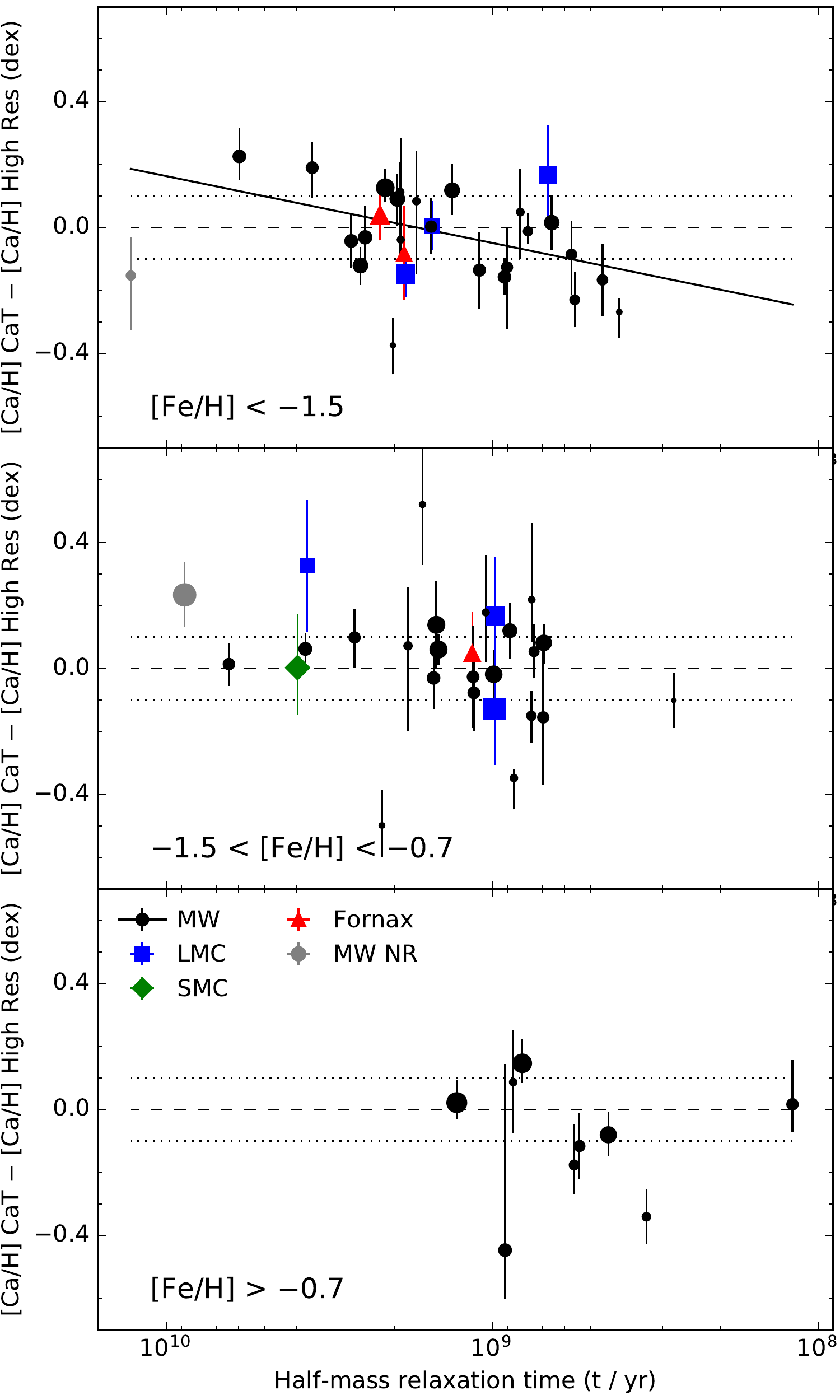}
\caption{Differences between literature and our CaT based [Ca/H] as a function of the half-mass radius relaxation time.
From top to bottom we show the difference for GCs with [Fe/H] $< -1.5$, $-1.5 <$ [Fe/H] $< -0.7$ and [Fe/H] $< -0.7$.
The black line in the top panel shows our fitted relationship between $\Delta$[Ca/H] and the relaxation time for the lowest metallicity GCs.
The colours, shapes and size of the points are the same as in Figure \ref{fig:CaT_Fe_H}.
Since the slope of the mass function and the relaxation time are anti-correlated, the relaxation time increases from right to left to match the trend seen between the CaT and the mass function slope.
As in Figure \ref{fig:delta_Fe_H_Ca_Fe}, the dashed line shows $\Delta$[Ca/H] $= 0$ while the dotted lines show $\Delta$[Ca/H] $= \pm 0.1$.
The lowest metallicity GCs with longer relaxation times show stronger CaT strengths than GCs with shorter relaxation times and the same metallicity.
However, intermediate and high metallicity GCs show no relationship between CaT strength and relaxation time at fixed metallicity.}
\label{fig:log_hrt_Ca_H}
\end{center}
\end{figure}

We break our high enclosed mass, high resolution abundances sample into three metallicity bins ([Fe/H] $< -1.5$, $-1.5 <$ [Fe/H] $< -0.7$ and [Fe/H] $> -0.7$) and plot the difference between CaT based and literature [Ca/H] as a function of half mass relaxation time for each bin in Figure \ref{fig:log_hrt_Ca_H}.
For the low metallicity bin, Kendall's test provides weak evidence ($\tau = 0.26$, $p = 0.07$) for a correlation.
Using \textsc{pymc3} we fit a significant slope ($0.202_{-0.058}^{+0.051}$) and find that a linear relationship is favoured over a constant value ($\Delta$BIC $= 9.2$) for 24 GCs in the low-metallicity bin.
The nuclear remnant NGC 5139 appears to be a significant outlier from the low-metallicity GC relationship.
For both the 20 GCs in the intermediate metallicity bin and the 8 GCs in the high metallicity bin, Kendall's test provides no evidence ($\tau = 0.18$, $p = 0.27$ and $\tau = 0.00$, $p = 1.0$ respectively) for a correlation.
The \textsc{pymc3} fits finds an insignificant slope for the intermediate metallicity subsample ($0.035_{-0.60}^{+0.061}$) but a marginally significant slope for the metal rich sub sample ($0.226_{-0.103}^{+0.113}$).
Neither subsample shows strong evidence that a linear relationship is preferred over a constant value ($\Delta$BIC $= -5.7$ and $2.3$ respectively).
A significant relationship between the CaT and the half mass relaxation time only at low metallicity is consistent with the behaviour of the CaT with the slope of the stellar mass function.
For both the mass function slope relation and for the relaxation time relation, the relationship for all GCs in our sample is driven by the low-metallicity GCs.

The observed qualitative behaviour of the CaT with the stellar mass function slope and metallicity is consistent with the predictions of stellar population synthesis models.
The \citet{2003MNRAS.340.1317V, 2012MNRAS.424..157V} models predict that the slope of the relationship between the CaT strength and the slope of the stellar mass function as a strong function of metallicity, with a strong anti-correlation between the two at low metallicities, little or no mass function dependence at intermediate metallicities and a strong correlation at solar metallicities.
In the models of \citet{2000ApJ...532..453S}, the CaT shows no dependence on the mass function slope at subsolar metallicities ([Fe/H] $= -0.5$) but a weak correlation at supersolar metallicities.
We note that most of the work on using the CaT as an IMF indicator \citep[e.g.][]{2012ApJ...747...69C, 2013MNRAS.433.3017L} has focused on metallicities close to solar.
These models predict that the CaT responds more weakly to the mass function slope for models shallower than that of a \citet[$\alpha = -2.35$]{1955ApJ...121..161S} IMF although we note that none makes predictions for mass functions as shallow as observed in some MW GCs ($\alpha \sim 0$, e.g. \citealt{2010AJ....139..476P, 2017MNRAS.471.3668S}).
The predicted anti-correlation at low metallicities and the lack of significant correlation at intermediate metallicities is consistent with our observations.

The lack of an observed correlation between the mass function and the strength of the CaT at near-solar metallicity is likely due to the small number of high-metallicity GCs in our sample (only NGC 6528 and NGC 6553 are close to solar) and the challenges associated with reliably measuring both the integrated CaT strengths and the chemical abundances of these GCs.
We note that the metal-rich GCs have on average shorter relaxation times than the metal-poor GCs due to their smaller mean half-light radii.
The lack of a significant relationship between the slope of the stellar mass function and the CaT for most metallicities may explain why we did not find a significant dependence on the fraction of half light radius observed (Section \ref{sec:stochastic}).

Dynamical effects should be less serious for extragalactic studies as they typically target the most massive GCs that have longer relaxation times.
Galaxies are unaffected by internal dynamical effects due to their much longer relaxation time scales.
However, the effects of a varying IMF remain a concern when using the CaT to measure the metallicity of galaxies although the IMF likely only varies significantly in the centres of some massive galaxies \citep[e.g.][]{2015MNRAS.447.1033M, 2017ApJ...841...68V}.
We note that almost all of the GCs in our sample show mass functions shallower than the low mass slope of a \citet{2001MNRAS.322..231K} IMF much less the slope of a \citet{1955ApJ...121..161S} IMF so that the observed behaviour of the CaT with mass function slope in GCs should not be blindly extended to galaxies.
In summary we find evidence for a weak dependence of the CaT on the mass function slope only at low metallicities.

\subsection{Comparison with stellar population models}
\begin{figure*}
\begin{center}
\includegraphics[width=504pt]{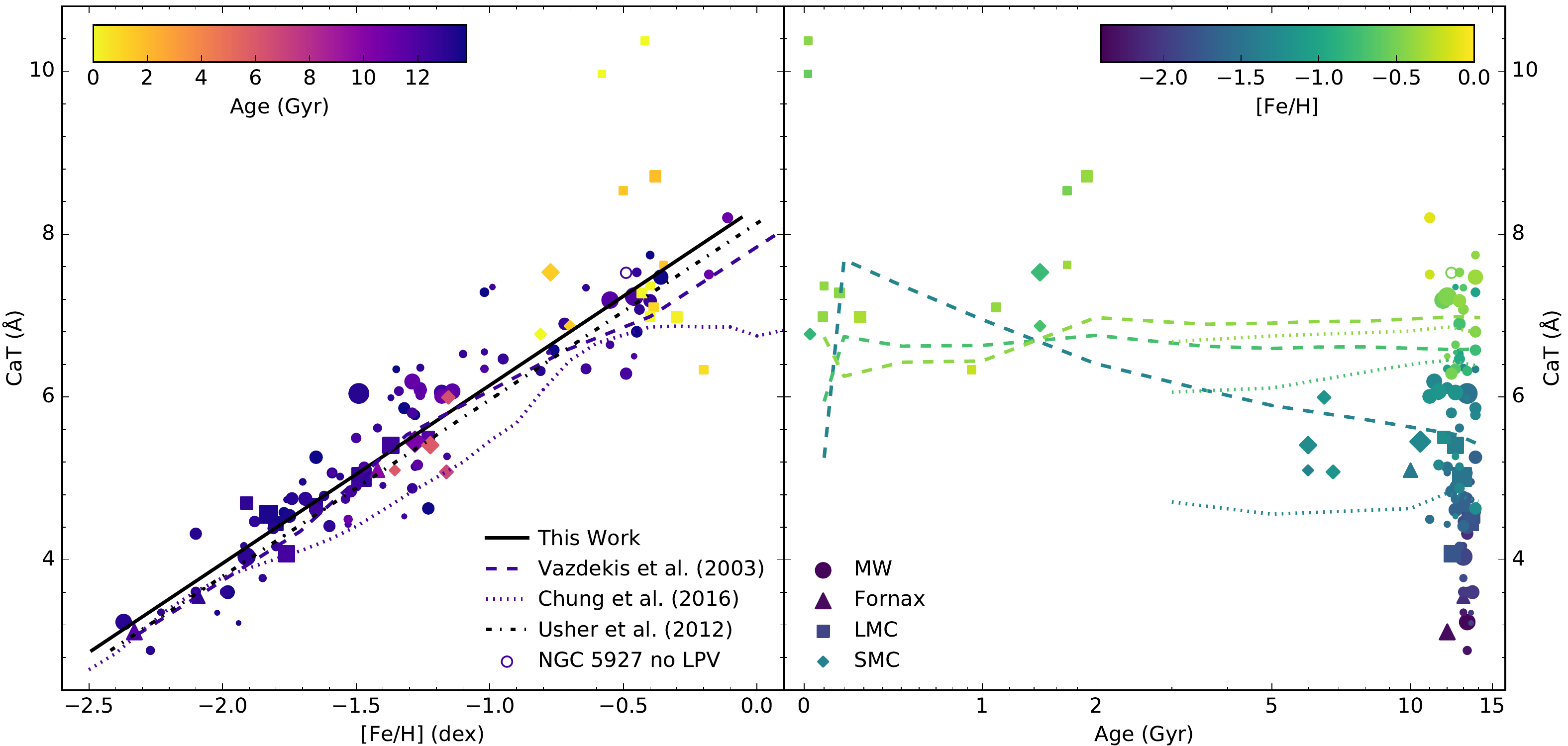}
\caption{Comparison of CaT strengths with SSP model predictions.
\emph{Left} CaT strength as a function of metallicity.
The points show our measurements colour coded by age with sizes proportional to the mass enclosed within the WiFeS field-of-view.
The dashed line is the 12.6 Gyr model predictions of \citet{2003MNRAS.340.1317V} the dotted line is the 12.5 Gyr model predictions of \citet{2016ApJ...818..201C}, the black dash-dotted line is the relationship adopted by \citet{2012MNRAS.426.1475U} based on the \citet{2003MNRAS.340.1317V} models and the black solid line is the empirical relationship found in this work (Equation \ref{eq:CaT_Fe_H}).
\emph{Right} CaT strength as a function of age.
The points and models are colour coded by metallicity.
The dashed lines show the \citet{2003MNRAS.340.1317V} models and the dotted lines show the \citet{2016ApJ...818..201C} models for metallicites of [Fe/H] $= -1.3$, $-0.7$ and $-0.4$ and a range of ages.
In both panels, the empty circle is the CaT measurement of NGC 5927 with the long period variable NGC 5927 V3 removed from the extracted spectra. }
\label{fig:CaT_all_models}
\end{center}
\end{figure*}

In Figure \ref{fig:CaT_all_models}, we also plot the predictions of version 11.0 of the \citet{2003MNRAS.340.1317V, 2010MNRAS.404.1639V} models using the \citet{2000A&AS..141..371G} isochrones.
For these models, we downloaded the predicted spectral energy distribution and measured the CaT strengths using the same techniques and code as for our spectra.
We note that spectral resolutions of the Vazdekis et al. models ($\delta \lambda / \lambda = 5700$) and our observations ($\delta \lambda / \lambda = 6800$) are quite similar so no correction is required to compare them. 
For old GCs, there is good agreement between the \citet{2003MNRAS.340.1317V, 2010MNRAS.404.1639V} models and observations at the $\sim 0.1$ dex level.
These models predict the observed weak evolution in CaT strength between $\sim 13$ and 6 Gyr but under predict the CaT strength at 2 Gyr and younger.
We also show in the left-hand panel of Figure \ref{fig:CaT_all_models} the CaT-metallicity relationship used by \citet{2012MNRAS.426.1475U} which is also based on the Vazdekis et al. models.
For extragalactic GCs, \citet{2012MNRAS.426.1475U} found good agreement between their metallicities measured using this relation and metallicities measured using Lick indices \citep{1994ApJS...94..687W, 1997ApJS..111..377W} and stellar population models such as those of \citet{2003MNRAS.339..897T}.
Unsurprisingly, the \citet{2012MNRAS.426.1475U} relation shows good agreement with our observations of Local Group GCs.

We also plot the model predictions of \citet{2016ApJ...818..201C}.
We use their predictions for the index definition of \citet{2010AJ....139.1566F} using the stellar library of \citet{2001MNRAS.326..959C}.
For old GCs, the \citet{2016ApJ...818..201C} models predict lower CaT values at higher metallicities which is likely due to the puzzling choice of \citet{2016ApJ...818..201C} to smooth their models to the low spectral resolution ($\delta \lambda / \lambda \lesssim 1000$) of the Lick index system \citep{1994ApJS...94..687W} prior to index measurement.
The lack of significant CaT strength variation with age is in line with the predictions of the models of both \citet{2012ApJ...747...69C} and \citet{2016ApJ...818..201C}.

\section{Conclusion}
\label{sec:conclusion}

Using spectra from the WAGGS project (\citealt{2017MNRAS.468.3828U}, Section \ref{sec:sample}), we have measured the strength of the near infrared CaT spectral feature in a sample of 113 GCs with a wide range of ages and metallicities in the MW and its satellite galaxies (Section \ref{sec:measurement}).
After refining the template-based measurement technique of \citet{2010AJ....139.1566F} and \citet{2012MNRAS.426.1475U}, we studied the systematic effects affecting our measurement of the CaT (Section \ref{sec:systematics}).
We find that masking sky lines in the template fitting process can introduce a strong bias when the radial velocity of the GC causes the CaT lines to overlap with the sky line mask (Figures \ref{fig:mask_on_off} and \ref{fig:rv_CaT_modelling}).
This effect is more important in the Local Group than in more distant galaxies because of their recessional velocities.
We also find that velocity dispersion or spectral resolution affects the CaT measurement process with our CaT measurements being less reliable above a velocity dispersion of 100 km s$^{-1}$ (or below a spectral resolution of $R = 1200$, Figure \ref{fig:sigma_CaT_modelling}).
The effects of velocity dispersion are stronger for metal-rich GCs but are unimportant for the current work given the relatively low velocity dispersions of GCs (1 to 20 km s$^{-1}$ e.g. \citealt{2018MNRAS.478.1520B}) and the intermediate resolution of our spectra ($R = 6800$).
In line with previous work \citep{2012MNRAS.426.1475U}, we find that our CaT measurements are biased to higher values at low metallicity and low S/N.
While a concern in extragalactic studies, this effect is not a concern for this study due to our generally high S/N (90 \% of our spectra have S/N $> 22$ \AA $^{-1}$, 50 \% S/N $> 116$ \AA $^{-1}$).

We also measured the CaT using classical spectral indices (Section \ref{sec:classic}).
While below a metallicity of [Fe/H] $= -1$ the CaT strength measured using the classical indices closely traces our template-based measurements, above this metallicity the classical indices become less sensitive to metallicity.
Classical index measurements also show a stronger dependence on age than template-based measurements.
Based on these tests, it is clear that the way the CaT is measured affects its behaviour with metallicity and age.
We strongly encourage others wanting to use the CaT to measure metallicity to download our spectra (available from \url{http://www.astro.ljmu.ac.uk/~astcushe/waggs/data.html}) to test their CaT measurement process and their adopted CaT-metallicity relation. 
We conclude that our template-based CaT measurement technique is the preferable method for measuring the metallicities of extragalactic GCs as it shows higher sensitivity to metallicity at high metallicity, less dependence on age, lower metallicity uncertainties at fixed S/N and less sensitivity to sky subtraction residuals.

We estimated the uncertainties of our measurements using both a Monte Carlo technique based on the uncertainties provided by the data reduction pipeline and by performing a block bootstrap on our datacubes (Section \ref{sec:repeats}).
At high stellar masses ($\gtrsim 10^{4}$ M$_{\sun}$) within the observed field-of-view, there is good agreement between repeated observations using the bootstrap-based uncertainties while the pipeline-based uncertainties appear to underpredict the differences between observations.
At low stellar observed masses ($\lesssim 10^{3}$ M$_{\sun}$), there is significant variation between different pointings of the same GCs.
Taken together, this demonstrates that stochastic variations in the number of bright giant stars is the dominant source of uncertainty for most of our CaT measurements.

In Section \ref{sec:CaT_Z} we derived a relationship between CaT strength and [Fe/H] using our measurements and iron abundances from high resolution spectroscopy of individual RGB stars (Figure \ref{fig:CaT_Fe_H}, Equation \ref{eq:CaT_Fe_H}).
We find that most of the residuals in this relation can be explained by the [Fe/Ca] ratio (Figure \ref{fig:delta_Fe_H_Ca_Fe}) with the relationship between [Ca/H] and CaT strength showing less scatter than the one between [Fe/H] and the CaT (Figure \ref{fig:CaT_Ca_H}, Equation \ref{eq:CaT_Ca_H}).
Together with previous work \citep{2012MNRAS.426.1475U, 2015MNRAS.446..369U, 2016MNRAS.456..831S}, the CaT appears to be a reliable metallicity indicator up to solar metallicity.

We find that our CaT-based metallicity measurements are insensitive to age for ages older than 6 Gyr (Figure \ref{fig:CaT_PaT}, Section \ref{sec:age}).
Younger than 6 Gyr, the effects of age on the CaT based metallicity are relatively small ($\lesssim 0.3$ dex in [Fe/H], Figure \ref{fig:delta_Fe_H_age}) except when there is a substantial contribution of light from C stars or red supergiants.
The observed behaviour of the CaT with age and metallicity is generally consistent with the predictions of the \citet{2003MNRAS.340.1317V, 2012MNRAS.424..157V} stellar population models but not the \citet{2016ApJ...818..201C} models.
Given the CaT's relatively low sensitivity to age and since our observations of $\omega$ Cen and M54 show that the CaT can be used to measure the mean metallicities of populations with significant metallicity spreads, the CaT can be reliably used to measure the mean metallicity of galaxy light.

Horizontal branch morphology has no effect on the CaT (Figure \ref{fig:milone_hb}, Section \ref{sec:hb}) implying that the CaT does not vary strongly with the size of the spread in light element abundances (multiple populations) in GCs.
We find no significant aperture bias in our observations but find that our CaT measurements are less reliable when there is a low amount of stellar mass in our observed field-of-view (Figure \ref{fig:coverage_delta_fe_h}, Section \ref{sec:stochastic}).
The effects of the stellar mass function on the CaT strength are more complicated (Section \ref{sec:smf}).
Below a metallicity of [Fe/H] $= -1.5$, the CaT strength decreases with the slope of the stellar mass function while at intermediate metallicities, the CaT strength is independent of the mass function slope (Figure \ref{fig:mass_func_ca_h}).
At metallicities close to solar, we do not have enough GCs in our sample with abundances from high resolution spectroscopy to properly assess what effect the mass function slope has on the CaT.
That the CaT is stronger in dwarf star rich stellar populations at low metallicity but not at intermediate metallicities is consistent with the predictions of stellar population models.

We note that our results are only for GCs in one galaxy and its satellites.
GCs in other galaxies or integrated galaxy light may have different chemistries or dynamical states at a given metallicity or age and thus slightly different CaT values for the same metallicity.
For populations with similar chemistry to the GCs in the MW and its satellite galaxies, we find that the CaT can be used to measure [Ca/H] reliably at the 0.1 dex level from intermediate resolution integrated spectra of stellar populations older than 3 Gyr.
At younger ages the CaT remains sensitive to metallicity but appears to be biased to higher [Ca/H] values.
The CaT is less reliable at high metallicities and low spectral resolutions/high velocity dispersions.

\section*{Acknowledgements}
The authors wish to thank the referee for their useful comments and suggestions which helped to improve this paper. 
We wish to thank Nate Bastian, Elena Pancino, Joel Pfeffer, Ted Mackereth and Lee Kelvin for helpful discussions and useful suggestions.
We wish to thank the staff at Siding Spring Observatory for their assistance with our observations with the ANU 2.3 m telescope and the WiFeS instrument.
C.U. and S.K. gratefully acknowledges financial support from the European Research Council (ERC-CoG-646928, Multi-Pop).
P.C. acknowledges the support provided by FONDECYT postdoctoral research grant no 3160375.
S.B. acknowledges the support of the AAO PhD Topup Scholarship.
This work was partially performed on the swinSTAR supercomputer at Swinburne University of Technology.
Part of this research was conducted by the Australian Research Council Centre of Excellence for All Sky Astrophysics in 3 Dimensions (ASTRO 3D), through project number CE170100013.
Based on data products from observations made with ESO Telescopes at the La Silla Paranal Observatory under programme ID 188.B-3002.

This work made use of \textsc{numpy} \citep{numpy}, \textsc{scipy} \citep{scipy}, and \textsc{matplotlib} \citep{matplotlib} as well as \textsc{astropy}, a community-developed core Python package for astronomy \citep{2013A&A...558A..33A}.
Additionally, this work made use of \textsc{TOPCAT} \citep{2005ASPC..347...29T} and \textsc{Aladin} \citep{2000A&AS..143...33B}.
This research has made use of the NASA/IPAC Extragalactic Database (NED), which is operated by the Jet Propulsion Laboratory, California Institute of Technology, under contract with the National Aeronautics and Space Administration.
This research has made use of the SIMBAD database,
operated at CDS, Strasbourg, France.

\bibliographystyle{mnras}
\bibliography{bib}{}

\appendix

\end{document}